\documentclass[12pt]{article}

\usepackage{graphicx}
\usepackage{amsfonts}

\def\gtwid{\mathrel{\raise.3ex\hbox{$>$\kern-.75em\lower1ex\hbox{$\sim$}}}}
\def\ltwid{\mathrel{\raise.3ex\hbox{$<$\kern-.75em\lower1ex\hbox{$\sim$}}}}
\def\square{\kern1pt\vbox{\hrule height 1.2pt\hbox{\vrule width 1.2pt\hskip 3pt
   \vbox{\vskip 6pt}\hskip 3pt\vrule width 0.6pt}\hrule height 0.6pt}\kern1pt}

\begin{document}

\begin{titlepage}

\begin{flushright}
UFIFT-QG-15-02
\end{flushright}

\vskip .5cm

\begin{center}
{\bf Graviton Loop Corrections to Vacuum Polarization in de Sitter in a General Covariant Gauge}
\end{center}

\vskip .5cm

\begin{center}
D. Glavan$^{1*}$, S. P. Miao$^{2\star}$, Tomislav Prokopec$^{1\dagger}$ 
and R. P. Woodard$^{3\ddagger}$
\end{center}

\vskip .5cm

\begin{center}
\it{$^{1}$ Institute for Theoretical Physics \& Spinoza Institute \\
Center for Extreme Matter and Emergent Phenomena \\
Utrecht University, Postbus 80195, 3508 TD Utrecht \\
THE NETHERLANDS}
\end{center}

\begin{center}
\it{$^{2}$ Department of Physics, National Cheng Kung University \\
No. 1, University Road, Tainan City 70101, TAIWAN}
\end{center}

\begin{center}
\it{$^{3}$ Department of Physics, University of Florida,\\
Gainesville, FL 32611, UNITED STATES}
\end{center}

\begin{center}
ABSTRACT
\end{center}

We evaluate the one-graviton loop contribution to the vacuum
polarization on de Sitter background in a 1-parameter family 
of exact, de Sitter invariant gauges. Our result is computed
using dimensional regularization and fully renormalized with
BPHZ counterterms, which must include a noninvariant owing to
the time-ordered interactions. Because the graviton propagator 
engenders a physical breaking of de Sitter invariance two 
structure functions are needed to express the result. In 
addition to its relevance for the gauge issue this is the 
first time a covariant gauge graviton propagator has been used 
to compute a noncoincident loop. A number of identities are 
derived which should facilitate further graviton loop 
computations.

\begin{flushleft}
\end{flushleft}

\noindent {\it Dedicated to Stanley Deser on the occasion of his
84th birthday.}

\vskip .5cm

\begin{flushleft}
$^{*}$ e-mail: D.Glavan@uu.nl \\
$^{\star}$ e-mail: spmiao5@mail.ncku.edu.tw \\
$^{\dagger}$ e-mail: T.Prokopec@uu.nl \\
$^{\ddagger}$ e-mail: woodard@phys.ufl.edu
\end{flushleft}

\end{titlepage}

\section{Introduction}\label{intro}

Observational evidence from cosmology has confronted theorists
with three crucial questions:
\begin{itemize}
\item{What drove primordial inflation?}
\item{What is causing the current phase of acceleration?}
\item{How can we define cosmological observables which are gauge 
independent, infrared finite, and BPHZ renormalizable (Bogoliubov,
Parasiuk, Hepp and Zimmerman \cite{Bogoliubov:1957gp,Hepp:1966eg,
Zimmermann:1968mu,Zimmermann:1969jj}) in the sense of low energy
effective field theory \cite{Donoghue:1993eb,Donoghue:1994dn}?}
\end{itemize}
This paper concerns the third question, whose answer in flat space would
be either the S-matrix (when only massive particles are present) or 
else inclusive rates and cross sections (when massless particles occur). 
Although a formal S-matrix can be constructed for massive scalars on de 
Sitter background \cite{Marolf:2012kh}, the causal structure of this
geometry precludes local observers from measuring this quantity, so 
it cannot serve as an observable. To the tree order accuracy which has 
so far been resolved \cite{Ade:2013zuv}, the primordial scalar power 
spectrum $\Delta^2_{\mathcal{R}}(k)$ seems to be well-represented by a 
2-point quantum gravitational correlation function. This correlator can 
be given a gauge independent expression at tree order 
\cite{Bardeen:1980kt,Bardeen:1983qw}, but no local extension of it can 
be gauge independent at higher orders \cite{Tsamis:1989yu}. Hence 
dependence upon the gravitational gauge has emerged as a central issue 
in loop corrections to cosmological observables \cite{Unruh:1998ic,
Abramo:2001db,Abramo:2001dc,Geshnizjani:2002wp,Garriga:2007zk,
Tsamis:2008zz,Urakawa:2010kr,Gerstenlauer:2011ti,Tanaka:2011aj,
Chialva:2011bg,Urakawa:2011fg,Giddings:2011zd,Marozzi:2011zb,
Giddings:2011ze,Miao:2012xc,Prokopec:2012ug,Tanaka:2012wi,
Marozzi:2012tp,Tanaka:2013xe,Prokopec:2013zya,Tanaka:2013caa,
Tsamis:2013cka,Tanaka:2014ina,Marozzi:2014xma}.

%\vspace{1cm} 
\begin{figure}[ht]
		\center
		\includegraphics[]{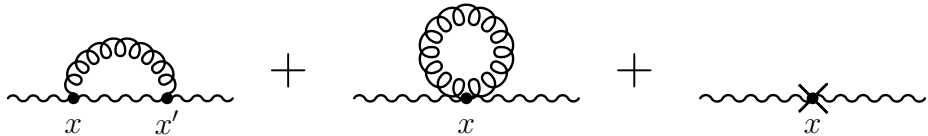}
%		\includegraphics{diagrams}
%\vspace{-2cm}
\caption{\label{fig:photon} Feynman diagrams relevant to the
one loop vacuum polarization from gravitons. Wavy lines are photons
and curly lines are gravitons.}
\label{vacpolgraphs}
\end{figure}

The aim of this paper is to explore gauge dependence in the one 
graviton loop correction to the vacuum polarization $i[\mbox{}^{\mu}
\Pi^{\nu}](x;x')$ on de Sitter background. This quantity, whose 
diagrammatic depiction is given in Fig. \ref{vacpolgraphs}, explores 
the same issues of gauge dependence as $\Delta^2_{\mathcal{R}}(k)$,
but is quite a bit simpler to compute and renormalize. It can be used
to quantum-correct Maxwell's equation \cite{Prokopec:2003bx},
\begin{equation}
\partial_{\nu} \Bigl[ \sqrt{-g} \, g^{\nu\rho} g^{\mu\sigma} 
F_{\rho\sigma}(x) \Bigr] + \int \!\! d^4x' \Bigl[\mbox{}^{\mu}
\Pi^{\nu}\Bigr](x;x') A_{\nu}(x') = J^{\mu}(x) \; , \label{maxeqn}
\end{equation}
where $A_{\mu}(x)$ is the vector potential, $F_{\mu\nu} \equiv 
\partial_{\mu} A_{\nu} - \partial_{\nu} A_{\mu}$ and $J^{\mu}(x)$ is
the current density. Quantum corrections to dynamical photons emerge 
from solutions with $J^{\mu}(x) = 0$, whereas quantum corrections to 
electrodynamic forces derive from the response to nonzero $J^{\mu}(x)$.

We define the graviton field operator $h_{\mu\nu}(x)$ by subtracting 
the de Sitter background $g_{\mu\nu}(x)$ from the full metric field
operator $\mathbf{g}_{\mu\nu}$,
\begin{equation}
\mathbf{g}_{\mu\nu}(x) \equiv g_{\mu\nu}(x) + \kappa h_{\mu\nu}(x) \qquad 
, \qquad \kappa^2 \equiv 16 \pi G \; . \label{hdef}
\end{equation}
We shall compute the vacuum polarization in the 1-parameter family of
exact covariant gauges that is a generalization of the de Donder condition,
\begin{equation}
g^{\rho\sigma} \Bigl[D_{\rho} h_{\sigma \mu} + \frac{b}2 D_{\mu} 
h_{\rho\sigma} \Bigr] = 0 \qquad , \qquad b > 2 \; , \label{gauge}
\end{equation}
where $D_{\mu}$ stands for the covariant derivative operator in de Sitter
background, treating $h_{\mu\nu}$ as a 2nd rank tensor. In gauges of the 
sort (\ref{gauge}) the graviton propagator breaks up into a 
trans\-verse-traceless ``spin two'' part, which does not depend upon $b$,
and a $b$-dependent, ``spin zero'' part \cite{Miao:2011fc,
Mora:2012zi},\footnote{Because covariant gauge propagators derive from 
both constrained and dynamical fields, the spin two part of (\ref{types})
--- whose zero components do not vanish --- actually comes from both the 
$\frac12 (D-3) D$ dynamical gravitons and $D-1$ of the constrained fields. 
The spin zero part derives from the remaining constrained field.} 
\begin{equation}
i\Bigl[\mbox{}_{\alpha\beta} \Delta_{\gamma\delta}\Bigr](x;x') =
i\Bigl[\mbox{}_{\alpha\beta} \Delta^2_{\gamma\delta}\Bigr](x;x') +
i\Bigl[\mbox{}_{\alpha\beta} \Delta^0_{\gamma\delta}\Bigr](x;x') \; .
\label{types}
\end{equation}
The graviton contribution to the vacuum polarization on de Sitter must 
depend upon $b$ because its flat space limit takes the form
\cite{Leonard:2012fs},
\begin{equation}
i\Bigl[\mbox{}^{\mu} \Pi^{\nu}_{\rm flat}\Bigr](x;x') = \frac{\kappa^2 
\Gamma(\frac{D}2) \Gamma(\frac{D}2 \!-\! 1)}{16 (D \!-\! 1) \pi^D}
\Bigl[ C_2 + C_0(b)\Bigr] \Bigl[ \eta^{\mu\nu} \partial' \!\cdot\!
\partial \!-\! \partial^{\prime \mu} \partial^{\nu} \Bigr] 
\frac1{\Delta x^{2D - 2}} \; . \label{flatvacpol} 
\end{equation}
Here $\eta^{\mu\nu}$ is the Lorentz metric and $\Delta x^2 \equiv 
\eta_{\mu\nu} (x-x')^{\mu} (x-x')^{\nu}$. The gauge independent, spin 
two term contributes a coefficient which vanishes in $D=4$ spacetime 
dimensions,
\begin{equation}
C_2 = \frac{(D \!-\! 4) (D \!-\! 2)^2 (D \!+\! 1) (D\!+\! 2)}{4 
(D \!-\! 1)} \; , \label{spin2C}
\end{equation}
and the gauge dependent, spin zero coefficient is,
\begin{equation}
C_0(b) = -\frac14 (D\!-\!2)^2 \Biggl[ \Bigl( \frac{D b \!-\! 2}{b 
\!-\! 2}\Bigr)^2 \!-\! 4 \Bigl( \frac{D \!-\! 4}{D \!-\! 2}\Bigr)
\Bigl( \frac{D b \!-\! 2}{b \!-\! 2}\Bigr) \!+\! \frac{2 (D \!-\!
4)^2}{(D \!-\! 2)(D \!-\! 1)} \Biggr] \; . \label{spin0C}
\end{equation}
Note that $C_0(b)$ is negative semi-definite for $D=4$.

It is commonplace to dismiss as unphysical any gauge dependent 
quantity like (\ref{flatvacpol}) \cite{Garriga:2007zk}; however, 
that view is simplistic. The gauge-independent S-matrix of flat 
space arises from combining gauge-dependent Green's functions, so 
the latter must possess legitimate physical information mixed in 
with artefacts of gauge fixing \cite{Tsamis:2008zz}. From the manner 
in which the flat S-matrix is constructed \cite{Itzykson:1980rh}, 
one realizes that this physical information is distinguished by 
possessing momentum space poles on each external leg. In position 
space these poles correspond to secular growth when the Green's 
function is integrated against tree order mode functions, as 
$[\mbox{}^{\mu} \Pi^{\nu}](x;x')$ necessarily is in the perturbative 
solution of (\ref{maxeqn}) for dynamical photons. It is therefore 
reasonable to expect that the leading secular growth factors of de 
Sitter Green's functions might be gauge independent 
\cite{Miao:2012xc}. That is what we seek to check for 
$[\mbox{}^{\mu} \Pi^{\nu}](x;x')$. We will of course be able to 
compare our results for different values of $b > 2$. We can also 
compare with the result previously obtained \cite{Leonard:2013xsa} 
in a noncovariant gauge \cite{Tsamis:1992xa,Woodard:2004ut}.

Some comments on the physics are worthwhile before commencing
this difficult computation. One might think an uncharged field like 
$h_{\mu\nu}(x)$ is not capable of contributing to vacuum
polarization but this ignores the role of electric and magnetic
fields in transferring momentum. Virtual gravitons which interact
with photons --- either real or virtual ones --- can alter this
momentum. There is no change in how dynamical photons propagate on
flat space background \cite{Leonard:2012fs}, essentially because 
virtual gravitons affect a single photon the same way throughout
space and time. However, the interaction between charged particles
on flat space background is slightly strengthened at short 
distances, as can be inferred from the gauge independent scattering 
amplitude \cite{BjerrumBohr:2002sx}. One way to understand this 
effect is that the virtual photons which transfer momentum between 
nearby charges do not survive long enough to experience the full 
effect of buffeting by the longest wave length (and hence longest 
lived) virtual gravitons.

On de Sitter background the vacuum polarization must of course 
show the same strengthening of force at short distance that is
encoded in its flat space limit \cite{Leonard:2012fs,
BjerrumBohr:2002sx}. However, it should also manifest new,
secular effects arising from the inflationary production of 
gravitons. For example, a tree order photon redshifts as it 
propagates, whereas the continual replenishment of Hubble-scale 
gravitons should lead to a relative one loop enhancement, as 
momentum tends to flow from inflationary gravitons into the ever
weaker photon. Similarly, the force between widely separated
sources should be relatively enhanced because the highly 
infrared virtual photons which mediate the force are more likely 
to acquire momentum from, rather than lose it to, the constant 
pool of Hubble-scale gravitons. Both effects have been seen 
\cite{Glavan:2013jca,Wang:2014tza} when the noncovariant gauge
$[\mbox{}^{\mu} \Pi^{\nu}](x;x')$ \cite{Leonard:2013xsa} is used 
in equation (\ref{maxeqn}), and it will be fascinating to learn
what happens in our covariant gauge (\ref{gauge}).

The one-loop effects of inflationary gravitons in the noncovariant 
gauge \cite{Tsamis:1992xa,Woodard:2004ut} have been studied for 
a variety of other particles over the years:
\begin{itemize}
\item{The graviton self-energy has been computed 
\cite{Tsamis:1996qk}, but has not yet been used to quantum-correct
the linearized Einstein equation. However, the Hartree approximation
has been used to show that the Weyl curvature of dynamical gravitons
experiences a secular enhancement \cite{Mora:2013ypa}.}
\item{The self-energy has been computed for massless fermions
\cite{Miao:2005am} and, to first order in the mass, for massive
fermions \cite{Miao:2012bj}. Quantum correcting the Dirac equation
reveals a secular enhancement of the field strength of massless
fermions \cite{Miao:2006gj,Miao:2007az}. The result for the 
massive case has not yet been derived.}
\item{The self-mass of massless, minimally coupled scalars has
been computed \cite{Kahya:2007bc}. However, quantum-correcting the
Klein-Gordon equation shows no secular enhancement of the scalar 
field strength \cite{Kahya:2007cm}.}
\item{A partial result has recently been obtained for the self-mass
of massless, conformally coupled scalars \cite{Boran:2014xpa}, but 
it has so far not been used to quantum-correct the linearized field
equation.}
\end{itemize}
The fact that inflationary gravitons give secular enhancements to 
the field strengths of massless fermions, gravitons and photons, 
but not to massless, minimally coupled scalars, seems to be due to 
spin \cite{Miao:2008sp}. In each case energy and momentum of the
physical particle under study redshifts as it propagates. If 
inflationary gravitons can only interact with the particle through
its redshifting energy-momentum then the interaction cuts off 
rapidly and there can be no growing effect at late times. The
presence of spin gives rise to a new interaction which does not
cut off, so that particles with spin are scattered more and more
as they propagate further through the sea of inflationary 
gravitons. Any other interaction which persists to late times 
should give rise to the same sort of secular enhancement, which 
is why it will be fascinating to see what happens to slightly
massive fermions and to conformally coupled scalars.

This paper consists of five sections, of which the first is this 
Introduction. In section \ref{prelim} we give those reductions of
the diagrams in Fig. \ref{vacpolgraphs} which do not depend upon
the form of the graviton propagator. Section \ref{spin2} derives
the contribution from the spin two part of the graviton propagator,
and section \ref{spin0} computes the contribution from the spin
zero part. Our conclusions are presented in section \ref{discuss}.

\section{Preliminary Reductions}\label{prelim}

The purpose of this section is to describe those parts of the 
computation which do not require a specific form for the graviton
propagator. We begin by expressing the two primitive diagrams of
Fig. \ref{vacpolgraphs} in terms of propagators and vertices. We
next point out that the form of these expressions lends itself to
a simple representation for $i[\mbox{}^{\mu} \Pi^{\nu}](x;x')$ as
the sum of two tensor differential operators acting on structure
functions. We express the BPHZ counterterms (the third diagram of
Fig. \ref{vacpolgraphs}) directly in terms of their contributions 
to these structure functions. The section closes with the 
derivation of an important identity concerning the photon 
propagator which permits a great simplification of the more 
difficult, first diagram of Fig. \ref{vacpolgraphs}.

\subsection{Notation and Primitive Diagrams}

Although we will cite original work, a unified treatment can be
found in section 5.2 of \cite{Woodard:2014jba}. We work on the 
spatially flat cosmological patch of de Sitter whose invariant 
element is,
\begin{equation}
ds^2 \equiv g_{\mu\nu} dx^{\mu} dx^{\nu} = a^2 (-d\eta^2 + d\vec{x} 
\cdot d\vec{x} ) \qquad , \qquad a(\eta) \equiv -\frac1{H \eta} \; ,
\end{equation}
where $H$ is the Hubble constant. Note that $g_{\mu\nu} = a^2 
\eta_{\mu\nu}$, where $\eta_{\mu\nu}$ is the Minkowski metric. We 
work in $D$ spacetime dimensions to facilitate the use of dimensional 
regularization. Whereas the $D-1$ spatial coordinates $-\infty < x^i < 
+\infty$ take their usual values, the conformal time $\eta$ runs from 
$\eta \rightarrow -\infty$ (the infinite past) to $\eta \rightarrow 
0^{-}$ (the infinite future).

The vacuum polarization $i[\mbox{}^{\mu} \Pi^{\nu}](x;x')$ is a 
bi-vector density which depends upon two spacetime points, $x^{\mu}$
and ${x'}^{\mu}$. In representing functions such as propagators which 
depend upon these two points, we will make extensive use of
the de Sitter length function,
\begin{equation}
y(x;x') \equiv a(\eta) a(\eta') H^2 \Bigl[ \Vert \vec{x} \!-\! 
\vec{x}' \Vert^2 - \Bigl( \vert\eta \!-\! \eta'\vert \!-\! i \epsilon
\Bigr)^2 \Bigr] \; . \label{ydef}
\end{equation}
We also need the de Sitter breaking product and ratio of the two scale 
factors,
\begin{equation}
u(x;x') \equiv \ln(aa') \qquad , \qquad v(x;x') \equiv \ln\Bigl(
\frac{a}{a'}\Bigr) \; , \label{uvdef}
\end{equation}
where $a' \equiv a(\eta')$.
The de Sitter metric at $x^{\mu}$ and ${x'}^{\mu}$, along with products 
of derivatives of $y$ (without the $i\epsilon$ term) and $u$ furnish a 
convenient basis for representing bi-tensor functions of $x^{\mu}$ and 
${x'}^{\mu}$,
\begin{equation}
\partial_{\mu} y \quad , \quad \partial'_{\nu} y \quad , \quad
\partial_{\mu} \partial'_{\nu} y \quad , \quad \partial_{\mu} u
\quad , \quad \partial'_{\nu} u \; . \label{basis}
\end{equation}
(We do not require derivatives of $v(x;x')$ because $\partial_{\mu}
v = + \partial_{\mu} u$ and $\partial'_{\mu} v = -\partial'_{\mu} u$.)
It turns out that either taking covariant derivatives of any of the
five derivatives (\ref{basis}), or contracting any two of them
into one another, produces more elements of the basis
\cite{Kahya:2005kj,Miao:2010vs}.

The Maxwell Lagrangian is $\mathcal{L}_{\rm Max} = -\frac14 F_{\mu\nu}
F_{\rho\sigma} \mathbf{g}^{\mu\rho} \mathbf{g}^{\nu\sigma} \sqrt{-
\mathbf{g}}$ and the only interactions we require descend from 
its second variation,
\begin{equation}
\frac{\delta^2 S_{\rm Max}}{\delta A_{\mu}(x) \delta A_{\rho}(x') }
= -\partial_{\kappa} \partial'_{\lambda} \Biggl\{ \sqrt{-\mathbf{g}} \, 
\Bigl[ \mathbf{g}^{\kappa\lambda} \mathbf{g}^{\mu \rho} \!-\! 
\mathbf{g}^{\kappa\rho} \mathbf{g}^{\lambda\mu} \Bigr] \delta^D(x \!-\! 
x') \Biggr\} \; .
\end{equation}
The necessary vertex functions are obtained by expanding the full metric
around the de Sitter background as in (\ref{hdef}). We can take advantage 
of the conformal flatness of the de Sitter background ($g_{\mu\nu} = 
a^2 \eta_{\mu\nu}$) to extract the scale factors and express the result
using the notation of previous work in flat space background 
\cite{Leonard:2012fs} and on de Sitter, with the conformally rescaled
graviton field, in the noncovariant gauge \cite{Leonard:2013xsa},
\begin{eqnarray}
\lefteqn{\sqrt{-\mathbf{g}} \, \Bigl( \mathbf{g}^{\kappa\lambda} 
\mathbf{g}^{\mu\rho} \!-\! \mathbf{g}^{\kappa\rho} \mathbf{g}^{\lambda\mu} 
\Bigr) \equiv a^{D-4} \Bigl(\eta^{\kappa\lambda} \eta^{\mu\rho} \!-\! 
\eta^{\kappa\rho} \eta^{\lambda\mu} \Bigr) } \nonumber \\
& & \hspace{2cm} + \kappa a^{D-6} V^{\mu\rho\kappa\lambda\alpha\beta} 
h_{\alpha\beta} + \kappa^2 a^{D-8} 
U^{\mu\rho\kappa\lambda\alpha\beta\gamma\delta} h_{\alpha\beta} 
h_{\gamma\delta} + O(\kappa^3) \; . \qquad
\end{eqnarray}
The tensor factors for the 3-point and 4-point vertices are
\cite{Leonard:2012fs,Leonard:2013xsa},
\begin{eqnarray}
\lefteqn{V^{\mu\rho\kappa\lambda\alpha\beta} = \eta^{\alpha\beta}
\eta^{\kappa [\lambda} \eta^{\rho ] \mu} \!+\! 4 \eta^{\alpha) [\mu}
\eta^{\kappa ] [\rho} \eta^{\lambda ] (\beta} \; , } \label{Vvert} \\
\lefteqn{U^{\mu\rho\kappa\lambda\alpha\beta\gamma\delta} = \Bigl[\frac14
\eta^{\alpha\beta} \eta^{\gamma\delta} \!-\! \frac12 \eta^{\alpha (\gamma}
\eta^{\delta) \beta} \Bigr] \eta^{\kappa [\lambda} \eta^{\rho ] \mu} +
\eta^{\alpha\beta} \eta^{\gamma) [\mu} \eta^{\kappa] [\rho}
\eta^{\lambda] (\delta} } \nonumber \\
& & \hspace{-.5cm} + \eta^{\gamma\delta} \eta^{\alpha) [\mu} \eta^{\kappa]
[\rho} \eta^{\lambda] (\beta} \!+\! \eta^{\kappa (\alpha} \eta^{\beta)
[\lambda} \eta^{\rho ] (\gamma} \eta^{\delta) \mu} \!+\! \eta^{\kappa (\gamma}
\eta^{\delta) [\lambda} \eta^{\rho ] (\alpha} \eta^{\beta) \mu}
\!+\! \eta^{\kappa (\alpha} \eta^{\beta) (\gamma} \eta^{\delta) [\lambda}
\eta^{\rho ] \mu} \nonumber \\
& & \hspace{1.5cm} + \eta^{\kappa (\gamma} \eta^{\delta) (\alpha}
\eta^{\beta) [\lambda} \eta^{\rho ] \mu} + \eta^{\kappa [\lambda}
\eta^{\rho ] (\alpha} \eta^{\beta) (\gamma} \eta^{\delta) \mu}
+ \eta^{\kappa [\lambda} \eta^{\rho ] (\gamma} \eta^{\delta) (\alpha}
\eta^{\beta) \mu} \; . \qquad \label{Uvert}
\end{eqnarray}
Parenthesized indices are symmetrized and indices enclosed in square
brackets are anti-symmetrized, and both are normalized.

We can express the first two diagrams of Fig. \ref{vacpolgraphs} using
the vertices (\ref{Vvert}-\ref{Uvert}), along with the graviton 
propagator $i[\mbox{}_{\alpha\beta} \Delta_{\gamma\delta}](x;x')$ and 
the photon propagator $i[\mbox{}_{\rho} \Delta_{\sigma}](x;x')$. The 
leftmost diagram is formed from two 3-point vertices,
\begin{eqnarray}
\lefteqn{i\Bigl[\mbox{}^{\mu} \Pi^{\nu}_{\rm 3pt}\Bigr](x;x') =
\partial_{\kappa} \partial_{\theta}' \Biggl\{ i\kappa a^{D-6}
V^{\mu\rho\kappa\lambda\alpha\beta} \, i\Bigl[
\mbox{}_{\alpha\beta} \Delta_{\gamma\delta} \Bigr](x;x') } \nonumber \\
& & \hspace{5.2cm} \times i\kappa {a'}^{D-6}
V^{\nu\sigma\theta\phi\gamma\delta} \, \partial_{\lambda}
\partial_{\phi}' i\Bigl[\mbox{}_{\rho} \Delta_{\sigma}\Bigr](x;x')
\Biggr\} \; . \qquad \label{3pt}
\end{eqnarray}
The middle diagram contains a single 4-point vertex,
\begin{equation}
i\Bigl[\mbox{}^{\mu} \Pi^{\nu}_{\rm 4pt}\Bigr](x;x') =
\partial_{\kappa} \partial'_{\lambda} \Biggl\{ -i\kappa^2 a^{D-8}
U^{\mu\nu\kappa\lambda\alpha\beta\gamma\delta}
\, i\Bigl[\mbox{}_{\alpha\beta} \Delta_{\gamma\delta} \Bigr](x;x)
\, \delta^D(x \!-\! x') \Biggr\} \; . \label{4pt}
\end{equation}

\subsection{Representing the Tensor Structure of $i[\mbox{}^{\mu} 
\Pi^{\nu}](x;x')$}
\label{subsec: Representing}

It can hardly escape notice that each of the two primitive
diagrams (\ref{3pt}-\ref{4pt}) takes the form of one primed and 
one unprimed derivative contracted into a bi-tensor 
density,\footnote{Note that our quantity $[\mbox{}^{\mu\rho} 
T^{\nu\sigma}](x;x')$ contains a factor of $\sqrt{-g(x)} \times 
\sqrt{-g(x')} = (a a')^{D}$ that was not part of the symbol of 
the same name  employed in Ref. \cite{Leonard:2012si,Leonard:2012ex}.}
\begin{equation}
i\Bigl[\mbox{}^{\mu} \Pi^{\nu}\Bigr](x;x') = \partial_{\rho}
\partial'_{\sigma} \Biggl\{ \Bigl[\mbox{}^{\mu\rho} T^{\nu\sigma}
\Bigr](x;x') \Biggr\} \; , \label{Piform}
\end{equation}
which is antisymmetric on each index group and symmetric under 
reflections,
\begin{equation}
\Bigl[\mbox{}^{\mu\rho} T^{\nu\sigma}\Bigr](x;x') =
-\Bigl[\mbox{}^{\rho\mu} T^{\nu\sigma}\Bigr](x;x') =
-\Bigl[\mbox{}^{\mu\rho} T^{\sigma\nu}\Bigr](x;x') =
+\Bigl[\mbox{}^{\nu\sigma} T^{\mu\rho}\Bigr](x';x) \; . 
\label{Tsyms}
\end{equation}
(Note that because the vacuum polarization is a bi-vector {\it 
density}, there is no distinction between divergences formed
with ordinary or covariant derivatives.) This form 
(\ref{Piform}-\ref{Tsyms}) is the necessary and sufficient 
condition for the vacuum polarization to be reflection 
symmetric and transverse,
\begin{equation}
\partial_{\mu} i\Bigl[\mbox{}^{\mu} \Pi^{\nu}\Bigr](x;x') = 0 =
\partial'_{\nu} i \Bigl[\mbox{}^{\mu} \Pi^{\nu}\Bigr](x;x') \; .
\label{transversality}
\end{equation}

The form (\ref{Piform}-\ref{Tsyms}) pertains to any background
geometry, not just to de Sitter. When the background metric
and the propagators possess isometries great restrictions on the 
form of the bi-tensor $[\mbox{}^{\mu\rho} T^{\nu\sigma}](x;x')$ 
can be imposed. For example, the flat space result 
(\ref{flatvacpol}) shows it can be reduced to the form of a single 
tensor times a scalar structure function,
\begin{equation}
\Bigl[\mbox{}^{\mu\rho} T_{\rm flat}^{\nu\sigma}\Bigr](x;x')
= \Bigl(\eta^{\mu\nu} \eta^{\rho\sigma} \!-\! \eta^{\mu\sigma}
\eta^{\nu\rho}\Bigr) \times \mathcal{F}\Bigl( (x \!-\! x')^2
\Bigr) \; . \label{flatform}
\end{equation}
This form (\ref{flatform}) is obviously more economical than 
acting the derivatives in (\ref{Piform}) and writing out all the 
resulting tensors. It is also more straightforward to employ in
the quantum-corrected Maxwell equation (\ref{maxeqn}). Not the 
least of this representation's advantages is that the one graviton 
loop contribution to the structure function $\mathcal{F}(
\Delta x^2)$ is only quadratically divergent, as opposed to the
quartic divergences in the primitive diagram.

The de Sitter geometry has as many isometries as flat space so one
might expect that a similar representation is possible in terms of
just one structure function. However, a second structure function is
required because the graviton propagator breaks de Sitter invariance 
\cite{Miao:2011fc,Mora:2012zi} down to just the cosmological 
symmetries of homogeneity and isometry. We have chosen to represent
the result in the same form which was first employed for the one
loop contribution from scalar quantum electrodynamics on de Sitter
background \cite{Prokopec:2002jn,Prokopec:2002uw},
\begin{equation}
\Bigl[\mbox{}^{\mu\rho} T^{\nu\sigma}\Bigr](x;x') = \Bigl(
\eta^{\mu\nu} \eta^{\rho\sigma} \!-\! \eta^{\mu\sigma} \eta^{\nu\rho}
\Bigr) \!\times\! F(x;x') + \Bigl(\overline{\eta}^{\mu\nu} 
\overline{\eta}^{\rho\sigma} \!-\! \overline{\eta}^{\mu\sigma} 
\overline{\eta}^{\nu\rho}\Bigr) \!\times\! G(x;x') \; . \label{ourrep}
\end{equation}
Here and henceforth, an overlined tensor represents the suppression 
of temporal components, for example, $\overline{\eta}^{\mu\nu} \equiv
\eta^{\mu\nu} + \delta^{\mu}_0 \delta^{\nu}_0$. Our representation 
(\ref{ourrep}) has a transparent physical interpretation
\cite{Prokopec:2003bx} and is simpler to use even in cases for which
a de Sitter invariant representation is possible 
\cite{Leonard:2012si}. There is also a straightforward procedure for 
passing from one representation to any other \cite{Leonard:2012ex}, 
so nothing is lost by employing the form (\ref{ourrep}).
 
The primitive diagrams (\ref{3pt}-\ref{4pt}) each permit one to read
off a contribution to the tensor $[\mbox{}^{\mu\rho} 
T^{\nu\sigma}](x;x')$, however, this contribution is not immediately 
in the form (\ref{ourrep}). The most general form consistent with 
homogeneity, isotropy and the symmetries (\ref{Tsyms}) can be 
expressed as a linear combination of the basis tensors formed from
differentiating $y(x;x')$ and $u(x;x')$
\cite{Leonard:2012si,Leonard:2012ex},
\begin{eqnarray}
\lefteqn{ \frac{[\mbox{}^{\mu\rho} T^{\nu\sigma}]}{(a a')^D} \equiv 
D^{\mu} {D'}^{[\nu} y \, {D'}^{\sigma]} D^{\rho} y \times f_1(y,u,v) 
+ D^{[\mu} y \, D^{\rho]} {D'}^{[\nu} y \, {D'}^{\sigma]} y \times 
f_2(y,u,v)} \nonumber \\
& & \hspace{0cm} + D^{[\mu} y \, D^{\rho]} {D'}^{[\nu} y \, 
{D'}^{\sigma]} u \times f_3(y,u,v) + D^{[\mu} u \, D^{\rho]} 
{D'}^{[\nu} y \, {D'}^{\sigma]} y \times \widetilde{f}_3(y,u,v)
\nonumber \\
& & \hspace{-.7cm} + D^{[\mu} u \, D^{\rho]} {D'}^{[\nu} y \, 
{D'}^{\sigma]} u \!\times\! f_4(y,u,v) + D^{[\mu} y \, D^{\rho]} u 
\, {D'}^{[\nu} y {D'}^{\sigma]} u \!\times\! f_5(y,u,v) \; , \qquad 
\label{fidefs}
\end{eqnarray}
where we define the $\widetilde{f}_3(y,u,v) \equiv f_3(y,u,-v)$.
(Acting on scalars as in (\ref{fidefs}) we have $D^{\mu} = 
g^{\mu\alpha}(x) \partial_{\alpha}$, ${D'}^{\nu} = 
g^{\nu\beta}(x') \partial'_{\beta}$.) Only two combinations of the 
$f_i(y,u,v)$ are independent. We call these the ``master structure 
functions'' $\Phi(y,u,v)$ and $\Psi(y,u,v)$. Given the values of 
the various $f_i(y,u,v)$, one constructs the master structure 
functions according to Table \ref{fitoPhiPsi} \cite{Leonard:2012ex}.
 
\begin{table}
\renewcommand{\arraystretch}{2}
\setlength{\tabcolsep}{8pt}
\centering
\begin{tabular}{|c||c|c|}
\hline 
$i$ & $4 H^{-4} \times \Phi_i(y,u,v)$ & $4 H^{-4} \times \Psi_i(y,u,v)$ \\ 
\hline\hline
1 & $2(D\!-\!1) f_1 - 2(2\!-\!y) \partial_y f_1$ & $2 (\partial_u^2 - 
\partial_v^2) f_1$ \\ 
\hline 
2 & $\!\!D(2\!-\!y) f_2 + (4y \!-\! y^2) \partial_y f_2$ & $(2-y) 
(\partial_u^2 - \partial_v^2) f_2\!\!$ \\ 
\hline 
3 & ${-(D-1) (f_3 + \widetilde{f}_3) + (2-y) \partial_y (f_3 + 
\widetilde{f}_3) \atop  - 2 \partial_y (e^v f_3 + e^{-v} 
\widetilde{f}_3)}$ 
& ${2 \partial_y \partial_u (e^{v} f_3 + e^{-v} \widetilde{f}_3) +
2 \partial_y \partial_v (-e^v f_3 + e^{-v} \widetilde{f}_3) \atop
-(\partial_u^2 - \partial_v^2) (f_3 + \widetilde{f}_3)}$ \\ 
\hline 
4 & $\partial_y f_4$ & $-(D-1) \partial_y f_4 + (2-y) 
\partial_y^2 f_4 - 2 \partial_y \partial_u f_4$ \\ 
\hline 
5 & ${-D f_5 + 2(2-y) \partial_y f_5 \atop - 4\cosh(v) \partial_y 
f_5}$ & 
$\!\!\!{(D-1) f_5 - (D+1) (2-y) \partial_y f_5 - (4y - y^2) 
\partial_y^2 f_5 + 2 \partial_u f_5 \atop -[2(2-y) - 4 \cosh(v)] 
\partial_y \partial_u f_5 - 4 \partial_y \partial_v [\sinh(v) f_5] 
- (\partial_u^2 - \partial_v^2) f_5}\!\!\!$ \\ 
\hline
\end{tabular} 
\caption{The contribution to master structure functions 
$\Phi(y,u,v) \equiv \sum_{i=1}^5 \Phi_i$ and $\Psi(y,u,v) \equiv
\sum_{i=1}^5 \Psi_i$ from each of the coefficient functions 
$f_i(y,u,v)$ of equation (\ref{fidefs}). Note that 
$\widetilde{f}_3(y,u,v) \equiv f_3(y,u,-v)$. \label{fitoPhiPsi}}
\end{table}

One constructs the structure functions $F(x;x)$ and $G(x;x')$ 
of our representation according to the rules 
\cite{Leonard:2012ex},\footnote{The formulae we give contain 
factors of $(a a')^{D-2}$ which were mistakenly omitted from
equations (38-39) of \cite{Leonard:2012ex}.}
\begin{eqnarray}
F(x;x') & = & (a a')^{D-2} \times I\Bigl[-2 \Phi\Bigr] \; , 
\label{Frule} \\
G(x;x') & = & (a a')^{D-2} \times I^2\Bigl[ (D \!-\! 1) \Phi +
y \partial_y \Phi + 2 \partial_u \Phi + \Psi\Bigr] \; . \qquad
\label{Grule}
\end{eqnarray}
Here and henceforth the symbol $I[f]$ represents the indefinite
integral with respect to $y$,
\begin{equation}
I[f] \equiv \int^y \!\! dy' \, f(y',u,v) \; . \label{Idef}
\end{equation}
Note from expressions (\ref{Frule}-\ref{Grule}) that the structure
function $G(x;x')$ is less divergent than $F(x;x')$. Whereas 
$F(x;x')$ must possess quadratic ultraviolet divergences which are
determined by the flat space limit, $G(x;x')$ is only logarithmically
divergent, and it vanishes in the flat space limit.

\subsection{Renormalization}

Einstein + Maxwell is not perturbatively renormalizable 
\cite{Deser:1974zzd,Deser:1974cz}, but we can still absorb the
divergences using BPHZ counterterms of higher dimension in the
standard sense of low energy effective field theory
\cite{Donoghue:1993eb,Donoghue:1994dn,BjerrumBohr:2002sx}.
Our gauge (\ref{gauge}) is covariant and we employ dimensional
regularization, so it might be thought that only invariant counterterms
are required. That turns out not to be true for three reasons:
\begin{itemize}
\item{Our interactions are time-ordered, as is apparent from the 
$i\epsilon$ prescription of the de Sitter length function (\ref{ydef})
which enters propagators;}
\item{Quantum gravitational interactions possess two derivatives, which
allows the noninvariant ordering to contaminate expression (\ref{3pt}); and}
\item{The coincidence limit of the graviton propagator diverges in de
Sitter background instead of vanishing like it does in flat space
\cite{Leonard:2012fs}.}
\end{itemize} 
We are loath to change the ordering prescription because it is so 
embedded in the Schwinger-Keldysh formalism \cite{Schwinger:1960qe,
Mahanthappa:1962ex,Bakshi:1962dv,Bakshi:1963bn,Keldysh:1964ud,
Chou:1984es,Jordan:1986ug,Calzetta:1986ey,Calzetta:1986cq} that 
must be employed to make the nonlocal part of equation (\ref{maxeqn}) 
both real and causal. The only alternative is to permit the same 
noninvariant counterterm that was necessary when using the 
noncovariant gauge \cite{Leonard:2013xsa},
\begin{eqnarray}
\lefteqn{\Delta \mathcal{L} = C_4 D_{\alpha} F_{\mu\nu} D_{\beta} 
F_{\rho\sigma} g^{\alpha\beta} g^{\mu\rho} g^{\nu\sigma} \sqrt{-g} }
\nonumber \\
& & \hspace{2.5cm} + \overline{C} H^2 F_{\mu\nu} F_{\rho\sigma} 
g^{\mu\rho} g^{\nu\sigma} \sqrt{-g} + \Delta C H^2 F_{ij} F_{k\ell}
g^{ik} g^{j\ell} \sqrt{-g} \; . \qquad \label{cterms}
\end{eqnarray}
The term proportional to $\overline{C}$ is the de Sitter specialization
of three counterterms for which the two field strengths are contracted 
into $g^{\mu\rho} g^{\nu\sigma} R$, $g^{\mu\rho} R^{\nu\sigma}$ and
$R^{\mu\nu\rho\sigma}$. The term proportional to $\Delta C$ is the 
noninvariant counterterm by virtue of its indices running only over
space. After some tedious tensor algebra one finds that $\Delta 
\mathcal{L}$ contributes to $F(x;x')$ and $G(x;x')$ as 
\cite{Leonard:2013xsa},
\begin{eqnarray}
\Delta F & \!\!\!\!=\!\!\!\! & 4 a^{D-4} \Biggl\{ \!\Bigl[\overline{C} 
\!-\! (3D \!-\! 8) C_4\Bigr] H^2 \!-\! \frac{C_4}{a a'} \partial^2 \!+\! 
\frac{(D\!-\! 4) C_4 H}{a} \partial_0 \! \Biggr\} i \delta^D(x \!-\! x') 
\; , \label{DF} \qquad \\
\Delta G & \!\!\!\!=\!\!\!\! & 4 a^{D-4} \Bigl[\Delta C - (D \!-\! 6) 
C_4 \Bigr] H^2 i \delta^D(x \!-\! x') \; . \label{DG}
\end{eqnarray}
Although the $\Delta C$ counterterm is strictly only required to 
renormalize divergences from the time-ordering we will also employ it
to simplify $G(x;x')$. The coefficient $C_4$ is fixed by the flat space 
limit \cite{Leonard:2012fs} in terms of the quantities $C_2$ and $C_0(b)$
defined in expressions (\ref{spin2C}-\ref{spin0C}),
\begin{equation}
C_4 = \frac{\kappa^2 \mu^{D-4}}{64 \pi^{\frac{D}2}} 
\frac{\Gamma(\frac{D}2)}{(D \!-\! 1) (D \!-\! 2)^2 (D \!-\! 3) 
(D \!-\! 4)} \Bigl[ C_2 + C_0(b)\Bigr] \; . \label{C4values}
\end{equation}
Note that the spin 2 contribution to $C_4$ is finite by virtue of the
factor of $D - 4$ in expression (\ref{spin2C}).

Those accustomed to modern techniques of renormalization in covariant
gauges sometimes find the appearance of noninvariant counterterms to 
be disconcerting. However, it is important to realize that they pose 
no problem of principle. The divergent part of the counterterm is of 
course fixed by the primitive divergences it is to remove, and the 
finite part can be determined to enforce physical symmetries. (In our 
case the focus on late times obviates the need for this as long as 
the finite part of the counterterm is assumed to be of order one.) 
The procedure is explained in older standard texts on quantum field 
theory, for example \cite{Jauch:1955}. And it is important to 
recognize that many of the classic computations of quantum
electrodynamics were in fact performed using noncovariant gauges
\cite{Schwinger:1958}.

\subsection{Reducing the Photon Propagator Term}

Our photon propagator is defined in exact Lorentz gauge,
\begin{equation}
D^{\rho} i\Bigl[\mbox{}_{\rho} \Delta_{\sigma}\Bigr](x;x') = 0 =
{D'}^{\sigma} i \Bigl[\mbox{}_{\rho} \Delta_{\sigma}\Bigr](x;x') \; .
\label{Lorentz}
\end{equation}
An early result \cite{Allen:1985wd} for this propagator contains a
small error which was corrected 20 years later 
\cite{Tsamis:2006gj}.\footnote{There have been some recent false 
claims about this in the mathematical physics literature 
\cite{Frob:2013qsa} so it is important to note that one really does 
need to employ the corrected propagator in computing standard things 
such as the effective potential of scalar quantum electrodynamics \cite{Allen:1983dg,Prokopec:2007ak}. Using the uncorrected propagator 
would not even recover the famous Coleman-Weinberg potential 
\cite{Coleman:1973jx} in the flat space limit.} The most useful 
form for our purposes was obtained in a study of the graviton 
propagator \cite{Miao:2011fc},
\begin{equation}
i\Bigl[\mbox{}_{\rho} \Delta_{\sigma}\Bigr](x;x') = -\frac1{2 H^2}
\Bigl[ \delta^{\alpha}_{\rho} \square \!-\! D^{\alpha} D_{\rho}
\Bigr] \Bigl[ \delta^{\beta}_{\sigma} \square' \!-\! {D'}^{\beta}
D'_{\sigma}\Bigr] \Bigl[ \partial_{\alpha} \partial'_{\beta} y 
\times i\Delta_{BBB}(x;x') \Bigr] \; . \label{photonprop}
\end{equation}
Here the $BBB$-type propagator is defined by inverting the 3rd
power of the scalar d'Alembertian with mass $M_S^2 = (D-2) H^2$,
\begin{eqnarray}
\Bigl[ \square - (D\!-\!2) H^2 \Bigr] i\Delta_{BBB}(x;x') = 
i\Delta_{BB}(x;x') \; , \label{3rdint} \\
\Bigl[ \square - (D\!-\!2) H^2 \Bigr] i\Delta_{BB}(x;x') = 
i\Delta_{B}(x;x') \; , \label{2ndint} \\
\Bigl[ \square - (D\!-\!2) H^2 \Bigr] i\Delta_{B}(x;x') = 
\frac{i \delta^D(x \!-\! x')}{\sqrt{-g(x)}} \; . \label{1stint}
\end{eqnarray}
Because $M_S^2$ is strictly positive, $i\Delta_{BBB}$, 
$i\Delta_{BB}$ and $i\Delta_{B}$ are all de Sitter invariant
functions of $y(x;x')$ \cite{Miao:2010vs} whose precise form can 
be found in \cite{Miao:2011fc}. We will give the expansion for 
$i\Delta_{B} = B(y)$ in Appendix A. We also give there the
expansion for the propagator $i\Delta_C(x;x') = C(y)$ of a scalar
with mass $M_S^2 = 2(D-3) H^2$. An important relation exists 
between them which we will use many times \cite{Miao:2009hb},
\begin{equation}
2 C'(y) = (2 \!-\! y) B'(y) - B(y) \; . \label{BCident}
\end{equation}

The differential operator used to construct the photon propagator
(\ref{photonprop}) has the key property of transversality 
\cite{Miao:2011fc},
\begin{equation}
D^{\rho} \Bigl[ \delta^{\alpha}_{\rho} \square \!-\! D^{\alpha} 
D_{\rho} \Bigr] = 0 = \Bigl[ \delta^{\alpha}_{\rho} \square \!-\! 
D^{\alpha} D_{\rho} \Bigr] D_{\alpha} \; .
\end{equation}
Exploiting this, the de Sitter invariance of $i\Delta_{BBB}$ and 
relation (\ref{2ndint}), it is straightforward to show 
\cite{Miao:2011fc},
\begin{eqnarray}
\lefteqn{ \Bigl[ \delta^{\alpha}_{\rho} \square \!-\! D^{\alpha} 
D_{\rho} \Bigr] \Bigl[ \delta^{\beta}_{\sigma} \square' \!-\! 
{D'}^{\beta} D'_{\sigma}\Bigr] \Bigl[ \partial_{\alpha} 
\partial'_{\beta} y \times i\Delta_{BBB}(x;x') \Bigr] } 
\nonumber \\
& & \hspace{4.5cm} = \Bigl[ \delta^{\beta}_{\sigma} \square' \!-\! 
{D'}^{\beta} D'_{\sigma}\Bigr] \Bigl[ \partial_{\rho} 
\partial'_{\beta} y \times i\Delta_{BB}(x;x') \Bigr] \; . \qquad
\label{photon1}
\end{eqnarray}
The next step is to act the remaining primed transverse projector,
\begin{eqnarray}
\lefteqn{ \Bigl[ \delta^{\alpha}_{\rho} \square \!-\! D^{\alpha} 
D_{\rho} \Bigr] \Bigl[ \delta^{\beta}_{\sigma} \square' \!-\! 
{D'}^{\beta} D'_{\sigma}\Bigr] \Bigl[ \partial_{\alpha} 
\partial'_{\beta} y \times i\Delta_{BBB} \Bigr] } \nonumber \\
& & \hspace{1cm} = \partial_{\rho} \partial'_{\sigma} y \times 
i\Delta_{B} - H^2 \partial_{\rho} \partial'_{\sigma} \Biggl\{
(2 \!-\! y) i\Delta_{BB} - (D\!-\!3) I\Bigl[i \Delta_{BB}\Bigr]
\Biggr\} \; . \qquad \label{photon2}
\end{eqnarray}

The final step is based on the fact that the 3-point vertex factors
in expression (\ref{3pt}) inherit an anti-symmetry from the Maxwell
field strength,
\begin{equation}
V^{\mu\rho\kappa\lambda\alpha\beta} = 
-V^{\mu\lambda\kappa\rho\alpha\beta} \qquad , \qquad
V^{\nu\sigma\theta\phi\gamma\delta} =
-V^{\nu\phi\theta\sigma\gamma\delta} \; . \label{antisym}
\end{equation}
This anti-symmetry can obviously be communicated to the differentiated
photon propagator in (\ref{3pt}),
\begin{equation}
D_{\lambda} D'_{\phi} i\Bigl[\mbox{}_{\rho} \Delta_{\sigma}\Bigr](x;x')
\longrightarrow D_{[\lambda} D'_{[[\phi} i\Bigl[\mbox{}_{\rho]} 
\Delta_{\sigma]]}\Bigr](x;x') \; .
\end{equation}
Here the single square brackets indicate anti-symmetrization under
$\lambda \leftrightarrow \rho$ while the double square brackets 
indicate anti-symmetrization under $\phi \leftrightarrow \sigma$. 
Because the double covariant derivative of a scalar is symmetric we
can use (\ref{photon2}) to conclude,
\begin{equation}
D_{[\lambda} D'_{[[\phi} i\Bigl[\mbox{}_{\rho]} \Delta_{\sigma]]}
\Bigr](x;x') = -\frac1{2 H^2} \partial_{[\rho} \partial'_{[[\sigma}
y(x;x') \times \partial_{\lambda]} \partial'_{\phi]]} B\Bigl(y(x;x')
\Bigr) \; . \label{photon3}
\end{equation}
Similar reductions of the original photon propagator structure 
functions down to the $i\Delta_{B}$ propagator have also been 
noted in explicit two loop computations involving the very different
interactions of scalar quantum electrodynamics \cite{Prokopec:2006ue,
Prokopec:2008gw}.

Identity (\ref{photon3}) allows us to re-express the 3-point diagram
(\ref{3pt}) as,
\begin{equation}
i\Bigl[\mbox{}^{\mu} \Pi^{\nu}_{\rm 3pt}\Bigr] = \frac{\kappa^2}{2 H^2}
\partial_{\kappa} \partial_{\theta}' \Biggl\{ (a a')^{D-6} 
V^{\mu\rho\kappa\lambda\alpha\beta} \, i\Bigl[\mbox{}_{\alpha\beta} 
\Delta_{\gamma\delta} \Bigr] V^{\nu\sigma\theta\phi\gamma\delta} \, 
\partial_{\rho} \partial'_{\sigma} y \times \partial_{\lambda} 
\partial'_{\phi} B \Biggr\} \; . \label{new3pt}
\end{equation}
This is as far as we can get without exploiting the explicit form of
the graviton propagator.

\section{Spin 2 Contributions}\label{spin2}

The purpose of this section is to work out the contributions to
the renormalized structure functions from the spin two part of 
the graviton propagator. We begin by describing how this part of
the propagator is expressed. We next work out the contributions
from the 4-point diagram and the local part of the 3-point diagram
which comes from the delta function part of the doubly differentiated
photon propagator. A much more involved analysis is necessary to 
work out the nonlocal contributions to the 3-point diagram from the
de Sitter breaking part of the propagator and from the de Sitter
invariant part. The section closes by giving the combined results 
for $F(x;x')$ and $G(x;x')$.

\subsection{Spin 2 Part of the Graviton Propagator}

The spin two part of the graviton propagator is transverse and
traceless and takes a form analogous to the transverse photon
propagator (\ref{photonprop}) \cite{Miao:2011fc,Mora:2012zi},
\begin{eqnarray}
\lefteqn{i\Bigl[\mbox{}_{\mu\nu} \Delta^2_{\rho\sigma}\Bigr](x;x') 
= \frac{2}{H^4} \Bigl( \frac{D \!-\! 2}{D \!-\! 3}\Bigr)^2 
\times \mathbf{P}_{\mu\nu}^{~~ \alpha\beta}(x) \times 
\mathbf{P}_{\rho\sigma}^{~~ \gamma\delta}(x') } \nonumber \\
& & \hspace{3cm} \times \Bigl[\partial_{\alpha} \partial'_{\gamma} 
y(x;x') \times \partial_{\beta} \partial'_{\delta} y(x;x') 
\times i\Delta_{AAABB}(x;x') \Bigr] \; . \qquad 
\label{gravpropspin2}
\end{eqnarray}
Here $\mathbf{P}_{\mu\nu}^{~~ \alpha\beta}(x)$ is a 4th order
differential operator which is transverse and traceless on both
index groups \cite{Miao:2011fc,Mora:2012zi},
\begin{eqnarray}
\lefteqn{\mathbf{P}_{\mu\nu}^{~~\alpha\beta} = \frac12
\Bigl(\frac{D\!-\!3}{D\!-\!2}\Bigr) \Biggl\{ -\delta^{\alpha}_{(\mu}
\delta^{\beta}_{\nu)} \Bigl[\square \!-\! D H^2\Bigr] \Bigl[ \square
\!-\! 2 H^2\Bigr] + 2 D_{(\mu} \Bigl[ \square \!+\! H^2\Bigr]
\delta^{(\alpha}_{\nu)} D^{\beta)} } \nonumber \\
& & \hspace{.3cm} - \Bigl( \frac{D \!-\!2}{D \!-\!1} \Bigr) D_{(\mu}
D_{\nu)} D^{(\alpha} D^{\beta)} + g_{\mu\nu} g^{\alpha\beta} \Bigl[
\frac{\square^2}{D \!-\! 1} \!-\! H^2 \square \!+\! 2 H^4\Bigr]
\qquad \nonumber \\
& & \hspace{.3cm} -\frac{D_{(\mu} D_{\nu)} }{D \!-\! 1} \Bigl[
\square \!+\! 2 (D \!-\! 1) H^2\Bigr] g^{\alpha\beta}
-\frac{g_{\mu\nu} }{D \!-\! 1} \Bigl[ \square \!+\! 2 (D \!-\! 1)
H^2\Bigr] D^{(\alpha} D^{\beta)} \Biggr\} . \qquad \label{spin2op}
\end{eqnarray}
The spin two structure function $i\Delta_{AAABB}(x;x')$ is 
constructed by inverting the kinetic operator for a massless scalar
three times and inverting the kinetic operator for an $M_S^2 = 
(D-2) H^2$ scalar twice. The order in which these inversions are
performed is irrelevant but we find it convenient to alternate,
\begin{eqnarray}
\square i\Delta_{AAABB}(x;x') & = & i\Delta_{AABB}(x;x') \; , 
\label{DAAABB} \\
\Bigl[ \square - (D\!-\!2) H^2\Bigr] i\Delta_{AABB}(x;x') & = & 
i \Delta_{AAB}(x;x') \; , \label{DAABB} \\
\square i\Delta_{AAB}(x;x') & = & i\Delta_{AB}(x;x') \; , 
\label{DAAB} \\
\Bigl[ \square - (D\!-\!2) H^2\Bigr] i\Delta_{AB}(x;x') & = & 
i \Delta_{A}(x;x') \; , \label{DAB} \\
\square i\Delta_{A}(x;x') & = & \frac{i \delta^D(x \!-\! x')}{
\sqrt{-g}} \; . \label{DA}
\end{eqnarray}

Scalars with $M_S^2 \leq 0$ inevitably break de Sitter invariance
\cite{Vilenkin:1982wt,Linde:1982uu,Starobinsky:1982ee,Allen:1987tz,
Miao:2010vs} so each of the integrated propagators in relations 
(\ref{DAAABB}-\ref{DA}) does as well.\footnote{The mathematical 
physics literature contains claims that there is no need for a de 
Sitter breaking part \cite{Higuchi:2011vw,Faizal:2011iv,
Higuchi:2012vy}. Morrison \cite{Morrison:2013rqa} has shown that 
constructions which purport to give a de Sitter invariant propagator 
differ from ours in two ways: (1) the propagator for a scalar with 
general mass-squared $M_S^2$ must be considered as both de Sitter 
invariant {\it and} well defined for all $M_S^2$, except for simple 
poles at $M^2_S = -N (N+D-1) H^2$ with $N=0,1,2,\dots$, and (2) it 
must be accepted that, for constructing the graviton propagator, an 
arbitrary constant can be added to equation (\ref{DA}). Both of 
these deviations are illegitimate, resulting in formal solutions to 
the propagator equation which are not true propagators in the sense 
of being the expectation values, in the presence of positive-normed 
states, of the time-ordered product of two graviton field operators 
\cite{Miao:2013isa}.} That is, we can express each of the integrated 
propagators as the sum of a de Sitter invariant function of $y(x;x')$ 
plus a de Sitter breaking term,
\begin{equation}
i\Delta_{AAABB}(x;x') = i\Delta_{AAABB}^{\rm inv}(x;x') + 
i\Delta_{AAABB}^{\rm brk}(x;x') \; . \label{invbrk}
\end{equation}
Explicit forms for both terms can be found in \cite{Kahya:2011sy}.
We refer to the ``de Sitter breaking'' and ``de Sitter invariant'' 
part of the spin two graviton propagator as derived from
$i\Delta_{AAABB}^{\rm brk}$ and $i\Delta_{AAABB}^{\rm inv}$,
respectively. The de Sitter breaking part is \cite{Kahya:2011sy},
\begin{equation}
i\Bigl[\mbox{}_{\mu\nu} \Delta^{\rm 2brk}_{\rho\sigma}\Bigr](x;x')
= (a a')^2 \Bigl[ 2 \overline{\eta}_{\mu(\rho} 
\overline{\eta}_{\sigma) \nu} \!-\! \frac2{D \!-\! 1} 
\overline{\eta}_{\mu\nu} \overline{\eta}_{\rho\sigma}\Bigr]
\times k \Bigl[ \ln(4 a a') \!+\! A_2 \Bigr] \; , \label{2brk}
\end{equation}
where we recall that $\overline{\eta}_{\mu\nu} \equiv \eta_{\mu\nu} 
+ \delta^0_{\mu} \delta^0_{\nu}$ and the constants $k$ and $A_2$ are,
\begin{equation}
k \equiv \frac{H^{D-2}}{(4\pi)^{\frac{D}2}} \frac{\Gamma(D\!-\!1)}{
\Gamma(\frac{D}2)} \qquad , \qquad A_2 \equiv 2 \psi\Bigl(\frac{D
\!-\!1}2\Bigr) \!-\! 4 \!+\! \frac1{D\!-\!1} \; . 
\end{equation}
(Here $\psi(z) \equiv \frac{d}{dz} \ln[\Gamma(z)]$ is the digamma 
function.) The de Sitter invariant part is much more complicated. 
The graviton analog of relation (\ref{photon1}) is 
\cite{Kahya:2011sy},
\begin{eqnarray}
\lefteqn{i\Bigl[\mbox{}_{\mu\nu} \Delta^{\rm 2inv}_{\rho\sigma}
\Bigr](x;x') = -H^{-4} \Bigl(\frac{D\!-\!2}{D\!-\!3}\Bigr) \times 
\mathbf{P}_{\rho\sigma}^{~~ \gamma\delta}(x') } \nonumber \\
& & \hspace{2cm} \times \Biggl[\partial_{\mu} \partial'_{\gamma} 
y \times \partial_{\nu} \partial'_{\delta} y \times \Bigl[\square 
\!-\! (D\!-\!2) H^2\Bigr] \square i\Delta^{\rm inv}_{AAABB}(x;x') 
\Biggr] \; . \qquad \label{graviton1}
\end{eqnarray}
In Appendix B we show that acting the derivatives gives,
\begin{equation}
\Bigl[\square \!-\! (D\!-\!2) H^2\Bigr] \square 
i\Delta^{\rm inv}_{AAABB}(x;x') = i\Delta^{\rm inv}_{AAB}(x;x') 
+ {\rm constant} \equiv J(y) \; . \label{Jzero}
\end{equation}
The analog of relation (\ref{photon2}) is,
\begin{eqnarray}
\lefteqn{i\Bigl[\mbox{}_{\mu\nu} \Delta^{\rm 2inv}_{\rho\sigma}
\Bigr] \!= \partial_{\mu} \partial'_{(\rho} y \, \partial'_{\sigma)} 
\partial_{\nu} y \, J_1(y) \!+\! D_{(\mu } D'_{((\rho} \Bigl[ 
\partial_{\nu)} \partial'_{\sigma))} y \, J_2(y)\Bigr] \!+\! D_{\mu} 
D_{\nu} D'_{\rho} D'_{\sigma} J_3(y) } \nonumber \\ 
& & \hspace{2.8cm} + H^2 \Bigl[ g_{\mu\nu} D'_{\rho} D'_{\sigma} 
\!+\! g'_{\rho\sigma} D_{\mu} D_{\nu} \Bigr] J_4(y) + H^4 
g_{\mu\nu} g'_{\rho\sigma} \, J_5(y) \; , \qquad \label{graviton2}
\end{eqnarray}
where Table \ref{Jays} gives the functions $J_i(y)$.

\begin{table}
\renewcommand{\arraystretch}{2}
\setlength{\tabcolsep}{8pt}
\centering
\begin{tabular}{|c||c|}
\hline 
$i$ & $J_i(y)$ \\ 
\hline\hline 
1 & $\frac12 \frac{\square}{H^2} [\frac{\square}{H^2} - (D\!-\!2)] 
J(y)$ \\ 
\hline 
2 & $I[-(2\!-\!y) (\frac{\square}{H^2} J)' + (D\!-\!1)
\frac{\square}{H^2} J + (D\!-\!2) (2\!-\!y) J' - (D\!-\!1)
(D\!-\!2) J]$ \\ 
\hline 
3 & $I^2[-\frac12 \frac{\square}{H^2} J + 2 (\frac{D-2}{D-1})
J'' - \frac12 (D\!-\!2)(2\!-\!y) J' + \frac12 (D\!-\!2)(D\!-\!1)
J]$ \\ 
\hline
4 & $\frac2{D-1} \frac{\square}{H^2} J(y) + I^2[-\frac12
(2 \!-\!y) (\frac{\square}{H^2} J)' + \frac12 (D\!-\!2)
\frac{\square}{H^2} J]$ \\ 
\hline
5 & $-\frac2{D-1} \frac{\square^2}{H^4} J(y) + \frac12 (2 \!-\!y)^2
\frac{\square}{H^2} J(y) - \frac12 (D\!-\!3) (2 \!-\!y) 
I[ \frac{\square}{H^2} J]$ \\ 
\hline 
\end{tabular} 
\caption{The five functions $J_i(y)$ in relation (\ref{graviton2})
for the de Sitter invariant part of the spin two graviton propagator. 
The function $J(y)$ is given in expression (\ref{Jzero}) and $I[f]$
stands for the indefinite integral of $f(y)$ as in (\ref{Idef}). 
\label{Jays}}
\end{table}

Relation (\ref{graviton2}) is remarkably similar to relation
(\ref{photon2}). In particular, the tensor structure of the first 
term is provided by second derivatives of $y(x;x')$, and the 
function $J_1(y)$ is $i\Delta_{A}^{\rm inv}/2H^4$ plus a constant. 
The remaining terms are all gradients and/or traces. Unfortunately 
for us, neither gradients nor traces drop out of the vacuum 
polarization the way the analogous photon gradient terms did in 
expression (\ref{photon3}). This leaves no alternative but to act 
the various covariant derivatives in expression (\ref{graviton2}) 
and express the result as a linear combination of the de Sitter 
invariant bi-tensors given in Table \ref{Tees},
\begin{equation}
i\Bigl[\mbox{}_{\mu\nu} \Delta^{\rm 2inv}_{\rho\sigma}\Bigr](x;x')
= \sum_{i=1}^5 \Bigl[\mbox{}_{\mu\nu} \mathcal{T}^i_{\rho\sigma}
\Bigr](x;x') \times K_i(y) \; . \label{graviton3}
\end{equation}
The coefficient functions $K_i(y)$ are given in Table \ref{Kays}.
It is worth noting that tracelessness implies two relations among
the $K_i(y)$,
\begin{eqnarray}
4 K_1 + (4y \!-\! y^2) K_4 + D K_5 & = & 0 \; , \\
-K_1 + (2 \!-\! y) K_2 + (4y \!-\! y^2) K_3 + D K_4 & = & 0 \; . 
\qquad
\end{eqnarray}

\begin{table}
\renewcommand{\arraystretch}{2}
\setlength{\tabcolsep}{8pt}
\centering
\begin{tabular}{|c||c|}
\hline 
$i$ & $[\mbox{}_{\mu\nu} \mathcal{T}^i_{\rho\sigma}](x;x')$ \\ 
\hline\hline
1 & $\partial{_\mu} \partial'_{(\rho} y(x;x') \, \partial'_{\sigma)}
\partial_{\nu} y(x;x')$ \\ 
\hline 
2 & $\partial_{(\mu} y(x;x') \, \partial_{\nu)} \partial'_{(\rho} 
y(x;x') \, \partial'_{\sigma)} y(x;x')$ \\ 
\hline 
3 & $\partial_{\mu} y(x;x') \, \partial_{\nu} y(x;x') \, 
\partial'_{\rho} y(x;x') \, \partial'_{\sigma} y(x;x')$ \\ 
\hline 
4 & $H^2 [g_{\mu\nu}(x) \, \partial'_{\rho} y(x;x') \, 
\partial'_{\sigma} y(x;x') + \partial_{\mu} y(x;x') \, 
\partial_{\nu} y(x;x') \, g_{\rho\sigma}(x')]$ \\ 
\hline 
5 & $H^4 g_{\mu\nu}(x) g_{\rho\sigma}(x')$ \\ 
\hline
\end{tabular} 
\caption{The de Sitter invariant basis bi-tensors in relation 
(\ref{graviton3}). As always, indices enclosed in parentheses are
symmetrized. \label{Tees}}
\end{table}

\begin{table}
\renewcommand{\arraystretch}{2}
\setlength{\tabcolsep}{8pt}
\centering
\begin{tabular}{|c||c|c|}
\hline 
$i$ & $K_i(y)$ & $K_i(y)$ \\ 
\hline\hline
1 & $J_1 \!+\! J_2' \!+\! 2 J_3''$ & ${\frac12 (\frac{\square^2}{H^4} 
J) - (2-y) (\frac{\square}{H^2} J)' \atop + \frac12 (D-2)
(\frac{\square}{H^2} J) + 4 (\frac{D-2}{D-1}) J''}$\\ 
\hline 
2 & $J_2'' \!+\! 4 J_3'''$ & ${-(2-y) (\frac{\square}{H^2} J)'' + 
(D-2) (\frac{\square}{H^2} J)' \atop + 8(\frac{D-2}{D-1}) J''' - 
(D-2) (2-y) J'' + (D-2) D J'}$\\ 
\hline
3 & $J_3''''$ & ${-\frac12 (\frac{\square}{H^2} J)''
+ 2 (\frac{D-2}{D-1}) J'''' \atop -\frac12 (D-2)(2-y) J''' + 
\frac12 (D-2) (D+1) J''}$\\ 
\hline
4 & $- J_2' \!+\! [(2\!-\!y) J_3']'' \!+\! J_4''$ & 
$\frac2{D-1} (\frac{\square}{H^2} J)'' \!+\! 2 (\frac{D-2}{D-1})
(2\!-\!y) J''' \!-\! \frac{2 (D-2)(D+1)}{D-1} J''$\\ 
\hline 
5 & ${-(2\!-\!y) J_2 + (2\!-\!y)[(2\!-\!y) J_3']' \atop
+ 2(2\!-y) J_4' + J_5}$ & ${-\frac2{D-1} (\frac{\square^2}{H^4} 
J) + \frac4{D-1} (2-y) (\frac{\square}{H^2} J)' \atop -2 
(\frac{D-2}{D-1}) (\frac{\square}{H^2} J) + 8 (\frac{D-2}{D-1}) 
J''}$\\ 
\hline 
\end{tabular} 
\caption{The coefficient functions $K_i(y)$ expressed first as
derivatives of the functions $J_i(y)$ from Table \ref{Jays}, then
in terms of the function $J(y)$ defined in expression (\ref{Jzero}).
\label{Kays}}
\end{table}

\subsection{The 4-Point Diagram}

The simplest diagram is the middle one of Fig. \ref{vacpolgraphs}.
The spin two contribution to it comes from substituting the spin 
two part of the graviton propagator into expression (\ref{4pt}).
Because the spin two part of the graviton propagator is traceless 
at each point, we can drop terms in the 4-point vertex
$U^{\mu\nu\kappa\lambda\alpha\beta\gamma\delta}$ which contain
either $\eta^{\alpha\beta}$ or $\eta^{\gamma\delta}$. Many of the
other terms are also related when contracted into $i[\mbox{}_{
\alpha\beta} \Delta_{\gamma\delta}](x;x)$ so that there are only
three distinct contributions,
\begin{eqnarray}
\lefteqn{ i\Bigl[\mbox{}^{\mu} \Pi^{\nu}_{\rm 4pt2}\Bigr](x;x') = 
\partial_{\rho} \partial'_{\sigma} \Biggl\{ -\kappa^2 a^{D-4} i
\delta^D(x \!-\! x') } \nonumber \\
& & \hspace{.2cm} \times \Biggl[ -\frac12 \eta^{\mu [\nu} 
\eta^{\sigma] \rho} i\Bigl[\mbox{}^{\alpha\beta} \Delta^2_{\alpha
\beta}\Bigr] + 2 a^4 i\Bigl[\mbox{}^{\mu [\nu} \Delta^{2 \sigma] 
\rho} \Bigr] + 4 a^2 \eta^{[\mu [[\nu} i\Bigl[\mbox{}^{\rho]}_{
~~\alpha} \Delta^{2 \sigma]] \alpha}\Bigr]\Biggr] \Biggl\} \; . 
\qquad \label{4pt2.1}
\end{eqnarray}
Recall that indices enclosed in square brackets are 
anti-symmetrized, and that the double square brackets in the final 
term of (\ref{4pt2.1}) serves to distinguish the 
anti-symmetrization on $\nu \leftrightarrow \sigma$ from that
$\mu \leftrightarrow \rho$.

The coincidence limit of the graviton propagator takes the form
\cite{Kahya:2011sy},
\begin{eqnarray}
\lefteqn{i\Bigl[\mbox{}^{\alpha\beta} \Delta^2_{\gamma\delta}
\Bigr](x;x) = \Bigl[ \delta^{\alpha}_{~\gamma} \delta^{\beta}_{
~\delta} \!+\! \delta^{\alpha}_{~\delta} \delta^{\beta}_{~\gamma} 
\!-\! \frac2{D} \eta^{\alpha\beta} \eta_{\gamma\delta} \Bigr] 
\times i\Delta_1 } \nonumber \\
& & \hspace{4cm} + \Bigl[ \overline{\delta}^{\alpha}_{~\gamma} 
\overline{\delta}^{\beta}_{~\delta} \!+\! \overline{\delta}^{
\alpha}_{~\delta} \overline{\delta}^{\beta}_{~\gamma} \!-\! 
\frac2{D\!-\!1} \overline{\eta}^{\alpha\beta} \overline{\eta}_{
\gamma\delta} \Bigr] \times i\Delta_2(x) \; . \qquad 
\label{coinc2}
\end{eqnarray}
Hence the three terms on the second line of (\ref{4pt2.1}) are,
\begin{eqnarray}
\lefteqn{-\frac12 \eta^{\mu [\nu} \eta^{\sigma] \rho} i\Bigl[
\mbox{}^{\alpha\beta} \Delta^2_{\alpha\beta}\Bigr] = -\frac12 
\Bigl[ (D\!-\!1)(D\!-\!2) i\Delta_1 } \nonumber \\
& & \hspace{5.5cm} + (D\!-\!2)(D\!+\!1) i\Delta_2(x)\Bigr] 
\times \eta^{\mu [\nu} \eta^{\sigma] \rho} \; , \qquad 
\label{trace1} \\
\lefteqn{2 a^4 i\Bigl[\mbox{}^{\mu [\nu} \Delta^{2 \sigma] 
\rho} \Bigr] = -2 \Bigl(\frac{D\!+\!2}{D}\Bigr) i\Delta_1 
\times \eta^{\mu [\nu} \eta^{\sigma] \rho} } \nonumber \\
& & \hspace{6.5cm} - 2 \Bigl( \frac{D\!+\!1}{D \!-\!1}\Bigr) 
i\Delta_2(x) \times \overline{\eta}^{\mu [\nu} 
\overline{\eta}^{\sigma] \rho} \; , \qquad \label{trace2} \\
\lefteqn{4 a^2 \eta^{[\mu [[\nu} i\Bigl[\mbox{}^{\rho]}_{
~~\alpha} \Delta^{2 \sigma]] \alpha}\Bigr] = \Bigl[ 
\frac{4 (D\!-\!1)(D\!+\!2)}{D} i\Delta_1 + \frac{2 (D\!-\!2)
(D\!+\!1)}{D\!-\!1} i\Delta_2(x) \Bigr] } \nonumber \\
& & \hspace{3.5cm} \times \eta^{\mu [\nu} \eta^{\sigma] 
\rho} + \frac{2 (D\!-\!2)(D\!+\!1)}{D\!-\!1} i\Delta_2(x) 
\times \overline{\eta}^{\mu [\nu} \overline{\eta}^{\sigma] 
\rho} \; . \qquad \label{trace3}
\end{eqnarray}
In deriving the last of these relations we have used,
\begin{equation}
\eta^{[\mu [[\nu} \overline{\eta}^{\sigma]] \rho]} = \frac12
\eta^{\mu [\nu} \eta^{\sigma] \rho} + \frac12 
\overline{\eta}^{\mu [\nu} \overline{\eta}^{\sigma] \rho} \; .
\end{equation}

Expressions (\ref{4pt2.1}) and (\ref{trace1}-\ref{trace3}) imply
that the 4-point contribution already takes the form described in
subsection 2.2, 
\begin{equation}
i\Bigl[\mbox{}^{\mu} \Pi^{\nu}_{\rm 4pt2}\Bigr](x;x') = 
\partial_{\rho} \partial'_{\sigma} \Biggl\{ 2 \eta^{\mu [\nu}
\eta^{\sigma] \rho} \times F_{2a}(x;x') + 2 \overline{\eta}^{\mu 
[\nu} \overline{\eta}^{\sigma] \rho} \times G_{2a}(x;x') \Biggr\}
\; . 
\end{equation}
The two structure functions are,
\begin{eqnarray}
\lefteqn{ F_{2a}(x;x') = \kappa^2 a^{D-4} i\delta^D(x \!-\! x') 
\Biggl\{ \frac{(D\!+\!2)(D\!-\!1)(D\!-\!4)}{4 D} i\Delta_1 } 
\nonumber \\
& & \hspace{2.3cm} -\frac{(D\!+\!2)(D\!-\!2)}{D} i\Delta_1 +
\frac{(D\!-\!5)(D\!-\!2)(D\!+\!1)}{4 (D\!-\!1)} i\Delta_2(x) 
\Biggr\} \; , \qquad \\
\lefteqn{ G_{2a}(x;x') = \kappa^2 a^{D-4} i\delta^D(x \!-\! x')
\Biggl\{ -\frac{(D\!-\!3)(D\!+\!1)}{D\!-\!1} i\Delta_2(x) 
\Biggr\} \; .}
\end{eqnarray}
From the previous subsection we find the constant $i\Delta_1$ 
and the time de\-pen\-dent function $i\Delta_2(x)$ to be,
\begin{eqnarray}
i\Delta_1 & = & \frac{4 (D\!-\!2) D (D\!+\!1)}{(D \!-\! 1)} \,
H^4 J''(0) \; , \qquad \\
i\Delta_2 & = & k \Bigl[ 2 \ln(2 a) + A_2\Bigr] \equiv k \Bigl[
2\ln(a) + \overline{A}_2 \Bigr] \; . \qquad
\end{eqnarray}
Comparison with expressions (\ref{DF}-\ref{DG}) suggests that we
choose the ``2a'' contributions to the $\overline{C}$ and 
$\Delta C$ counterterms to be,
\begin{eqnarray}
\overline{C}_{2a} & \!\!\!=\!\!\! & -\frac{\kappa^2 i \Delta_1}{H^2}  
\frac{(D\!+\!2)(D^2 \!-\! 9 D \!+\! 12)}{16D} - \frac{\kappa^2 k 
\overline{A}_2}{H^2} \frac{(D\!-\!5)(D\!-\!2)(D\!+\!1)}{16 
(D \!-\!1)} \; , \qquad \\
\Delta C_{2a} & \!\!\!=\!\!\! & \frac{\kappa^2 k \overline{A}_2}{H^2} 
\frac{(D\!-\!3)(D\!+\!1)}{4 (D\!-\!1)} \; .
\end{eqnarray}
The renormalized 4-point contributions to the structure functions are,
\begin{eqnarray}
F^{\rm ren}_{2a}(x;x') & = & -\frac{5 \kappa^2 H^2}{24 \pi^2} \, 
\ln(a) i\delta^4(x \!-\! x') \; , \label{F2aren} \\
G^{\rm ren}_{2a}(x;x') & = & -\frac{5 \kappa^2 H^2}{12 \pi^2} \, 
\ln(a) i\delta^4(x \!-\! x') \; . \label{G2aren}
\end{eqnarray}

\subsection{General Form of the 3-3 Diagram}

The 3-3 diagram is the leftmost part of Fig. \ref{vacpolgraphs}
and is by far the most difficult to evaluate. The first step is
to substitute the 3-point vertex factor (\ref{Vvert}) into 
expression (\ref{new3pt}). Because the spin two part of the
graviton propagator is traceless we retain only the second terms,
\begin{equation}
V^{\mu\rho\kappa\lambda\alpha\beta} \longrightarrow 4 
\eta^{\alpha [\mu} \eta^{\kappa] [\rho} \eta^{\lambda] \beta} 
\qquad , \qquad V^{\nu\sigma\theta\phi\gamma\delta} 
\longrightarrow 4 \eta^{\gamma [\nu} \eta^{\theta] [\sigma} 
\eta^{\phi] \delta} \; .
\end{equation} 
It is desirable to expand out the anti-symmetrizations over
$\rho \leftrightarrow \lambda$ and $\sigma \leftrightarrow \phi$,
and also to re-label the external derivatives from 
$\partial_{\kappa} \partial'_{\theta}$ to $\partial_{\rho}
\partial'_{\sigma}$,
\begin{eqnarray}
\lefteqn{i\Bigl[ \mbox{}^{\mu} \Pi^{\nu}_{\rm 3pt2}\Bigr](x;x') 
= \partial_{\rho} \partial'_{\sigma} \Biggl\{ \frac{2 \kappa^2
(a a')^D}{H^2} i \Bigl[\mbox{}^{[\mu}_{~~\alpha} 
\Delta^{2[[\nu}_{~~~\beta}\Bigr](x;x') \Biggl[ D^{\rho]} 
{D'}^{\sigma]]} y \times D^{\alpha} {D'}^{\beta} B} \nonumber \\
& & \hspace{-.5cm} -D^{\alpha} {D'}^{\sigma]]} y \!\times\! 
D^{\rho]} {D'}^{\beta} B \!-\! D^{\rho]} {D'}^{\beta} y \!\times\! 
D^{\alpha} {D'}^{\sigma]]} B \!+\! D^{\alpha} {D'}^{\beta} y 
\!\times\! D^{\rho]} {D'}^{\sigma]]} B \Biggr] \Biggr\} \; . 
\qquad \label{final3pt}
\end{eqnarray}
In expression (\ref{final3pt}) we have employed single and 
double square brackets to distinguish the anti-symmetrizations 
over $\mu \leftrightarrow \rho$ (single) from that over $\nu 
\leftrightarrow \sigma$ (double).

Recall from subsection 2.2 that the structure functions can be
computed directly from the portion of (\ref{final3pt}) within
the curly brackets. It is useful to give each of the four terms
its own symbol,
\begin{eqnarray}
\Bigl[\mbox{}^{\mu\rho}_{~1} T^{\nu\sigma}_{2}\Bigr](x;x') 
& \equiv &
+\frac{2 \kappa^2 (a a')^D}{H^2} \times i \Bigl[\mbox{}^{
[\mu}_{~~\alpha} \Delta^{2[[\nu}_{~~~\beta}\Bigr] \times
D^{\rho]} {D'}^{\sigma]]} y \times D^{\alpha} {D'}^{\beta} B
\; , \qquad \label{T1} \\
\Bigl[\mbox{}^{\mu\rho}_{~2} T^{\nu\sigma}_{2}\Bigr](x;x') 
& \equiv &
-\frac{2 \kappa^2 (a a')^D}{H^2} \times i \Bigl[\mbox{}^{
[\mu}_{~~\alpha} \Delta^{2[[\nu}_{~~~\beta}\Bigr] \times
D^{\alpha} {D'}^{\sigma]]} y \times D^{\rho]} {D'}^{\beta} B
\; , \qquad \label{T2} \\
\Bigl[\mbox{}^{\mu\rho}_{~3} T^{\nu\sigma}_{2}\Bigr](x;x') 
& \equiv &
-\frac{2 \kappa^2 (a a')^D}{H^2} \times i \Bigl[\mbox{}^{
[\mu}_{~~\alpha} \Delta^{2[[\nu}_{~~~\beta}\Bigr] \times
D^{\rho]} {D'}^{\beta} y \times D^{\alpha} {D'}^{\sigma]]} B
\; , \qquad \label{T3} \\
\Bigl[\mbox{}^{\mu\rho}_{~4} T^{\nu\sigma}_{2}\Bigr](x;x') 
& \equiv &
+\frac{2 \kappa^2 (a a')^D}{H^2} \times i \Bigl[\mbox{}^{
[\mu}_{~~\alpha} \Delta^{2[[\nu}_{~~~\beta}\Bigr] \times
D^{\alpha} {D'}^{\beta} y \times D^{\rho]} {D'}^{\sigma]]} B
\; . \qquad \label{T4}
\end{eqnarray}
Our notation is that the left hand subscript denotes which
of the four permutations is intended, while the right hand 
subscript ``2'' indicates that these are all contributions 
from the spin two part of the graviton propagator.

The next step is to act the derivatives on $B(y)$
\cite{Onemli:2002hr},
\begin{equation}
D^{\alpha} {D'}^{\beta} B(y) = \frac{\delta_0^{\alpha} 
\delta_0^{\beta} i \delta^D(x \!-\! x')}{a^{D+2}} +
D^{\alpha} y \, {D'}^{\beta} y \times B''(y) + D^{\alpha} 
{D'}^{\beta} y \times B'(y) \; .
\end{equation}
It is natural to distinguish the ``2b'' local delta function terms
(subsection 3.4) from the nonlocal terms. We further distinguish
the (nonlocal terms between the ``2c'' ones from the de Sitter 
breaking part of the graviton propagator (subsection 3.5) and 
the ``2d'' ones from the de Sitter invariant part (subsection 3.6).

\subsection{Local Contributions from the 3-3 Diagram}

The ``2b'' contributions come from the replacement in expressions
(\ref{T1}-\ref{T4}),\footnote{Note that the analytic continuation of
Barvinsky and Vilkovisky \cite{Barvinsky:1985an} would give a 
covariant result instead,
\begin{eqnarray}
D^{\rho} {D'}^{\sigma} y \times D^{\alpha} {D'}^{\beta} B 
\longrightarrow +\frac{2 H^2}{D} \, g^{\rho\sigma} g^{\alpha\beta} 
\, \frac{i\delta^D(x \!-\! x')}{\sqrt{-g}} \; . \nonumber
\end{eqnarray}
Because physics is ultimately based on the Minkowski signature of our
expression (\ref{ydef}) we feel it is safer to work with 
(\ref{locreplace}), even though it entails a noncovariant counterterm. 
Some of the problems which can arise from analytic continuation are 
explained in \cite{Miao:2013isa}.}
\begin{equation}
D^{\rho} {D'}^{\sigma} y \times D^{\alpha} {D'}^{\beta} B 
\longrightarrow -\frac{2 H^2}{a^{D+4}} \, \eta^{\rho\sigma}
\delta^{\alpha}_0 \delta^{\beta}_0 \, i\delta^D(x \!-\! x') \; .
\label{locreplace}
\end{equation}
The delta function re-introduces the coincident graviton 
propagator expression (\ref{coinc2}). With replacement 
(\ref{locreplace}) the four contractions in (\ref{final3pt}) 
give,
\begin{eqnarray}
\Bigl[\mbox{}^{\mu\rho}_{~1} T^{\nu\sigma}_{2b}\Bigr] 
&\!\!\!\!\!=\!\!\!\!\!& 4 \kappa^2 a^{D-4} i\delta^D(x\!-\!x') 
\Biggl\{\eta^{\mu [\nu} \eta^{\sigma] 
\rho} \, i\Delta_1 - \delta^{[\mu}_0 \delta^{[[\nu}_0 \eta^{\rho] 
\sigma]]} \Bigl(\frac{D\!-\!2}{D}\Bigr) i\Delta_1 \Biggr\} , 
\qquad \\
\Bigl[\mbox{}^{\mu\rho}_{~2} T^{\nu\sigma}_{2b}\Bigr] 
&\!\!\!\!\!=\!\!\!\!\!& 4 \kappa^2 a^{D-4} i\delta^D(x\!-\!x')
\Biggl\{ \delta^{[\mu}_0 
\delta^{[[\nu}_0 \eta^{\rho] \sigma]]} \Bigl( 
\frac{D\!+\!2}{D}\Bigr) i \Delta_1 \Biggr\} , \qquad \\
\Bigl[\mbox{}^{\mu\rho}_{~3} T^{\nu\sigma}_{2b}\Bigr] 
&\!\!\!\!\!=\!\!\!\!\!& 4 \kappa^2 a^{D-4} i\delta^D(x\!-\!x')
\Biggl\{ \delta^{[\mu}_0 
\delta^{[[\nu}_0 \eta^{\rho] \sigma]]} \Bigl( 
\frac{D\!+\!2}{D}\Bigr) i \Delta_1 \Biggr\} , \qquad \\
\Bigl[\mbox{}^{\mu\rho}_{~4} T^{\nu\sigma}_{2b}\Bigr] 
&\!\!\!\!\!=\!\!\!\!\!& 4 \kappa^2 a^{D-4} i\delta^D(x\!-\!x') 
\nonumber \\
& & \hspace{-.2cm} \times \Biggl\{ \delta^{[\mu}_0 
\delta^{[[\nu}_0 \eta^{\rho] \sigma]]} \Biggl[ -\frac{(D\!+\!2)
(D\!-\!1)}{D}\, i \Delta_1 -\frac{(D\!+\!1)(D\!-\!2)}{D\!-\!1} \, 
i \Delta_2 \Biggr] \Biggr\} . \qquad
\end{eqnarray}
We can read off the structure functions using the relation,
\begin{equation}
\delta^{[\mu}_0 \delta^{[[\nu}_0 \eta^{\rho] \sigma]]} =
-\frac12 \eta^{\mu [\nu} \eta^{\sigma] \rho} + \frac12
\overline{\eta}^{\mu [\nu} \overline{\eta}^{\sigma] \rho} \; .
\end{equation}
The results are,
\begin{eqnarray}
F_{2b} &\!\!\!\!=\!\!\!\!& \kappa^2 a^{D-4} i\delta^D(x\!-\!x')
\Biggl\{ \frac{(D\!+\!4) (D\!-\!2)}{D} \, i\Delta_1 
\!+\! \frac{(D\!+\!1)(D\!-\!2)}{(D\!-\!1)} \, 
i\Delta_2(x) \Biggl\} , \qquad \label{F2b} \\
G_{2b} &\!\!\!\!=\!\!\!\!& \kappa^2 a^{D-4} i\delta^D(x\!-\!x')
\Biggl\{ -\frac{(D^2\!-\!8)}{D} \, i\Delta_1
- \frac{(D\!+\!1)(D\!-\!2)}{(D\!-\!1)} \,
i\Delta_2(x) \Biggl\} . \qquad \label{G2b} 
\end{eqnarray}

The constant $i\Delta_1$ is divergent, which is why this
computation requires a noninvariant counterterm. Comparing
expressions (\ref{F2b}-\ref{G2b}) with (\ref{DF}-\ref{DG})
suggests that we take the ``2b'' contributions to $\overline{C}$
and $\Delta C$ as,
\begin{eqnarray}
\overline{C}_{2b} & = & -\frac{\kappa^2 i\Delta_1}{H^2}
\frac{(D\!+\!4)(D\!-\!2)}{4D} - \frac{\kappa^2 k 
\overline{A}_2}{H^2} \frac{(D\!+\!1)(D\!-\!2)}{4(D\!-\!1)} \; , 
\qquad \\
\Delta C_{2b} & = & \frac{\kappa^2 i\Delta_1}{H^2} 
\frac{(D^2 \!-\! 8)}{4 D} + \frac{\kappa^2 k \overline{A}_2}{H^2} 
\frac{(D\!+\!1)(D\!-\!2)}{4 (D\!-\!1)} \; . \label{C2b}
\end{eqnarray}
Our final results for the renormalized structure functions are,
\begin{eqnarray}
F_{2b}^{\rm ren}(x;x') & = & \frac{5 \kappa^2 H^2}{6 \pi^2} \,
\ln(a) i\delta^4(x \!-\! x') \; , \label{F2bren} \\
G_{2b}^{\rm ren}(x;x') & = & -\frac{5 \kappa^2 H^2}{6 \pi^2} \,
\ln(a) i\delta^4(x \!-\! x') \; . \label{G2bren}
\end{eqnarray}
Note that the noncovariant divergence in expression (\ref{G2b}) is
exactly cancelled by the noncovariant counterterm (\ref{C2b}), so
there are no spurious finite terms.

\subsection{Nonlocal de Sitter Breaking 3-3 Contributions}

The ``2c'' contributions derive from making the following
replacements in expressions (\ref{T1}-\ref{T4}),
\begin{eqnarray}
i\Bigl[\mbox{}^{\mu}_{~\alpha} \Delta^{2 \nu}_{~~\beta}
\Bigr](x;x') & \longrightarrow & \Bigl[ \overline{\eta}^{\mu\nu}
\overline{\eta}_{\alpha\beta} \!+\! \overline{\delta}^{\mu}_{\beta}
\overline{\delta}^{\nu}_{\alpha} \!-\! \frac{2}{D\!-\!1} 
\overline{\delta}^{\mu}_{\alpha} \overline{\delta}^{\nu}_{\beta}
\Bigr] \times k \Bigl[ \ln(4 a a') \!+\! A_2\Bigr] \; , \qquad 
\label{gravpropbrk} \\
D^{\alpha} {D'}^{\beta} B(y) & \longrightarrow & D^{\alpha}
{D'}^{\beta} y \!\times\! B'(y) + D^{\alpha} y \, {D'}^{\beta} y
\!\times\! B''(y) \; . \qquad \label{Bderivs}
\end{eqnarray}
Recall that an overlined tensor indicates the suppression of its
temporal components, $\overline{\delta}^{\mu}_{\alpha} \equiv
\delta^{\mu}_{\alpha} - \delta^{\mu}_0 \delta^0_{\alpha}$. These
overlined tensors in expression (\ref{gravpropbrk}) can all be
represented using the standard basis described in section 2.2 
\cite{Kahya:2011sy,Mora:2012kr},
\begin{equation}
\overline{\eta}^{\mu\nu} = \frac{a a'}{2 H^2} \Biggl\{-D^{\mu} 
{D'}^{\nu} y \!+\! D^{\mu} y \, {D'}^{\nu} u \!+\! D^{\mu} u \,
{D'}^{\nu} y \!+\! (2 \!-\! y) D^{\mu} u \, {D'}^{\nu} u \Biggr\}
\; . \label{bardef}
\end{equation}

\begin{table}
\renewcommand{\arraystretch}{2}
\setlength{\tabcolsep}{8pt}
\centering
\begin{tabular}{|c||c|c|c|c|}
\hline 
$f_i$ & $[\mbox{}^{\mu\rho}_{~1} T^{\nu\sigma}_{2c}]$ & 
$[\mbox{}^{\mu\rho}_{2+3} T^{\nu\sigma}_{2c}]$ &
$[\mbox{}^{\mu\rho}_{~4} T^{\nu\sigma}_{2c}]$ &
$[\mbox{}^{\mu\rho} T^{\nu\sigma}_{2c}]$ \\ 
\hline\hline 
$f_1$ & $-\frac{(D-2)(D+1)}{2(D-1)}$ & $(\frac{D+1}{D-1})$ & 
$-\frac{(D-2)(D+1)}{2(D-1)}$ & $\frac{(D-3)(D+1)}{D-1}$ \\ 
\hline 
$f_2$ & $0$ & $0$ & $0$ & $0$ \\ 
\hline 
$f_3$ & $-\frac{(D-2)(D+1)}{2(D-1)}$ & $(\frac{D+1}{D-1})$ 
& $-\frac{(D-2)(D+1)}{2(D-1)}$ & $-\frac{(D-3)(D+1)}{D-1}$ \\ 
\hline 
$\widetilde{f}_3$ & $-\frac{(D-2)(D+1)}{2(D-1)}$ & 
$(\frac{D+1}{D-1})$ & $-\frac{(D-2)(D+1)}{2(D-1)}$ & 
$-\frac{(D-3)(D+1)}{D-1}$ \\ 
\hline 
$f_4$ & $-\frac{(D-2)(D+1)}{2(D-1)} (2\!-\!y)$ & ${2(2-y)
\atop + \frac8{D-1} \cosh(v)}$ & $-\frac{(D-2)(D+1)}{2(D-1)} (2\!-\!y)$ 
& ${-\frac{(D-3)D}{D-1} (2-y) \atop + \frac8{D-1} 
\cosh(v)}$ \\ 
\hline 
$f_5$ & $0$ & $(\frac{D-3}{D-1})$ & $0$ & $(\frac{D-3}{D-1})$ \\ 
\hline 
\end{tabular}
\caption{The contribution proportional to $B'(y)$ from each 
permutation type to the coefficient functions $f_i(y,u,v)$ which
were defined in expression (\ref{fidefs}). Each contribution 
should be multiplied by $B'(y) \times \frac{4 \kappa^2 k}{H^2} 
[u + 2 \ln(2) + A_2]$. \label{B'fs}}
\end{table}

The first step is to substitute relations (\ref{gravpropbrk}), 
(\ref{Bderivs}) and (\ref{bardef}) into the standard permutations
(\ref{T1}-\ref{T4}) and read off the contributions to the five
coefficient functions $f_i(y,u,v)$ which were defined in
expression (\ref{fidefs}). Relation (\ref{Bderivs}) contains
a term proportional to $B'(y)$, whose contributions to each
$f_i$ is given in Table \ref{B'fs}. The $B''(y)$ contributions
are listed in Table \ref{B''fs}.

\begin{table}
\renewcommand{\arraystretch}{2}
\setlength{\tabcolsep}{8pt}
\centering
\begin{tabular}{|c||c|c|c|c|}
\hline 
$f_i$ & $[\mbox{}^{\mu\rho}_{~1} T^{\nu\sigma}_{2c}]$ & 
$[\mbox{}^{\mu\rho}_{2+3} T^{\nu\sigma}_{2c}]$ &
$[\mbox{}^{\mu\rho}_{~4} T^{\nu\sigma}_{2c}]$ &
$[\mbox{}^{\mu\rho} T^{\nu\sigma}_{2c}]$ \\ 
\hline\hline 
$f_1$ & ${(2-y) \atop - 2\cosh(v)}$ & $0$ & $0$ 
& ${(2-y) \atop - 2\cosh(v)}$ \\ 
\hline
$f_2$ & $\frac12 (\frac{D-3}{D-1})$ & $-(\frac{D+1}{D-1})$ 
& $\frac{(D-2)(D+1)}{2(D-1)}$ 
& $\frac{(D^2 -2D -7)}{2(D-1)}$ \\ 
\hline
$f_3$ & ${\frac12 (\frac{3D-5}{D-1}) (2-y) \atop 
-2(\frac{D-2}{D-1}) e^{v} - e^{-v}}$ 
& ${\frac12 (\frac{D-3}{D-1}) (2-y) \atop \frac2{D-1} e^{v}
- e^{-v}}$ & $0$ 
& ${2(\frac{D-2}{D-1})(2-y) \atop -2(\frac{D-3}{D-1}) e^{v}
- 2 e^{-v}}$ \\ 
\hline
$\widetilde{f}_3$ & ${\frac12 (\frac{3D-5}{D-1}) (2-y) \atop 
-e^{v} -2(\frac{D-2}{D-1} e^{-v}}$ 
& ${\frac12 (\frac{D-3}{D-1}) (2-y) \atop -e^{v} + \frac2{D-1} 
e^{-v}}$ & $0$
& ${2(\frac{D-2}{D-1})(2-y) \atop -2 e^{v} - 2(\frac{D-3}{D-1})
e^{-v}}$ \\ 
\hline
$f_4$ & $\!{2(\frac{D-3}{D-1}) + \frac{(3D-5)}{2(D-1)} (2-y)^2 
\atop - 4(\frac{D-2}{D-1}) (2-y) \cosh(v)}\!$ & $0$ & $0$ 
& $\!{2(\frac{D-3}{D-1}) + \frac{(3D-5)}{2(D-1)} (2-y)^2 
\atop - 4(\frac{D-2}{D-1}) (2-y) \cosh(v)}\!$ \\ 
\hline 
$f_5$ & $0$ 
& ${-(\frac{3D-5}{D-1}) (2-y) \atop -4 (\frac{D-3}{D-1}) \cosh(v)}$ 
& $\frac{(D-2)(D+1)}{2(D-1)} (2\!-\!y)$ 
& ${\frac{(D^2-7D+8)}{2(D-1)} (2-y) \atop + 4 (\frac{D-3}{D-1})
\cosh(v)}$ \\ 
\hline 
\end{tabular} 
\caption{The contribution proportional to $B''(y)$ from each 
permutation type to the coefficient functions $f_i(y,u,v)$ which
were defined in expression (\ref{fidefs}). Each contribution 
should be multiplied by $B''(y) \times \frac{4 \kappa^2 k}{H^2} 
[u + 2 \ln(2) + A_2]$. \label{B''fs}}
\end{table}

The next step is to compute the master structure functions 
$\Phi(y,u,v)$ and $\Psi(y,u,v)$ according to the rules which
were originally derived in Ref. \cite{Leonard:2012ex} and which
are summarized in Table \ref{fitoPhiPsi} of section 2.2. The
intermediate results can be substatially simplified using the
$B$-type propagator equation,
\begin{equation}
(4y \!-\! y^2) B''(y) + D(2\!-\!y) B'(y) - (D\!-\!2) B(y) =
0 \; . \label{Bprop}
\end{equation}
The final result for the ``2c'' contribution to $\Phi(y,u,v)$ is,
\begin{eqnarray}
\lefteqn{ \Phi_{2c}(y,u,v) = \kappa^2 H^2 k \Bigl( \frac{D}2 
\!-\! 1\Bigr) \Bigl[ u \!+\! 2\ln(2) \!+\! A_2\Bigr] } 
\nonumber \\
& & \hspace{2.2cm} \times \Biggl\{ 4\Bigl[ (2\!-\!y) B\Bigr]''' 
\cosh(v) - 8 B''' - (D\!-\!3) \Bigl[ (2\!-\!y) B\Bigr]'' \Biggr\} . 
\qquad \label{Phi2c}
\end{eqnarray}
The ``2c'' contribution to $\Psi(y,u,v)$ is more complicated
because it involves derivatives with respect to $u$,
\begin{eqnarray}
\lefteqn{ \Psi_{2c}(y,u,v) = \frac{\kappa^2 H^2 k}{D \!-\! 1} 
\Biggl\{ 2 (D^2 \!-\!3D \!+\!4) \Bigl[-(2\!-\!y)B \cosh(v) 
\!+\! 2 B\Bigr]''' } \nonumber \\
& & \hspace{-.5cm} + (D\!-\!3)(D\!-\!2)(D\!-\!1) \Bigl[ 
(2\!-\!y) B\Bigr]'' \Biggr\} \!+\! 2 \kappa^2 H^2 k 
\Bigl(\frac{D\!-\!2}{D\!-\!1}\Bigr) \Bigl[u \!+\! 2\ln(2) \!+\! 
A_2\Bigr] \nonumber \\
& & \hspace{2.2cm} \times \Biggl\{ \Bigl[ 8 B'
+ (D\!-\!1) (2\!-\!y) B\Bigr]''' \cosh(v) - 4 \Bigl[(2\!-\!y) B''
\Bigr]'' \Biggr\} . \qquad \label{Psi2c}
\end{eqnarray}

The structure function $F(x;x')$ is constructed by integrating 
$\Phi(y,u,v)$ with respect to $y$ according to the expression
(\ref{Frule}). This is simple to do because each term in 
(\ref{Phi2c}) is either a 2nd or 3rd derivative with respect to
$y$,
\begin{eqnarray}
\lefteqn{ F_{2c}(x;x') = (D\!-\!2) \kappa^2 H^2 k (a a')^{D-2} 
\Bigl[ u \!+\! 2\ln(2) \!+\! A_2\Bigr] } \nonumber \\
& & \hspace{2.2cm} \times \Biggl\{ -4\Bigl[ (2\!-\!y) B\Bigr]'' 
\cosh(v) + 8 B'' + (D\!-\!3) \Bigl[ (2\!-\!y) B\Bigr]' \Biggr\} . 
\qquad \label{F2c0}
\end{eqnarray}
The other structure function follows from substituting
(\ref{Phi2c}-\ref{Psi2c}) in (\ref{Grule}),
\begin{eqnarray}
\lefteqn{ G_{2c}(x;x') = \frac{2 (D\!-\!3) D}{D\!-\!1} \, \kappa^2
H^2 k (a a')^{D-2} \Biggl\{ \Bigl[ (2\!-\!y) B\Bigr]' \cosh(v)
- 2 B'\Biggr\} } \nonumber \\
& & \hspace{-.5cm} + \Bigl( \frac{D\!-\!2}{D\!-\!1}\Bigr) \kappa^2
H^2 k (a a')^{D-2} \Bigl[ u \!+\! 2\ln(2) \!+\! A_2\Bigr] \Biggl\{
\Bigl[ 16 B - 4 (D\!-\!1) y B\Bigr]'' \cosh(v) \nonumber \\
& & \hspace{.9cm} - \Bigl[16 B + 4 (D\!-\!3) y B\Bigr]'' - 4
(D\!-\!3)^2 B' + (D\!-\!3) (D\!-\!1) (y B)' \Biggr\} . \qquad
\label{G2c0}
\end{eqnarray}

These results (\ref{F2c0}-\ref{G2c0}) are valid in $D$ dimensions.
Because the quantum-corrected Maxwell equation (\ref{maxeqn}) 
involves integrals of $F(x;x')$ and $G(x;x')$ with respect to 
${x'}^{\mu}$, we can set $D=4$ for any part of the structure 
functions which diverges less strongly than $1/(x-x')^4$ as 
$x' \rightarrow x$. Expression (\ref{B(y)}) of Appendix A gives 
the expansion of $B(y)$,
 \begin{equation}
B(y) = \frac{H^{D-2}}{4 \pi^{\frac{D}2}} 
\frac{\Gamma(\frac{D}2\!-\!1)}{y^{\frac{D}2-1}} + O\Bigl( 
(D\!-\!4) y^0\Bigr) \; . \label{Bexp}
\end{equation}
Because $y = a a' H^2 \Delta x^2$, it might seem that the terms 
in (\ref{F2c0}-\ref{G2c0}) which involve $B''(y)$ harbor 
quadratic divergences, while those involving $y B''(y)$ and 
$B'(y)$ diverge logarithmically. In fact all divergences cancel. 
To see this first note that the factor of $\cosh(v)$ can be 
rewritten as,
\begin{equation} 
\cosh(v) = \frac12 \Bigl[ \frac{a}{a'} \!+\! \frac{a'}{a}\Bigr]
= 1 + \frac12 a a' H^2 (\eta \!-\! \eta')^2
\equiv 1 + \frac12 a a' H^2 \Delta \eta^2 \; . \label{cosh}
\end{equation}
This demonstrates that the quadratic divergences of $F_{2c}$ and 
$G_{2c}$ cancel,
\begin{eqnarray}
\lefteqn{ -4\Bigl[ (2\!-\!y) B\Bigr]'' \cosh(v) + 8 B'' + 
(D\!-\!3) \Bigl[ (2\!-\!y) B\Bigr]' = -4 a a' H^2 \Delta 
\eta^2 B'' } \nonumber \\
& & \hspace{1.5cm} + 4 (y B)'' + 2(D\!-\!3) B' + 2 a a' H^2 
\Delta \eta^2 (y B)'' - (D \!-\!3) (y B)' \; , \qquad 
\label{Fcancel1} \\
\lefteqn{\Bigl[ 16 B - 4 (D\!-\!1) y B\Bigr]'' \cosh(v)
- \Bigl[16 B + 4 (D\!-\!3) y B\Bigr]'' - 4 (D\!-\!3)^2 B' }
\nonumber \\
& & \hspace{0cm} + (D\!-\!3) (D\!-\!1) (y B)' = 8 a a' H^2 
\Delta \eta^2 B'' -8 (D\!-\!2) (y B)'' - 4 (D\!-\!3)^2 B' 
\nonumber \\
& & \hspace{3.2cm} - 2(D\!-\!1) a a' H^2 \Delta \eta^2 
(y B)'' + (D\!-\!3) (D\!-\!1) (y B)' \; . \qquad 
\label{Gcancel1}
\end{eqnarray}

Cancelling the logarithmic divergences is more subtle. 
Relations (\ref{Bexp}) and (\ref{cosh}) suffice for the
$u$-independent part of $G_{2c}$,
\begin{equation}
\Bigl[ (2\!-\!y) B\Bigr]' \cosh(v) - 2 B' = a a' H^2 \Delta
\eta^2 B' - (y B)' \cosh(v) \longrightarrow -\frac1{4\pi^2}
\frac{\Delta \eta^2}{a a' \Delta x^4} \; . \label{Gcancel1.5}
\end{equation}
In the $u$-dependent parts we can ignore $(yB)'$ and 
$\Delta \eta^2 (y B)''$ which are both integrable and 
vanish in $D=4$ dimensions. Further, we need only retain 
the first term of (\ref{Bexp}) in evaluating the combinations
of $\Delta \eta^2 B''$, $(yB)''$ and $B'$ which appear in
expressions (\ref{Fcancel1}) and (\ref{Gcancel1}),
\begin{eqnarray}
\lefteqn{-4 a a' H^2 \Delta \eta^2 B'' + 4 (y B)'' + 2
(D\!-\!3) B' } \nonumber \\
& & \hspace{3.8cm} = \frac{H^{D-2} \Gamma(\frac{D}2)}{4 
\pi^{\frac{D}2}} \Biggl\{ -\frac{2 D a a' H^2 \Delta 
\eta^2}{y^{\frac{D}2+1}} - \frac{2}{y^{\frac{D}2}} + \ldots 
\Biggr\} , \qquad \label{Fcancel2} \\
\lefteqn{8 a a' H^2 \Delta \eta^2 B'' -8 (D\!-\!2) (y B)'' 
- 4 (D\!-\!3)^2 B' } \nonumber \\
& & \hspace{4cm} =\frac{H^{D-2} \Gamma(\frac{D}2)}{4 
\pi^{\frac{D}2}} \Biggl\{ \frac{4 D a a' H^2 \Delta 
\eta^2}{y^{\frac{D}2+1}} + \frac{4}{y^{\frac{D}2}} + \ldots 
\Biggr\} . \qquad \label{Gcancel2}
\end{eqnarray}
Expressions (\ref{Fcancel2}) and (\ref{Gcancel2}) are
proportional to the same function which can be reduced to
a form that is integrable in $D=4$,
\begin{eqnarray}
\lefteqn{\frac{D a a' H^2 \Delta \eta^2}{y^{\frac{D}2+1}} + 
\frac{1}{y^{\frac{D}2}} = \frac1{(H^2 a a')^{\frac{D}2}}
\Biggl\{ \frac{D \Delta \eta^2}{\Delta x^{D+2}} + 
\frac1{\Delta x^D}\Biggr\} \; , } \\
& & \hspace{2cm} = \frac1{(D\!-\!2) (H^2 a a')^{\frac{D}2}} 
\Biggl\{\partial_0^2 \Bigl( \frac1{\Delta x^{D-2}}\Bigr) +
\frac{4 \pi^{\frac{D}2} i\delta^D(x \!-\! x')}{
\Gamma(\frac{D}2 \!-\!1)} \Biggr\} \; . \qquad
\end{eqnarray}
After some simplifications our final unregulated forms are,
\begin{eqnarray}
F_{2c}^{\rm ren}(x;x') & = & -\frac{\kappa^2 H^2}{16 \pi^4}
\Bigl[ \ln\Bigl(\frac{a a'}4\Bigr) \!+\! \frac13 \!-\! 2\gamma\Bigr] 
\, \nabla^2 \Bigl( \frac1{\Delta x^2} \Bigr) \; , 
\qquad \label{F2cren} \\
G_{2c}^{\rm ren}(x;x') & = & \frac{\kappa^2 H^2}{24 \pi^4}
\Biggl\{ \frac{H^2 a a'}{4} (\partial_0^2 \!+\! \nabla^2) 
\ln(H^2 \Delta x^2) \nonumber \\
& & \hspace{4cm} + \Bigl[ \ln\Bigl(\frac{a a'}4\Bigr) \!+\! \frac13 
\!-\! 2 \gamma\Bigr] \, \nabla^2 \, \frac1{\Delta x^2} \Biggr\}
. \label{G2cren} \qquad
\end{eqnarray}

\subsection{Nonlocal de Sitter Invariant 3-3 Contributions}

The ``2d'' contributions derive from making the following
replacements in expressions (\ref{T1}-\ref{T4}),
\begin{eqnarray}
i\Bigl[\mbox{}^{\mu}_{~\alpha} \Delta^{2 \nu}_{~~\beta}
\Bigr](x;x') & \longrightarrow & \sum_{i=1}^5 \Bigl[
\mbox{}^{\mu}_{~\alpha} \mathcal{T}^{i\nu}_{~~\beta}\Bigr](x;x') 
\times K_i(y) \; , \qquad \label{gravpropinv} \\
D^{\alpha} {D'}^{\beta} B(y) & \longrightarrow & D^{\alpha}
{D'}^{\beta} y \!\times\! B'(y) + D^{\alpha} y \, {D'}^{\beta} y
\!\times\! B''(y) \; . \qquad
\end{eqnarray}
Recall that the basis tensors $[\mbox{}_{\mu\nu} \mathcal{T}^i_{
\rho\sigma}](x;x')$ are listed in Table \ref{Tees} and the 
functions $K_i(y)$ are given in Table \ref{Kays}. Because 
everything is de Sitter invariant, the only coefficient 
functions $f_i$ from expression (\ref{fidefs}) which occur 
are $f_1(y)$ and $f_2(y)$. It is simplest to extract a factor of
$\kappa^2 H^2$, and to report results from each of the basis
tensors $[\mbox{}^{\mu}_{~\alpha} \mathcal{T}^{i\nu}_{~~\beta}
](x;x')$ in the form,
\begin{equation}
f_1(y) \equiv -\kappa^2 H^2 \sum_{i=1}^5 (\Delta \phi_1)_i(y) 
\times K_i(y) \; , \; f_2(y) \equiv -\kappa^2 H^2 
\sum_{i=1}^5 (\Delta \phi_2)_i(y) \times K_i(y) . \label{dSfis}
\end{equation}
Our results for the functions $(\Delta \phi_1)_i(y)$ and
$(\Delta \phi_2)_i(y)$ are given in Table \ref{Dphi}. Relations
(\ref{BCident}) and (\ref{Bprop}) are helpful in simplifying 
these expressions.

\begin{table}
\renewcommand{\arraystretch}{2}
\setlength{\tabcolsep}{8pt}
\centering
\begin{tabular}{|c||c|c|}
\hline 
$i$ & $(\Delta \phi_1)_{i}(y)$ & $(\Delta \phi_2)_i(y)$ \\ 
\hline\hline 
1 & $-8(D\!-\!1) B' + 2 (D\!-\!2) (2\!-\!y) C'$ 
& $4D B''$ \\ 
\hline 
2 & $(D\!-\!2) (4y \!-\!y^2) C'$ 
& $4(D\!-\!2) B' - 2(D\!-\!2) (2\!-\!y) C'$ \\ 
\hline 
3 & $0$ & $-4 (D\!-\!2) (4y \!-\! y^2) C'$ \\ 
\hline 
4 & $0$ & $-16 (D\!-\!2) C'$ \\ 
\hline
5 & $-8 B'$ & $8 B''$ \\ 
\hline
\end{tabular} 
\caption{The functions $(\Delta \phi_1)_i(y)$ and $(\Delta 
\phi_2)_i(y)$ defined in equation (\ref{dSfis}).
\label{Dphi}}
\end{table}

The next step is to construct the master structure functions.
Only $\Phi(y)$ is nonzero because everything is de Sitter 
invariant. We have found it useful to extract a factor of
$-\frac14 \kappa^2 H^6$, and to distinguish the terms which
are proportional to $K_i(y)$ from those which are proportional
to $K_i'(y)$,
\begin{eqnarray}
\lefteqn{ \Phi_{2d}(y) = -\frac14 \kappa^2 H^6 \sum_{i=1}^5
\Biggl\{ \Bigl[ 2 (D\!-\!1) (\Delta \phi_1)_i \!-\! 2(2\!-\!y)
(\Delta \phi_1)'_i \!+\! D (2\!-\!y) (\Delta \phi_2)_i }
\nonumber \\
& & \hspace{-.2cm} +(4y \!-\! y^2) (\Delta \phi_2)'_i\Bigr] 
\!\times\! K_i + \Bigl[-2 (2\!-\!y) (\Delta \phi_1)_i 
\!+\! (4y \!-\!y^2) (\Delta \phi_2)_i\Bigr] \!\times\! K'_i
\Biggr\} . \qquad \label{Phi2ddef}
\end{eqnarray}
Our results for the coefficients of each $K_i(y)$ and $K'_i(y)$
are reported in Table \ref{KandK'}. Substituting the $K_i(y)$ 
from Table \ref{Kays} and adding everything up gives,
\begin{eqnarray}
\lefteqn{\Phi_{2d}(y) = \frac{(D\!-\!2)(D\!+\!1)}{D\!-\!1} \,
\kappa^2 H^6 } \nonumber \\
& & \hspace{-.5cm} \times \Biggl\{ 8 B' \Bigl[ -\Bigl( 
\frac{\square}{H^2} J \Bigr)'' \!+\! (D\!-\!2) (2\!-\!y) J''' 
\!-\! (D\!-\!2)(D\!-\!1) J''\Bigr] \nonumber \\
& & \hspace{0cm} + (D\!-\!2) C' \Bigl[ \Bigl( \frac{\square^2}{H^4}
J \Bigr)' \!\!-\! 2 (2\!-\!y) \Bigl( \frac{\square}{H^2} J\Bigr)''
\!\!+\! D \Bigl( \frac{\square}{H^2} J\Bigr)' \!\!+\! 8 (D\!-\!2) 
J''' \Bigr] \Biggr\} . \qquad \label{Phi2dJ}
\end{eqnarray}

\begin{table}
\renewcommand{\arraystretch}{2}
\setlength{\tabcolsep}{8pt}
\centering
\begin{tabular}{|c||c|c|}
\hline 
$i$ & ${2(D-1) (\Delta \phi_1)_{i} - 2(2-y) (\Delta \phi_1)'_i
\atop + D (2-y) (\Delta \phi_2)_i + (4y - y^2) (\Delta \phi_2)'_i}$ 
& $\!-2(2\!-\!y) (\Delta \phi_1)_i + (4y\!-\!y^2) (\Delta \phi_2)_i\!$ \\ 
\hline\hline
1 & $-8(D\!-\!1) B' + 8 (2\!-\!y) C'$ 
& $\!8 (2\!-\!y) B' \!-\! 8(D\!+\!2) C' \!+\! 4 (4y\!-\!y^2) C'\!$ \\ 
\hline 
2 & ${4 (D+2) (2-y) B' \atop - 32 C' + 12 (4y-y^2) C'}$ 
& $\!\!4(4y\!-\!y^2) B' \!-\! 4 (2\!-\!y) (4y \!-\!y^2) C'\!\!$ \\ 
\hline 
3 & $\!16 (4y\!-\!y^2) B' \!-\! 16 (2\!-\!y) (4y \!-\! y^2) C'\!$ 
& $-4 (4y \!-\! y^2)^2 C'$ \\ 
\hline 
4 & $64 B' - 32 (2\!-\!y) C'$ & $-16 (4y\!-\!y^2) C'$ \\ 
\hline
5 & $0$ & $-16 C'$ \\ 
\hline
\end{tabular} 
\caption{The coefficients of $K_i(y)$ and $K'_i(y)$ in equation
(\ref{Phi2ddef}). An overall factor of $(D\!-\!2)$ has been extracted 
from each term. \label{KandK'}}
\end{table}

The expansions of $B'(y)$ and $C'(y)$ are given in expressions
(\ref{Bprim}-\ref{Cprim}) of Appendix A. The function $J(y)$ was 
defined in expression (\ref{Jzero}) and its expansion is given in 
expression (\ref{Jexp}) of Appendix B. When the various factors
in equation (\ref{Phi2dJ}) are combined the result is,
\begin{eqnarray}
\lefteqn{ \Phi_{2d}(y) = \frac{\kappa^2 H^{2D-2} \Gamma^2(
\frac{D}2)}{16 \pi^D} \frac{(D\!-\!2)(D\!+\!1)}{(D \!-\! 1)} 
\Biggl\{ \frac{(D \!-\! 4) (D \!+\! 2)}{4 y^D} } \nonumber \\
& & \hspace{3cm} + \frac{(3 D^2 \!-\! 26 D \!+\! 52) D^2}{48 (D 
\!-\! 2) y^{D-1}} + \frac{\frac43 \ln(\frac{y}{4})}{y^2 (4 \!-\! 
y)^2} + \frac{\frac13}{y^2 (4 \!-\! y)} \Biggr\} . \qquad 
\label{Phi2d}
\end{eqnarray}
For this case the master structure function $\Psi(y,u,v)$
vanishes so we have,
\begin{eqnarray}
F_{2d}(x;x') & = & - 2 (a a')^{D-2} I\Bigl[\Phi_{2d}(y) \Bigr] 
\; , \label{F2dformula} \\
G_{2d}(x;x') & = & (a a')^{D-2} I^2\Bigl[(D \!-\! 1) \Phi_{2d}(y)
+ y \Phi_{2d}'(y)\Bigr] \; . \label{G2dformula}
\end{eqnarray}
Note that the $1/y^{D-1}$ term in (\ref{Phi2d}) drops out of
expression (\ref{G2dformula}).

Substituting (\ref{Phi2d}) into (\ref{F2dformula}), performing the 
integration and recalling that $y = a a' H^2 \Delta x^2$ gives, 
\begin{eqnarray}
\lefteqn{ F_{2d}(x;x') = \frac{\kappa^2 \Gamma^2(\frac{D}2)}{32 \pi^D}
\frac{(D \!-\! 4) (D\!-\! 2) (D \!+\!1) (D \!+\! 2)}{(D \!-\! 1)^2 a a' 
\Delta x^{2D-2} } } \nonumber \\
& & \hspace{1.5cm} + \frac{\kappa^2 H^2 \Gamma^2(\frac{D}2)}{384 \pi^D}
\frac{(3D^2 \!-\! 26D \!+\! 52) D^2 (D \!+\! 1)}{(D\!-\!2) (D \!-\!1)
\Delta x^{2D-4}} \nonumber \\ 
& & \hspace{2.5cm} + \frac{5 \kappa^2 H^6 (a a')^2}{144 \pi^4} \Biggl\{ 
-\frac12 \mathcal{L}(y) - \Bigl[\frac1{y} \!-\! \frac1{4 \!-\! y}\Bigr] 
\ln\Bigl(\frac{y}4\Bigr) - \frac{2}{y} \Biggr\} . \qquad
\end{eqnarray}
where we define the function $\mathcal{L}(y)$ as,
\begin{equation}
\mathcal{L}(y) \equiv {\rm Li}_2\Bigl(\frac{y}4\Bigr) + 
\ln\Bigl(1 \!-\! \frac{y}4\Bigr) \ln\Bigl(\frac{y}4\Bigr)
- \frac12 \ln^2\Bigl(\frac{y}4\Bigr) \; . \label{Ldef}
\end{equation}
The dilogarithm function ${\rm Li}_2(z)$ is defined in (\ref{dilog}).

Renormalization is accomplished by first localizing the ultraviolet
divergence by a combination of partial integration, adding zero in
the form of a delta function identity, and taking $D=4$ in the finite,
integrable remainder \cite{Onemli:2002hr},
\begin{eqnarray}
\lefteqn{\frac1{\Delta x^{2D-4}} = \frac{\partial^2}{2 (D\!-\!3)(D\!-\!4)}
\Bigl[ \frac1{\Delta x^{2D-6}} \Bigr] \; , } \\
& & = \frac{4 \pi^{\frac{D}2} \mu^{D-4} i \delta^D(x \!-\! x')}{2 
(D \!-\! 3) (D \!-\! 4) \Gamma( \frac{D}2 \!-\! 1)} + 
\frac{\partial^2}{2 (D \!-\! 3)(D\!-\! 4)} \Bigl[ 
\frac1{\Delta x^{2D-6}} - \frac{\mu^{D-4}}{\Delta x^{D-2}}\Bigr] 
\; , \qquad \\
& & \longrightarrow \frac{4 \pi^{\frac{D}2} \mu^{D-4} i 
\delta^D(x \!-\! x')}{2 (D \!-\! 3) (D \!-\! 4) \Gamma( \frac{D}2 
\!-\! 1)} -\frac{\partial^2}4 \Bigl[ \frac{ \ln(\mu^2 \Delta x^2)}{
\Delta x^2} \Bigr] \; . \label{Dx2D-4}
\end{eqnarray}
We also employ the relation,
\begin{equation}
\frac1{\Delta x^{2D-2}} \longrightarrow \frac{4 \pi^{\frac{D}2} 
\mu^{D-4} \partial^2 i \delta^D(x \!-\! x')}{4 (D\!-\!2)^2 (D \!-\! 3) 
(D \!-\! 4) \Gamma(\frac{D}2 \!-\! 1)} -\frac{\partial^4}{32} \Bigl[ 
\frac{ \ln(\mu^2 \Delta x^2)}{\Delta x^2} \Bigr] \; . \label{Dx2D-2}
\end{equation}
The localized ultraviolet divergence is absorbed by a counterterm of 
the form (\ref{DF}) with the $C_4$ and $\overline{C}$ coefficients,
\begin{eqnarray}
C_4^{2d} & = & \frac{\kappa^2 \mu^{D-4} \Gamma(\frac{D}2)}{64 
\pi^{\frac{D}2}} \frac{(D \!+\!1) (D \!+\! 2)}{(D \!-\! 1)^2
(D \!-\! 3)} \; , \\
\overline{C}^{2d} & = & \frac{\kappa^2 H^2 \mu^{D-4} 
\Gamma(\frac{D}2)}{96 \pi^{\frac{D}2}} \Biggl\{ \frac{D^2 
(D \!+\!1)}{(D \!-\! 1)(D\!-\! 3)(D \!-\! 4)} \nonumber \\
& & \hspace{2cm} -
\frac{(3D \!-\! 14) D^2 (D\!+\!1)}{4 (D \!-\! 1) (D\!-\! 3)} +
\frac{3 (3D \!-\! 8) (D \!+\! 1) (D \!+\! 2)}{2 (D\!-\! 1)^2 
(D \!-\! 3)} \Biggr\} . \qquad 
\end{eqnarray}
The final renormalized result is,
\begin{eqnarray}
\lefteqn{ F^{\rm ren}_{2d}(x;x') = \frac{5 \kappa^2 H^2}{3^2 \pi^2} \, 
\ln(a) i\delta^4( x \!-\! x') + \frac{5 \kappa^2 H^2}{2^4 3^2 \pi^4} \, 
\partial^2 \Bigl[ \frac{ \ln(\mu^2 \Delta x^2)}{\Delta x^2} \Bigr] } 
\nonumber \\
& & \hspace{2.5cm} + \frac{5 \kappa^2 H^6 (a a')^2}{144 \pi^4} \Biggl\{ 
-\frac12 \mathcal{L}(y) - \Bigl[\frac1{y} \!-\! \frac1{4 \!-\! y}\Bigr] 
\ln\Bigl(\frac{y}4\Bigr) - \frac{2}{y} \Biggr\} . \qquad \label{F2dren}
\end{eqnarray}

The second structure function is,
\begin{eqnarray}
\lefteqn{ G_{2d}(x;x') = -\frac{\kappa^2 H^2 \Gamma^2(\frac{D}2)}{64 
\pi^D} \frac{(D \!-\!4) (D\!+\!1) (D\!+\!2)}{(D \!-\!1)^2 \Delta 
x^{2D-4}} } \nonumber \\
& & \hspace{2.5cm} + \frac{5 \kappa^2 H^6 (a a')^2}{72 \pi^4} \Biggl\{ 
\frac14 (1 \!-\!y) \mathcal{L}(y) + \Bigl[ \frac1{4 \!-\! y} \!-\! 1
\Bigr] \ln\Bigl(\frac{y}4\Bigr) \Biggr\} \; . \qquad
\end{eqnarray} 
We go through the same procedure of localization as before, and add a
counterterm of the form (\ref{DG}) to produce the final renormalized 
result,
\begin{equation}
G^{\rm ren}_{2d}(x;x') = \frac{5 \kappa^2 H^6 (a a')^2}{72 \pi^4} 
\Biggl\{ \frac14 (1 \!-\!y) \mathcal{L}(y) +  \Bigl[ \frac1{4 \!-\! y} 
\!-\! 1\Bigr] \ln\Bigl(\frac{y}4\Bigr) \Biggr\} \; . \label{G2dren}
\end{equation}

\subsection{The Full Spin 2 Contribution}

It remains to combine the various spin two contributions to the
two structure functions. Our results for $F_2(x;x')$ are relations
(\ref{F2aren}), (\ref{F2bren}), (\ref{F2cren}) and (\ref{F2dren}),
\begin{eqnarray}
\lefteqn{ F_2(x;x') = \frac{85 \kappa^2 H^2}{72 \pi^2} \, \ln(a) 
i\delta^4(x \!-\! x') - \frac{\kappa^2 H^2}{ 16 \pi^4} \Bigl[
\ln\Bigl(\frac{a a'}4\Bigr) \!+\! \frac13 \!-\! 2\gamma\Bigr] \, 
\nabla^2 \Bigl( \frac1{\Delta x^2} \Bigr) } \nonumber \\
& & \hspace{-.5cm} + \frac{5 \kappa^2 H^2}{144 \pi^4} \, \partial^2
\Bigl( \frac{ \ln(\mu^2 \Delta x^2)}{\Delta x^2} \Bigr) 
- \frac{5 \kappa^2 H^6 (a a')^2}{144 \pi^4} \Biggl\{ 
\frac{\mathcal{L}(y)}{2} \!+\! \frac{2 (2 \!-\! y) \ln(\frac{y}4)}{
4 y \!-\! y^2} \!+\! \frac{2}{y} \Biggr\} . \qquad \label{F2ren}
\end{eqnarray}
Our results for the other structure function are given in equations
(\ref{G2aren}), (\ref{G2bren}), (\ref{G2cren}) and (\ref{G2dren}),
\begin{eqnarray}
\lefteqn{ G_2(x;x') = -\frac{5 \kappa^2 H^2}{4 \pi^2} \, \ln(a) 
i\delta^4(x \!-\! x') + \frac{\kappa^2 H^2}{ 24 \pi^4} \Bigl[
\ln\Bigl(\frac{a a'}4\Bigr) \!+\! \frac13 \!-\! 2 \gamma\Bigr] \, 
\nabla^2 \Bigl( \frac1{\Delta x^2} \Bigr) } \nonumber \\
& & \hspace{-.7cm} + \frac{\kappa^2 H^4 a a'}{96 \pi^4} (\partial^2_0 
\!\!+\!\! \nabla^2) \!\ln(\!H^2\! \Delta x^2\!) \!+\! \frac{5 
\kappa^2 H^6 (a a')^2}{72 \pi^4} \Biggl\{\!\! \frac{(1 \!-\! y) 
\mathcal{L}(y)}{4} \!+\! \frac{(y \!-\! 3) \ln(\frac{y}4)}{4 \!-\! y}
\! \Biggr\}\! . \qquad \label{G2ren}
\end{eqnarray}
Note that there can also be arbitrary finite contributions of the 
form (\ref{DF}-\ref{DG}).

\section{Spin 0 Contributions}\label{spin0}

This section follows the analysis of Section \ref{spin2} applied to spin
zero part of the graviton propagator. Its purpose is to work out the
contributions to the renormalized vacuum polarization structure functions 
coming from this part of the propagator.
Firstly, the description of the spin 0 part of the graviton propagator is given.
Next, we work out the contributions from the 4-point diagram ("0a") and 
the local part of the 3-3 diagram ("0b"). Working out the contribution from the 
(de Sitter invariant) nonlocal part of the 3-3 diagram 
("0d") comprises the largest part of this section.
In the end the full results for $F(x;x')$ and $G(x;x')$ are given.

Many of the tensor structure contractions performed in this section were
checked with the help of the {\it xTensor} part of the
tensor calculus package {\it xAct} \cite{xAct} written for {\it Mathematica}.

\subsection{Spin 0 Part of the Graviton Propagator}

The spin 0 part of the graviton propagator depends on the 
gauge parameter $b$ introduced in the gauge condition 
(\ref{gauge}). Instead of working with $b$, 
we find it more convenient to use parameter
$\beta$, defined to be
\begin{equation}
\beta = \frac{D b \!-\! 2}{b \!-\! 2} ,
\end{equation}
Note that for the range of gauges $b>2$ considered in this work,
$\beta$ is a monotonically decreasing function of $b$, with $\beta>D$.

The spin zero part of the graviton propagator
can be written as two projectors acting on a scalar structure function
\cite{Mora:2012zi},
\begin{equation}
i\Bigl[\mbox{}_{\mu\nu} \Delta^0_{\rho\sigma}\Bigr](x;x') 
	= \frac{-2 \beta^2}{(D \!-\! 1)(D \!-\! 2)} 
		\times \mathcal{P}_{\mu\nu}(x)
		\times \mathcal{P}_{\rho\sigma}(x') 
		\times \Bigl[ i \Delta_{WNN}(x;x') \Bigr] \; ,
\label{gravpropspin0}
\end{equation}
where the projector $\mathcal{P}_{\mu\nu}(x)$ is a 2nd order differential
operator,
\begin{equation}
\mathcal{P}_{\mu\nu}(x) = D_{\mu} D_{\nu} 
	- \frac{g_{\mu\nu}(x)}{\beta} \Bigl[ \square 
		- (\beta \!-\! D)H^2 \Bigr] \; .
\end{equation}
The spin 0 scalar structure function $i\Delta_{WNN}(x;x')$
is obtained by inverting twice the kinetic operator for an
$M_S^2=(\beta \!-\! D)H^2$ scalar, and once for an $M_S^2=-DH^2$
scalar. The order of these inversions is irrelevant, we find 
the following one convenient,
\begin{eqnarray}
\Bigl[ \square - (\beta \!-\! D)H^2 \Bigr] i\Delta_{WNN}(x;x') 
	& = & i\Delta_{WN}(x;x') \; , \label{DWNN} \\
\Bigl[ \square - (\beta \!-\! D)H^2 \Bigr] i\Delta_{WN}(x;x') 
	& = & i \Delta_{W}(x;x') \; , \label{DWN} \\
\Bigl[ \square + DH^2 \Bigr] i\Delta_{W}(x;x') 
	& = & \frac{i \delta^D(x \!-\! x')}{\sqrt{-g}(x)} \; . \label{DW} 
\end{eqnarray}
Since one of the masses in the scalar kinetic operators is tachyonic,
the scalar structure function $i\Delta_{WNN}(x;x')$ must break de Sitter invariance.
We separate it into a de Sitter invariant and a de Sitter breaking part,
\begin{equation}
i \Delta_{WNN}(x;x') = W\!N\!N(y) + \delta W\!N\!N(y,u,v) \; ,
\label{DWNNsplit}
\end{equation}
the details of which are given in Appendix C.

The de Sitter breaking part $\delta W\!N\!N$ of the scalar 
structure function (\ref{DWNNsplit}) is given in 
(\ref{deltaWNN}). 
The projectors in (\ref{gravpropspin0}),
when acted on this de Sitter breaking part
 give zero,
\begin{equation}
\mathcal{P}_{\mu\nu}(x)
		\times \mathcal{P}_{\rho\sigma}(x') 
		\times \Bigl[ \delta W\!N\!N(y,u,v) \Bigr] = 0 \; ,
\end{equation}
which means that the spin 0 part of the graviton propagator is de Sitter
invariant for gauges $b>2$ (even though the scalar structure
function is not). 
Therefore, we can drop the de Sitter breaking part 
in (\ref{gravpropspin0}), and write 
\begin{equation}
i\Bigl[\mbox{}_{\mu\nu} \Delta^0_{\rho\sigma}\Bigr](x;x') 
	= \frac{-2 \beta^2}{(D \!-\! 1)(D \!-\! 2)} 
		\times \mathcal{P}_{\mu\nu}(x)
		\times \mathcal{P}_{\rho\sigma}(x') 
		\times \Bigl[ W\!N\!N(y) \Bigr] \; .
\label{gravpropspin0dS}
\end{equation}
By making use of relations (\ref{BoxWNN}-\ref{BoxWN})
from Appendix C, the action of the projectors can be written out as 
\begin{eqnarray}
\lefteqn{ i\Bigl[\mbox{}_{\mu\nu} \Delta^0_{\rho\sigma}\Bigr](x;x') 
	= -\frac{2\beta^2}{(D \!-\! 1)(D \!-\! 2)} 
		D_{\mu} D_{\nu} D'_{\rho} D'_{\sigma} W\!N\!N(y) }
\nonumber \\
& &	+ \frac{2\beta}{(D \!-\! 1)(D \!-\! 2)} \Bigl[
		g_{\mu\nu}(x) D'_{\rho} D'_{\sigma} W\!N(y)
		+ g_{\rho\sigma}(x') D_{\mu} D_{\nu} W\!N(y) \Bigr]
\nonumber \\
& &	- \frac{2}{(D \!-\! 1)(D \!-\! 2)} 
		g_{\mu\nu}(x) g_{\rho\sigma}(x') W(y) \; .
\end{eqnarray}
It is convenient to expand this propagator in a de Sitter invariant tensor 
basis defined in Table \ref{Tees},
\begin{equation}
i\Bigl[\mbox{}_{\mu\nu} \Delta^0_{\rho\sigma}\Bigr](x;x') 
	= \sum_{i=1}^{5} [\mbox{}_{\mu\nu} \mathcal{T}^i_{\rho\sigma}](x;x')
		\times C_i(y) \; , \label{spin0expansion}
\end{equation}
where the coefficient functions in this expansion are given in Table \ref{Cees}.

\begin{table}
\renewcommand{\arraystretch}{2}
\setlength{\tabcolsep}{8pt}
\centering
\begin{tabular}{|c||c|}
\hline
$i$ 	& $ \left[ \frac{-2\beta^2}{(D \!-\! 1)(D \!-\! 2)} \right]^{-1}
			\times C_i(y)$  \\ 
\hline \hline
1 	& $2 W\!N\!N''$ \\ 
\hline 
2 	& $4 W\!N\!N'''$ \\ 
\hline
3 	& $W\!N\!N''''$ \\ 
\hline
4 	& $(2 \!-\! y) W\!N\!N''' - 2 W\!N\!N'' 
		- \frac{1}{\beta} \frac{W\!N''}{H^2}$ \\ 
\hline 
5 	& $ (2 \!-\! y)^2 W\!N\!N'' - (2 \!-\! y) W\!N\!N'
		- \frac{2}{\beta}(2 \!-\! y) \frac{W\!N'}{H^2}
		+ \frac{1}{\beta^2} \frac{W}{H^4}$ \\ 
\hline 
\end{tabular} 
\caption{The coefficient functions $C_i(y)$ in the expansion
(\ref{spin0expansion}) of the spin 0 graviton propagator in
the tensor basis given in Table \ref{Tees}.
\label{Cees}}
\end{table}

\subsection{The 4-Point Diagram}

The contribution to the vacuum polarization from the spin 0
part of the 4-point diagram
(middle one in Fig. \ref{fig:photon}) is
\begin{equation}
i\Bigl[\mbox{}^{\mu} \Pi^{\nu}_{\rm 4pt0}\Bigr](x;x') =
\partial_{\kappa} \partial'_{\lambda} \Biggl\{ -i\kappa^2 a^{D-8}
U^{\mu\nu\kappa\lambda\alpha\beta\gamma\delta}
\, i\Bigl[\mbox{}_{\alpha\beta} \Delta^0_{\gamma\delta} \Bigr](x;x)
\, \delta^D(x \!-\! x') \Biggr\} \; , 
\label{4pt0}
\end{equation}
Here the spin 0 part of the graviton propagator was substituted in 
the full contribution (\ref{4pt}), 
and the 4-point vertex function is given in (\ref{Uvert}).
The coincidence limit of the spin 0 graviton propagator is
\begin{equation}
i\Bigl[\mbox{}_{\alpha\beta} \Delta^0_{\gamma\delta}\Bigr](x;x) 
	= 4(aH)^4 \eta_{\alpha(\gamma} \eta_{\delta)\beta} C_1(0)
	+ (aH)^4 \eta_{\alpha\beta} \eta_{\gamma\delta} C_5(0) \; ,
\label{coincidenceSpin0}
\end{equation}
where the coincidence limits of the relevant coefficient functions
from Table \ref{Cees} are
\begin{eqnarray}
\lefteqn{C_1(0) = \frac{H^{D-6}}{(4\pi)^{\frac{D}{2}}} 
	\times \frac{\Gamma(D\!+\!2)}
		{4(D\!-\!1)(D\!-\!2) \Gamma(\frac{D}{2} \!+\! 2)}  }
\nonumber \\
& & \hspace{1.5cm} \times \Biggl\{ 1 + \frac{\Gamma( \frac{D}{2} ) 
		\Gamma(1 \!-\! \frac{D}{2})}{\Gamma(D\!+\!2)}
		\frac{\Gamma(\frac{D+3}{2} \!+\! b_N) 
		\Gamma(\frac{D+3}{2}\!-\!b_N)}
		{\Gamma(\frac{1}{2}\!+\!b_N) \Gamma(\frac{1}{2}\!-\!b_N)}
\nonumber \\
& &	 \hspace{3cm} \times \Biggl[ 1 + \frac{\beta}{2b_N} 
	\Bigl[ \psi\Bigl( \frac{D\!+\!3}{2} \!+\! b_N \Bigr) 
	- \psi\Bigl( \frac{D\!+\!3}{2} \!-\! b_N \Bigr) 
\nonumber \\
& &	\hspace{6cm}- \psi\Bigl( \frac{1}{2} \!+\! b_N \Bigr)
		+ \psi\Bigl( \frac{1}{2} \!-\! b_N \Bigr) \Bigr] \Biggr] 
	\Biggr\} , \qquad
\label{C1(0)}
\\
\lefteqn{C_5(0) = 2C_1(0) +
	\frac{H^{D-6}}{(4\pi)^{\frac{D}{2}}} \times 
	\frac{\Gamma(D\!+\!1)}{(D\!-\!1)(D\!-\!2)\Gamma(\frac{D\!+\!1}{2})} }
\nonumber \\
& & \hspace{1.5cm} \times \Biggl\{ - \frac{D\!+\!1}{D}
	+ \frac{\Gamma(\frac{D}{2})\Gamma(1\!-\!\frac{D}{2})}
		{\Gamma(D\!+\!1)} \frac{\Gamma(\frac{D\!+\!1}{2}\!+\!b_N)
			\Gamma(\frac{D\!+\!1}{2}\!-\!b_N)}
			{\Gamma(\frac{1}{2}\!+\!b_N)
			\Gamma(\frac{1}{2}\!-\!b_N)}
\nonumber \\
& & \hspace{3cm} \Biggl[ 1 - \frac{\beta}{2b_N}	
	\Bigl[ \psi\Bigl( \frac{D\!+\!1}{2}\!+\!b_N \Bigr)
	- \psi\Bigl( \frac{D\!+\!1}{2}\!-\!b_N \Bigr) 
\nonumber \\
& & \hspace{6cm} - \psi\Bigl( \frac{1}{2}\!+\!b_N \Bigr)
		+ \psi\Bigl( \frac{1}{2}\!-\!b_N \Bigr) \Bigr] \Biggr] 
\Biggr\} . \qquad
\label{C5(0)}
\end{eqnarray}
They are easily calculated from explicit forms of scalar 
propagators (\ref{W(y)}) and (\ref{N(y)}), and expressions (\ref{WNN}-\ref{NN})
in Appendix C, where, by rules of dimensional regularization, 
all $D$-dependent powers of $y$ vanish at coincidence.

The contractions of the tensor structures  of the
spin 0 graviton propagator coincident limit 
(\ref{coincidenceSpin0}) with the 4-vertex (\ref{Uvert}) are
\begin{eqnarray}
U^{\mu\nu\kappa\lambda\alpha\beta\gamma\delta}
	\eta_{\alpha(\gamma} \eta_{\delta)\beta}
	&=& \frac{1}{4} [12 \!-\! (D\!-\!4)^2] 
		\eta^{\mu[\nu} \eta^{\lambda]\kappa},
\\
U^{\mu\nu\kappa\lambda\alpha\beta\gamma\delta}
	\eta_{\alpha\beta} \eta_{\gamma\delta}
	&=& \frac{1}{4}(D\!-\!4)(D\!-\!6)
		\eta^{\mu[\nu} \eta^{\lambda]\kappa} ,
\end{eqnarray}
so that,
\begin{eqnarray}
\lefteqn{U^{\mu\nu\kappa\lambda\alpha\beta\gamma\delta}
	i\Bigl[\mbox{}_{\alpha\beta} \Delta^0_{\gamma\delta}\Bigr](x;x)
	= \frac{(aH)^4}{2} \Bigl( \eta^{\mu\nu} \eta^{\kappa\lambda}
			- \eta^{\mu\lambda} \eta^{\nu\kappa} \Bigr) }
\nonumber \\
& & \hspace{2.5cm} \times \Biggl\{ \left[ 12 - (D \!-\! 4)^2 \right] C_1(0)
		+ \frac{(D \!-\! 6)(D \!-\! 4)}{4} C_5(0) \Biggr\} . \qquad
\end{eqnarray}
From here it is straightforward to see that the vacuum polarization is
in the form given by (\ref{Piform}) and (\ref{ourrep}) in subsection 2.2,
\begin{equation}
i\Bigl[\mbox{}^{\mu} \Pi^{\nu}_{\rm 4pt0}\Bigr](x;x') = 
	\partial_{\kappa} \partial'_{\lambda} 
	\Biggl\{ 2 \eta^{\mu [\nu}
		\eta^{\lambda] \kappa} \!\times\! F_{0a}(x;x') 
	+ 2 \overline{\eta}^{\mu [\nu} \overline{\eta}^{\lambda] \kappa} 
		\!\times\! G_{0a}(x;x') \Biggr\} ,
\end{equation}
where the structure functions are
\begin{eqnarray}
\lefteqn{ F_{0a}(x;x') = \kappa^2 a^{D-4} i\delta^D(x \!-\! x') }
\nonumber \\
	& & \hspace{0.4cm} \times \Biggl\{ - 6 H^4 C_1(0)
		- \frac{(D \!-\! 4)(D \!-\! 6)}{8} H^4 C_5(0) 
		+ \frac{(D \!-\! 4)^2}{2} H^4 C_1(0) \Biggr\}  , \qquad 
\label{F0a}
\\
\lefteqn{ G_{0a}(x;x') = 0 \; . } 
\label{G0a}
\end{eqnarray}
In the $D\!=\!4$ limit the first term in the brackets in (\ref{F0a}) diverges
as $1/(D \!-\! 4)$, the second term is finite, and the third term vanishes 
in this limit. This is a consequence of the coefficient functions (\ref{C1(0)}) and 
(\ref{C5(0)}) diverging as $1/(D \!-\! 4)$.
Comparing  with (\ref{DF}-\ref{DG}) we see that
the entire contribution of (\ref{F0a}-\ref{G0a}) can be absorbed into the
counterterms by choosing the coefficient
\begin{equation}
\overline{C}_{0a} = - \frac{\kappa^2}{4} \Biggl\{ - 6 H^2 C_1(0)
		+ \frac{(D \!-\! 6)(D \!-\! 4)}{4} H^2 C_5(0)
		+ \frac{(D \!-\! 4)^2}{2} H^2 C_1(0)  \Biggr\} \; .
\end{equation}
Therefore, the renormalized 4-point
contributions to the structure functions are,
\begin{equation}
F_{0a}^{\rm ren} = 0 \; , \qquad G_{0a}^{\rm ren} = 0 \; .
\label{F0aG0a}
\end{equation}

\subsection{Local contributions from the 3-3 diagram}

The "0b" local part of the 3-3 diagram (left one in Fig. \ref{vacpolgraphs})
descends from isolating the delta function coming from two derivatives
acting on the $B$-type scalar propagator as in (\ref{locreplace}),
which we reproduce here (with all indices lowered),
\begin{equation}
D_{\rho} {D'}_{\sigma} y \times D_{\lambda} {D'}_{\phi} B 
\longrightarrow -\frac{2 H^2}{a^{D-4}} \, \eta_{\rho\sigma}
\delta_{\lambda}^0 \delta_{\phi}^0 \, i\delta^D(x \!-\! x') \; .
\label{locB}
\end{equation}
 and by making the substitution in (\ref{new3pt}),
\begin{equation}
i\Bigl[\mbox{}^{\mu} \Pi^{\nu}_{\rm 3pt0b}\Bigr](x;x') 
	= \partial_{\kappa} \partial_{\theta}' \Biggl\{ -i\kappa^2
	a^{D-8} \, \widetilde{U}^{\mu\nu\kappa\theta\alpha\beta\gamma\delta} 
	\, i\Bigl[\mbox{}_{\alpha\beta} \Delta^0_{\gamma\delta} \Bigr](x;x)  \, 
	\delta^D(x\!-\!x') \Biggr\} \; .
\end{equation}
Note that this contribution has the same structure as the 4-point diagram 
contribution (\ref{4pt0}). The effective 4-vertex
$\widetilde{U}^{\mu\nu\kappa\theta\alpha\beta\gamma\delta}$ 
is constructed by contracting the tensor structure in (\ref{locB}) 
with the two 3-vertices in (\ref{new3pt}),
\begin{eqnarray}
\lefteqn{\widetilde{U}^{\mu\nu\kappa\theta\alpha\beta\gamma\delta}
	= V^{\mu\rho\kappa\lambda\alpha\beta}
	\eta_{\rho\sigma} \delta_{\lambda}^{0} \delta_{\phi}^{0}
	V^{\nu\sigma\theta\phi\gamma\delta}}
\nonumber \\
& & = - \delta_0^{[\mu} \eta^{\kappa][\nu} \delta_0^{\theta]} 
		\eta^{\alpha\beta} \eta^{\gamma\delta}
	+ 2 \Bigl[ \delta_0^{[\mu} \eta^{\kappa](\gamma}
		\eta^{\delta)[\nu} \delta_0^{\theta]} \eta^{\alpha\beta}
		+ \delta_0^{[\mu} \eta^{\kappa](\alpha}
		\eta^{\beta)[\nu} \delta_0^{\theta]} \eta^{\gamma\delta} \Bigr]
\nonumber \\
& & +2 \Bigl[ \delta_0^{[\mu} \eta^{\kappa][\nu} 
		\eta^{\theta](\gamma} \delta_0^{\delta)} \eta^{\alpha\beta}
	+ \delta_0^{(\alpha} \eta^{\beta)[\mu} \eta^{\kappa][\nu} 
		\delta_0^{\theta]} \eta^{\gamma\delta} \Bigr]
	- 4 \delta_0^{[\mu} \eta^{\kappa](\alpha} \eta^{\beta)(\gamma}
		\eta^{\delta)[\nu} \delta_0^{\theta]}
\nonumber \\
& &	- 4 \delta_0^{(\alpha} \eta^{\beta)[\mu} \eta^{\kappa][\nu}
		\eta^{\theta](\gamma} \delta_0^{\delta)}
	-4 \Bigl[ \delta_0^{[\mu} \eta^{\kappa](\alpha} \eta^{\beta)[\nu}
		\eta^{\theta](\gamma} \delta_0^{\delta)}
		+ \delta_0^{(\alpha} \eta^{\beta)[\mu} \eta^{\kappa](\gamma}
		\eta^{\delta)[\nu} \delta_0^{\theta]} \Bigr] . \qquad \quad
\end{eqnarray}
Contractions of this effective 4-vertex with the tensor structures in the 
coincident graviton propagator (\ref{coincidenceSpin0}) are
\begin{eqnarray}
\widetilde{U}^{\mu\nu\kappa\theta\alpha\beta\gamma\delta}
	\eta_{\alpha(\gamma} \eta_{\delta)\beta}
	& = & -2 \eta^{\mu[\nu} \eta^{\theta]\kappa}
		- (3D \!-\! 8) \delta_0^{[\mu} \eta^{\kappa][\nu} 
		\delta_0^{\theta]} \; ,
\\
\widetilde{U}^{\mu\nu\kappa\theta\alpha\beta\gamma\delta}
	\eta_{\alpha\beta} \eta_{\gamma\delta}
	& = & - (D\!-\!4)^2 \delta_0^{[\mu} \eta^{\kappa][\nu} 
		\delta_0^{\theta]} \; .
\end{eqnarray}

Using the identity
\begin{equation}
\delta_0^{[\mu} \eta^{\kappa][\nu} \delta_0^{\theta]}
	= \frac{1}{2} \eta^{\mu[\nu} \eta^{\theta]\kappa}	
		- \frac{1}{2} \overline{\eta}^{\mu[\nu} 
		\overline{\eta}^{\theta]\kappa} \; ,
\end{equation}
we can write the contribution to vacuum polarization as
described in subsection 2.2,
\begin{equation}
i\Bigl[\mbox{}^{\mu} \Pi^{\nu}_{\rm 3pt0b}\Bigr](x;x') 
	= \partial_\kappa \partial'_\theta \Biggl\{
		2 \eta^{\mu[\nu} \eta^{\theta]\kappa} 
			\!\times\! F_{0b}(x;x')
		+ 2 \overline{\eta}^{\mu[\nu} 
		\overline{\eta}^{\theta]\kappa} 
			\!\times\! G_{0b}(x;x') \Biggr\} ,
\end{equation}
where the structure functions are
\begin{eqnarray}
\lefteqn{ F_{0b}(x;x') = \kappa^2 a^{D-4} i \delta^D(x\!-\!x')  }
\nonumber \\
& & \hspace{2cm} \times \Biggl\{ (3D\!-\!4) H^4C_1(0) 
	+ \frac{(D\!-\!4)^2}{4}H^4C_5(0) \Biggr\} \; ,  \qquad \quad 
\label{F0bBare}
\\
\lefteqn{ G_{0b}(x;x') = \kappa^2 a^{D-4} i \delta^D(x\!-\!x') }
\nonumber \\
& & \hspace{2cm} \times \Biggl\{ -(3D\!-\!8) H^4C_1(0) 
	- \frac{(D\!-\!4)^2}{4} H^4C_5(0) \Biggr\} \; ,
\label{G0bBare}
\end{eqnarray}
and $C_1(0)$ and $C_5(0)$ are given in (\ref{C1(0)}-\ref{C5(0)})
(recall that they diverge as $1/(D\!-\!4)$ in $D\rightarrow4$ limit).
Again, we can completely absorb these contributions into the counterterms
(\ref{DF}-\ref{DG}) by choosing the coefficients
\begin{eqnarray}
\overline{C}_{0b} &=& -\frac{\kappa^2}{4}
	\Biggl\{ (3D\!-\!4)H^2 C_1(0)
	+ \frac{(D \!-\! 4)^2}{4} H^2 C_5(0) \Biggr\} \; ,
\\
\Delta C_{0b} &=& \frac{\kappa^2}{4}
	\Biggl\{ (3D\!-\!8)H^2C_1(0) 
	+ \frac{(D \!-\! 4)^2}{4}H^2 C_5(0) \Biggr\} \; ,
\end{eqnarray}
making the renormalized contributions to vacuum polarization scalar structure
functions vanish,
\begin{equation}
F_{0b}^{\rm ren} = 0 \; , \qquad G_{0b}^{\rm ren} = 0 \; .
\label{F0bG0b}
\end{equation}
Note that had we not introduced a noninvariant counterterm
in (\ref{cterms}) we would not have been able to remove the divergence in
(\ref{G0bBare}).

\subsection{Nonlocal contributions from the 3-3 diagram}

The non-local "0d" contribution from the 3-3 diagram 
derives from making the replacement
for the two derivatives acting on the $B$-propagator,
\begin{equation}
D_\lambda D'_\phi B \longrightarrow 
	D_\lambda D'_\phi y \times B' 
	+ D_\lambda y \, D'_\phi y \times B'' \; ,
\end{equation}
in expression (\ref{new3pt}). This makes the bi-tensor density
defined in (\ref{Piform}),
\begin{eqnarray}
\lefteqn{ \Bigl[\mbox{}^{\mu\kappa} T_{0d}^{\nu\theta} \Bigr](x;x') 
	= \frac{\kappa^2}{2H^2} (aa')^{D-6}
	V^{\mu\rho\kappa\lambda\alpha\beta}
	i\Bigl[ \mbox{}_{\alpha\beta}\Delta^0_{\gamma\delta} \Bigr](x;x')
	V^{\nu\sigma\theta\phi\gamma\delta}
	D_\rho D'_\sigma y }
\nonumber \\
& &	\hspace{6cm} \times \Bigl[ D_\lambda D'_\phi y \!\times\! B' 
	+ D_\lambda y \, D'_\phi y \!\times\! B'' \Bigr]  . \qquad
\end{eqnarray}
We use the recipe developed in \cite{Leonard:2012ex}
and outlined in subsection \ref{subsec: Representing} to find the "0d"
contribution to $F(x;x')$ and $G(x;x')$ structure functions. 
The first thing is to calculate coefficient functions $f_i$
from (\ref{fidefs}). We do this by contracting the tensor structure
of the graviton propagator with the vertices and the tensor structure of
the photon part. The contraction rules can be found in
\cite{Kahya:2011sy}. This contribution is de Sitter invariant so only $f_1$ and 
$f_2$ can appear (the de Sitter breaking part "0c" is zero in this case), 
and we find it convenient to express them as a sum
over spin 0 graviton coefficient functions defined in Table \ref{Cees},
\begin{equation}
f_1(y) \equiv -\kappa^2H^2 \sum_{i=1}^{5}
	(\Delta\varphi_1)_i(y)\times C_i(y)	\;, 
\; f_2(y) \equiv -\kappa^2H^2 \sum_{i=1}^{5}
	(\Delta\varphi_2)_i(y)\times C_i(y) ,
\label{fiExpandedInC}
\end{equation}
where factors $-\kappa^2H^2$ are extracted for convenience.
The coefficient functions $(\Delta\varphi_1)_i$ and $(\Delta\varphi_2)_i$ are 
given in Table \ref{Dvarphi}, where identities (\ref{BCident}) and (\ref{Bprop})
have been used to simplify the expressions.

\begin{table}
\renewcommand{\arraystretch}{2}
\setlength{\tabcolsep}{8pt}
\centering
\begin{tabular}{|c||c|c|}
\hline 
$i$ 	& $(\Delta\varphi_1)_i(y)$ & $(\Delta\varphi_2)_i(y)$ \\ 
\hline \hline
1 	& $-2(5D\!-\!13)B'  - \frac{1}{2}(2\!-\!y)^2B'
		\atop + 2(D\!-\!2)(2\!-\!y)C'$ 
	& $2(3D\!-\!8)B'' \!+\! (2\!-\!y)B' \!-\! 3(D\!-\!2)C'$ \\ 
\hline 
2 	& $-\frac{1}{2}(2\!-\!y)(4y\!-\!y^2)B'
		\atop +(D\!-\!2)(4y\!-\!y^2)C'$ 
	& $4(D\!-\!3)B' + (4y\!-\!y^2)B'
		\atop+(D\!-\!2)(2\!-\!y)C'$ \\ 
\hline
3 	& $-\frac{1}{2}(4y\!-\!y^2)^2B'$ 
	& $\frac{1}{2}(4y\!-\!y^2)^2B''$ \\ 
\hline
4 	& $-(D\!-\!4)(4y\!-\!y^2)B'$ 
	& $-2(D\!-\!4)[(2\!-\!y)B' \!-\! (D\!-\!2)C']$ \\ 
\hline 
5 	& $-\frac{1}{2}(D\!-\!4)^2B'$ 
	& $\frac{1}{2}(D\!-\!4)^2B''$ \\ 
\hline 
\end{tabular} 
\caption{The functions $(\Delta\varphi_1)_i(y)$ and 
$(\Delta\varphi_2)_i(y)$ defined in equation (\ref{fiExpandedInC}). 
Function $B(y)$ is defined in (\ref{B(y)}),
and $C'(y)$ in (\ref{BCident}).
\label{Dvarphi}}
\end{table}

The next step is to construct the master structure functions
defined in Table \ref{fitoPhiPsi}. Because
of de Sitter invariance $\Psi(y,u,v)$ has to be zero, and
$\Phi(y)$ is, according to Table \ref{fitoPhiPsi} and expression 
(\ref{fiExpandedInC}),
\begin{eqnarray}
\lefteqn{ \Phi_{0d}(y) = - \frac{1}{4}\kappa^2H^6 \sum_{i=1}^{5}
	\Biggl\{ \Bigl[ 2(D\!-\!1)(\Delta\varphi_1)_i 
		\!-\! 2(2\!-\!y)(\Delta\varphi_1)_i' 
		\!+\! D(2\!-\!y)(\Delta\varphi_2)_i }
\nonumber \\
& &	+\! (4y\!-\!y^2)(\Delta\varphi_2)_i' \Bigr] \! \times \! C_i 
	+ \Bigl[ -2(2\!-\!y)(\Delta\varphi_1)_i 
	\!+\! (4y\!-\!y^2)(\Delta\varphi_2)_i \Bigr] \! \times \! C_i' \Biggr\} . 
\ \qquad  \label{PhiExpandedInC}
\end{eqnarray}
The results for coefficients of $C_i$ and $C_i'$
in this expansion are presented in Table \ref{CC'coeff}.
\begin{table}
\renewcommand{\arraystretch}{2}
\setlength{\tabcolsep}{8pt}
\centering
\begin{tabular}{|c||c|c|}
\hline 
$i$ & $2(D\!-\!1)(\Delta\varphi_1)_i
		\!-\! 2(2\!-\!y)(\Delta\varphi_1)_i' 
		\atop \!+\! D(2\!-\!y)(\Delta\varphi_2)_i 
		+\! (4y\!-\!y^2)(\Delta\varphi_2)_i'$ 
	& $-2(2\!-\!y)(\Delta\varphi_1)_i 
		\!+\! (4y\!-\!y^2)(\Delta\varphi_2)_i$ \\ 
\hline\hline
1 	& $2(2\!-\!y)C' \!-\! 8(D\!-\!3)B'$ 
	& $8(2\!-\!y)B' \!+\! (4y\!-\!y^2)C' \!-\! 4(3D\!-\!4)C'$ \\ 
\hline 
2	& $4(D\!-\!2)(2\!-\!y)B' 
		\atop + 3(4y\!-\!y^2)C' - 8C'$ 
	& $(4y\!-\!y^2)[4B' \!-\! (2\!-\!y)C']$ \\ 
\hline
3	& $-4(2\!-\!y)(4y\!-\!y^2)C'$ 
	& $-(4y\!-\!y^2)^2C'$ \\ 
\hline
4	& $4(D\!-\!4)[(2\!-\!y)C' \!-\! 4B']$ 
	& $2(D\!-\!4)(4y\!-\!y^2)C'$ \\ 
\hline 
5	& $0$
	& $-(D\!-\!4)^2C'$ \\ 
\hline 
\end{tabular} 
\caption{The coefficient functions of $C_i(y)$ and $C_i'(y)$ from 
expression (\ref{PhiExpandedInC}). An overall factor of $(D\!-\!2)$
has been extracted from each term.
\label{CC'coeff}}
\end{table}
Plugging in $C_i(y)$ from Table \ref{Cees} and 
coefficient functions from Table \ref{CC'coeff} into 
(\ref{PhiExpandedInC}) we get for the master function,
\begin{eqnarray}
\lefteqn{\Phi_{0d}(y) = 
	\frac{\kappa^2H^6 \beta^2B'}{2(D\!-\!1)} \times \Biggl\{ 
	16(4y\!-\!y^2)W\!N\!N'''' + 48(2\!-\!y)W\!N\!N''' } 
\nonumber \\
& & \hspace{2cm}
	+16(D\!-\!5)W\!N\!N'' 
	+ \frac{16}{\beta}(D\!-\!4) \frac{W\!N''}{H^2} \Biggr\}
\nonumber \\
& & + \frac{\kappa^2H^6\beta^2 C'}{2(D\!-\!1)} \times \Biggl\{ 
	 -  (4y\!-\!y^2)^2 W\!N\!N''''' 
	+ 2(D\!-\!8) (2\!-\!y)(4y\!-\!y^2) W\!N\!N''''
\nonumber \\
& &	\hspace{1.7cm} 
	+ \Bigl[ (D^2\!-\!18D\!+\!70)(4y\!-\!y^2) 
	- 4(D^2\!-\!6D\!+\!32) \Bigr]W\!N\!N'''
\nonumber \\
& & \hspace{1.7cm}
	+(D\!-\!6)(3D\!-\!14)(2\!-\!y) W\!N\!N'' 
	- (D\!-\!4)^2  W\!N\!N'
\nonumber \\
& &	\hspace{1.7cm}
	- \frac{2}{\beta}(D\!-\!4)(4y\!-\!y^2) \frac{W\!N'''}{H^2}
	 + \frac{2}{\beta} (D\!-\!4)(D\!-\!6)(2\!-\!y) \frac{W\!N''}{H^2}	
\nonumber \\
& & \hspace{1.7cm} 
	- \frac{2}{\beta} (D\!-\!4)^2 \frac{W\!N'}{H^2}	
	- \frac{(D\!-\!4)^2}{\beta^2} \frac{W'}{H^4} \Biggr\} .
\end{eqnarray}
Using the identity for the d'Alembertian operator acting on a scalar function
that depends only on $y$,
\begin{equation}
\frac{\square}{H^2} S(y) = (4y\!-\!y^2)S''(y) + D(2\!-\!y)S'(y) \; ,
\end{equation}
and derivatives of it with respect to $y$, we can rewrite the above expression 
for master function as
\begin{eqnarray}
\lefteqn{ \Phi_{0d}(y) = \frac{\kappa^2H^6\beta^2B'}{2(D\!-\!1)}
	\times \Biggl\{ 16 \partial_y^2 \Bigl[ \frac{\square}{H^2}W\!N\!N \Bigr]
	-16(D\!+\!1) (2\!-\!y)W\!N\!N''' }
\nonumber \\
& &	\hspace{2cm}+ 48(D\!-\!1) W\!N\!N'' 
	+ 16\frac{(D\!-\!4)}{\beta} \frac{W\!N''}{H^2}\Biggr\}
\nonumber \\
& & + \frac{\kappa^2H^6\beta^2C'}{2(D\!-\!1)} \times \Biggl\{ 
	- \partial_{y} \Bigl[ \Bigl(\frac{\square}{H^2}\Bigr)^2 W\!N\!N \Bigr]
	+4(D\!-\!2)(2\!-\!y) \partial_y^2 \Bigl[ \frac{\square}{H^2}W\!N\!N \Bigr]
\nonumber \\
& & \hspace{1.7cm}+4(D\!-\!2)(D\!-\!3) 
		\partial_y \Bigl[ \frac{\square}{H^2}W\!N\!N \Bigr]
	- \frac{2}{\beta}(D\!-\!4)\partial_y 
	\Bigl[ \frac{\square}{H^2}\frac{W\!N}{H^2} \Bigr]
\nonumber \\
& &	\hspace{1.7cm} 
	-16(D\!-\!2)(D\!+\!1)W\!N\!N''' 
	-4(D\!-\!2)^2(D\!-\!1)(2\!-\!y)W\!N\!N''
\nonumber \\
& & \hspace{1.7cm}	
	+ 4(D\!-\!2)^2(D\!-\!1)W\!N\!N'
	+ \frac{4}{\beta}(D\!-\!4)(D\!-\!2)(2\!-\!y) \frac{W\!N''}{H^2}
\nonumber \\
& &	\hspace{1.7cm} 	
	- \frac{4}{\beta}(D\!-\!4)(D\!-\!2)\frac{W\!N'}{H^2}
	- \frac{(D\!-\!4)^2}{\beta^2} \frac{W'}{H^4} \Biggr\} \; .
\end{eqnarray}
Next, using identities (\ref{BoxWNN}-\ref{BoxWN}) 
from Appendix C, we can further simplify this expression,
\begin{eqnarray}
\lefteqn{ \Phi_{0d}(y) = \frac{\kappa^2H^6\beta^2 B'}{2(D\!-\!1)}
	\times \Biggl\{ -16(D\!+\!1)(2\!-\!y)W\!N\!N''' } 
\nonumber \\
& & \hspace{2cm}+ 16 \Bigl[ \beta \!+\! (2D\!-\!3) \Bigr] W\!N\!N''  
	+ 16\Bigl[ 1 + \frac{(D\!-\!4)}{\beta} \Bigr] \frac{W\!N''}{H^2} \Biggr\}
\nonumber \\
& & \hspace{0.6cm} + \frac{\kappa^2H^6\beta^2 C'}{2(D\!-\!1)}
	\times \Biggl\{
	-16(D\!-\!2)(D\!+\!1)W\!N\!N'''
\nonumber \\
& & \hspace{2cm}
	+ 4(D\!-\!2) \Bigl[ \beta \!-\! (D^2\!-\!2D\!+\!2) \Bigr]
		(2\!-\!y) W\!N\!N''
\nonumber \\
& & \hspace{2cm}
	- \Bigl[ \beta^2 \!-\! 2(2D^2\!-\!9D\!+\!12)\beta
		\!+\! (D\!-\!4)^2 \Bigr] W\!N\!N'
\nonumber \\
& & \hspace{2cm}
	+4(D\!-\!2) \Bigl[ 1 \!+\! \frac{D\!-\!4}{\beta} \Bigr]
		(2\!-\!y) \frac{W\!N''}{H^2}
\nonumber \\
& & \hspace{2cm}
	-2 \Bigl[ \beta \!-\! 2(D^2\!-\!5D\!+\!8)
		+ \frac{(D\!-\!4)^2}{\beta} \Bigr] \frac{W\!N'}{H^2}
\nonumber \\
& & \hspace{2cm}- \Bigl[ 1 \!+\! \frac{D\!-\!4}{\beta} \Bigr]^2 \frac{W'}{H^4}
	- \frac{4}{\beta^2}(D\!-\!2)^2 \frac{w'}{H^4}
	\Biggr\}
\end{eqnarray}

As argued in subsection 3.5, we can set $D=4$ in any part of the structure
functions $F(x;x')$ and $G(x;x')$ that diverges less strongly than
$1/\Delta x^4$, as $\Delta x\rightarrow 0$. Since, according to 
(\ref{Frule}-\ref{Grule}), $F$ is an integral of $\Phi$ with respect to $y$,
and $G$ is a double integral of $\Phi$ and $\Psi$ with respect to $y$,
in the master functions we can set $D=4$ in any of the parts
which diverge less strongly than $y^{-3}$ as $y\rightarrow0$
(recall that $y=H^2aa'\Delta x^2$). Also, we can throw away terms
containing $(D\!-\!4)^2$.
Taking into account the expansion of $B'$ and $C'$ given in 
(\ref{Bprim}-\ref{Cprim}) in Appendix A, and functions $W$, $N$ 
and $N\!N$ defined in Appendix C, we get for the master function
\begin{equation}
\Phi_{0d}(y) = \frac{\kappa^2H^{2D-2}\Gamma^2(\frac{D}{2})}{8\pi^D}
	\Biggl\{ \frac{\ell_1(D)}{y^D} + \frac{\ell_2(D)}{y^{D-1}}
	+ \frac{\mathcal{N}(y)}{y^2} \Biggr\} ,
\end{equation}
where the two $D$-dependent coefficients are
\begin{eqnarray}
\ell_1(D) &=& - \frac{1}{8} \Bigl[ (D\!-\!2)\beta^2
	- 4(D\!-\!4)\beta + \frac{2(D\!-\!4)^2}{(D\!-\!1)} \Bigr] \; ,
\\
\ell_2(D) &=& \frac{1}{72} \Bigl[ \beta^2(5\!-\!\beta)
	- \frac{\beta}{6} (4\beta^2\!-\!83\beta\!+\!180)(D\!-\!4) \Bigr] 
	+ \mathcal{O}\Bigl((D\!-\!4)^2\Bigr) \, . \qquad
\end{eqnarray}
%We note right away that $\ell_1(D)=C_0(b)/2(D\!-\!2)$, where 
%$C_0(b)$ was defined in (\ref{spin0C}).
The de Sitter invariant function $\mathcal{N}(y)$ is defined to be
\begin{eqnarray}
\mathcal{N}(y) &\equiv&
	\frac{\beta^2}{48} \overline{N}_1' + \frac{5}{3}y \overline{N}_1''
	- \frac{5}{3}\overline{N}_2'' + \frac{5}{3}y \overline{N}_2'''
	- \frac{20}{3}\overline{N}_3'''
 + \frac{\beta^2(\beta-16)}{48} \overline{NN}_1' 
\nonumber \\
& &
	+ \frac{\beta(\beta-10)}{6} y \overline{NN}_1''
	- \frac{\beta(2\beta-5)}{3} \overline{NN}_2''
	- \frac{5\beta}{3} y \overline{NN}_2'''
	+ \frac{20\beta}{3} \overline{NN}_3'''  
\nonumber \\
&=&\frac{\beta^2}{6} \frac{\partial}{\partial\beta} \Biggl\{ 
	\Bigl[ \frac{\beta}{8}-2 \Bigr] \overline{N}_1'
	+ \Bigl[ -\frac{10}{\beta} +1 \Bigr] y\overline{N}_1'' 
\nonumber\\
& &	\hspace{3cm} 
	+ \Bigl[ \frac{10}{\beta}-4 \Bigr] \overline{N}_2''
	- \frac{10}{\beta} y\overline{N}_2'''
	+ \frac{40}{\beta} \overline{N}_3'''\Biggr\} , \quad
\end{eqnarray}
where the definitions in $D$ dimensions of functions $\overline{N}_i(y)$ 
and $\overline{NN}_i(y)$ are given in (\ref{Ni}-\ref{NNi}) 
in Appendix C, and the limit $D\rightarrow4$ in (\ref{NiD=4}),
which is taken here. The power series representation of this
function is
\begin{equation}
\mathcal{N}(y) =
	\frac{\beta^2}{6} \frac{\partial}{\partial\beta}
	\sum_{n=0}^{\infty} q_n \, y^n
	\Bigl[ A_n \ln\Bigl( \frac{y}{4} \Bigr) + B_n \Bigr]  \; ,
\end{equation}
where the coefficients are
\begin{eqnarray}
q_n &=& \frac{\Gamma(\frac{5}{2}\!+\!b_N\!+\!n)
			\Gamma(\frac{5}{2}\!-\!b_N\!+\!n)}
		{4^{n+1}(n\!+\!1)!(n\!+\!2)! \,
			\Gamma(\frac{1}{2}\!+\!b_N)\Gamma(\frac{1}{2}\!-\!b_N)} \; ,
\label{q_n}
\\
A_n &=& \frac{(n\!+\!1)}{8(n\!+\!3)(n\!+\!4)\beta}
	\Bigl[ n(n\!-\!1)\beta^2 - 4(n\!-\!1)(3n\!+\!2)\beta + 40n(n\!+\!1) \Bigr]  , \qquad
\label{A_n}
\\
B_n &=& \frac{(n\!+\!1)}{8\beta(n\!+\!3)(n\!+\!4)}
	\Bigl[ n(n\!-\!1)\beta^2 - 4(n\!-\!1)(3n\!+\!2)\beta + 40n(n\!+\!1) \Bigr]
\nonumber \\
& & \hspace{1cm}\times \Bigl[ \psi\Bigl( \frac{5}{2}\!+\!b_N\!+\!n \Bigr)
		+ \psi\Bigl( \frac{5}{2}\!-\!b_N\!+\!n \Bigr) 
		- \psi(n\!+\!2) - \psi(n\!+\!3) \Bigr]
\nonumber \\
& & + \frac{1}{8(n\!+\!3)^2(n\!+\!4)^2\beta} \Bigl[ 
	\beta^2 (n^4\!+\!14n^3\!+\!37n^2\!-\!12)
\nonumber \\
& & \hspace{4.3cm} -4\beta (3n^4\!+\!42n^3\!+\!125n^2\!+\!52n\!-\!22)
\nonumber \\
& & \hspace{4.3cm}+40(n\!+\!1)(n^3\!+\!13n^2\!+\!36n\!+\!12) \Bigr] . \qquad
\label{B_n}
\end{eqnarray}

Since the master function $\Psi(y,u,v)$ vanishes, 
according to (\ref{Frule}-\ref{Grule}), we have
for the vacuum polarization structure functions,
\begin{eqnarray}
F_{0d}(x;x') &=& -2(aa')^{D-2} \times I\Bigl[ \Phi_{0d}(y) \Bigr] \; ,
\\
G_{0d}(x;x') &=& (aa')^{D-2} \times I^2 \Bigl[ (D\!-\!1)\Phi_{0d}(y)
	+ y\Phi_{0d}'(y) \Bigr] \; .
\end{eqnarray}
Performing the integrals above gives the following,
\begin{eqnarray}
F_{0d}(x;x') &=& \frac{\kappa^2\Gamma^2(\frac{D}{2})}{8\pi^D}
	\Biggl\{ \frac{2 \ell_1(D)}{(D\!-\!1)aa' \Delta x^{2D-2}}
	+ \frac{2 \ell_2(D)H^2}{(D\!-\!2)\Delta x^{2D-4}} \Biggr\}
\nonumber \\
& &	\hspace{1cm} - \frac{\kappa^2H^6}{4\pi^4}(aa')^2
	\, I \Bigl[ y^{-2} \mathcal{N}(y) \Bigr] \; ,	\label{F0dBare}
\\
G_{0d}(x;x') &=&\frac{\kappa^2\Gamma^2(\frac{D}{2})}{8\pi^D}
	\Biggl\{ - \frac{\ell_1(D) H^2}{(D\!-\!1)(D\!-\!2)\Delta x^{2D-4}} \Biggr\}
\nonumber \\
& & \hspace{1cm} + \frac{\kappa^2H^6}{8\pi^4} (aa')^2
	\Biggl\{ 2I^2\Bigl[ y^{-2}\mathcal{N}(y) \Bigr]
		+ I\Bigl[ y^{-1}\mathcal{N}(y) \Bigr] \Biggr\} \; ,	\qquad
\label{G0dBare} 
\end{eqnarray}
where we have plugged in the definition $y=aa'H^2\Delta x^2$.

Next we need to localize the divergences using relations 
(\ref{Dx2D-4}) and (\ref{Dx2D-2}),
which we reproduce here,
\begin{eqnarray}
\frac{1}{\Delta x^{2D-4}} &\longrightarrow&
	\frac{4 \pi^{\frac{D}2} \mu^{D-4} i 
\delta^D(x \!-\! x')}{2 (D \!-\! 3) (D \!-\! 4) \Gamma( \frac{D}2 
\!-\! 1)} -\frac{\partial^2}4 \Bigl[ \frac{ \ln(\mu^2 \Delta x^2)}{
\Delta x^2} \Bigr] \; ,
\\
\frac1{\Delta x^{2D-2}} &\longrightarrow& \frac{4 \pi^{\frac{D}2} 
\mu^{D-4} \partial^2 i \delta^D(x \!-\! x')}{4 (D\!-\!2)^2 (D \!-\! 3) 
(D \!-\! 4) \Gamma(\frac{D}2 \!-\! 1)} -\frac{\partial^4}{32} \Bigl[ 
\frac{ \ln(\mu^2 \Delta x^2)}{\Delta x^2} \Bigr] \; . \qquad
\end{eqnarray}
The divergences in (\ref{F0dBare}) and (\ref{G0dBare}) 
are absorbed into the counterterms (\ref{cterms}) by choosing the
following coefficients
\begin{eqnarray}
C_4^{0d} &=&
	\frac{\kappa^2\mu^{D-4}}{32\pi^{\frac{D}{2}}}
	\frac{\Gamma(\frac{D}{2}) \ell_1(D)}{(D\!-\!1)(D\!-\!2)(D\!-\!3)(D\!-\!4)} \; ,
\\
\overline{C}^{0d} &=& (3D\!-\!8) C_4^{0d}
	- \frac{\kappa^2\mu^{D\!-\!4}}{16\pi^{\frac{D}{2}}}
	\frac{\Gamma(\frac{D}{2}) \ell_2(D)}{(D\!-\!3)(D\!-\!4)} \; .
\end{eqnarray}
Note that, since $\ell_1(D)=C_0(b)/2(D\!-\!2)$, where 
$C_0(b)$ was defined in (\ref{spin0C}), the $C_4^{0d}$ coefficient indeed coincides with  the
spin 0 part of the coefficient (\ref{C4values}) as inferred from the flat space 
limit \cite{Leonard:2012fs}. What remains is to perform 
explicitly the integrals in second lines of 
(\ref{F0dBare}-\ref{G0dBare}),
\begin{eqnarray}
\lefteqn{\frac{\beta^2}{6} \mathcal{N}_F(y) \equiv
	I\Bigl[ y^{-2} \mathcal{N}(y) \Bigr] }
\nonumber \\
&=& \frac{\beta^2}{6} \frac{\partial}{\partial\beta} \Biggl\{
	- \frac{q_0A_0}{y} \ln\Bigl( \frac{y}{4} \Bigr)
	- \frac{q_0(A_0\!+\!B_0)}{y}
	+ \frac{q_1A_1}{2} \ln^2\Bigl( \frac{y}{4} \Bigr)
	+ q_1B_1 \ln\Bigl( \frac{y}{4} \Bigr)
\nonumber \\
& & \hspace{2cm}+ \sum_{n=0}^{\infty} \frac{q_{n+2}}{(n\!+\!1)}y^{n+1}
	\Bigl[ A_{n+2} \ln\Bigl( \frac{y}{4} \Bigr)
	+B_{n+2} - \frac{A_{n+2}}{(n\!+\!1)} \Bigr] \Biggr\} . \qquad
\label{N_F}
\end{eqnarray}
\begin{eqnarray}
\lefteqn{ \frac{\beta^2}{6} \mathcal{N}_G(y) \equiv
	2I\Bigl[ y^{-2}\mathcal{N}(y) \Bigr] 
	+ I\Bigl[ y^{-1}\mathcal{N}(y) \Bigr]}
\nonumber \\
& &\hspace{0.5cm} = \frac{\beta^2}{6} \frac{\partial}{\partial\beta} \Biggl\{
	- \frac{q_0A_0}{2}\ln^2\Bigl( \frac{y}{4} \Bigr)
	- q_0(2A_0\!+\!B_0) \ln\Bigl( \frac{y}{4} \Bigr)
	+ q_1A_1 y\ln^2\Bigl( \frac{y}{4} \Bigr)
\nonumber \\
& & \hspace{2.5cm}+q_1(2B_1\!-\!A_1) y\ln\Bigl( \frac{y}{4} \Bigr)
	+ q_1(A_1\!-\!B_1) y
\nonumber \\
& & \hspace{1.5cm}+ \sum_{n=0}^{\infty} \frac{q_{n+2}y^{n+2}}{(n\!+\!1)(n\!+\!2)} 
	\Bigl[ (n\!+\!3) A_{n+2} \ln\Bigl( \frac{y}{4} \Bigr)
	+ (n\!+\!3)B_{n+2}
\nonumber \\
& & \hspace{6.5cm} - \frac{n^2\!+\!6n\!+\!7}{(n\!+\!1)(n\!+\!2)} A_{n+2} \Bigr]
	\Biggr\} . \qquad
\label{N_G}
\end{eqnarray}
The following integrals have been used to calculate the two functions above,
\begin{eqnarray}
&& I\Bigl[ y^n \ln\Bigl( \frac{y}{4} \Bigr) \Bigr] = 
	\frac{y^{n+1}}{(n\!+\!1)} 
	\Bigl[ \ln\bigl( \frac{y}{4} \Bigr) - \frac{1}{(n\!+\!1)} \Bigr]
	\quad (n\!\neq\!-1) \; ,
\\
&& I\Bigl[ y^{-1} \ln\Bigl( \frac{y}{4} \Bigr) \Bigr]
	= \frac{1}{2} \ln^2\Bigl( \frac{y}{4} \Bigr) \ ,
\\
&&I\Bigl[ \ln^2\Bigl( \frac{y}{4} \Bigr) \Bigr] = 
	2y \Bigl[ 1 - \ln\Bigl( \frac{y}{4} \Bigr)
	+ \frac{1}{2}\ln^2\Bigl( \frac{y}{4} \Bigr) \Bigr] , \quad 
\\
&& I\Big[ y^n \Bigr] = \frac{y^{n+1}}{(n\!+\!1)} \quad 
	(n\!\neq\!-1) \ ,
\qquad I\Bigl[y^{-1}\Bigr] =  \ln\Bigl( \frac{y}{4} \Bigr) \ ,
\end{eqnarray}
where the choice of integration constants is immaterial since any dependence
on them drops out when the derivatives in (\ref{Piform}) are acted with.
Here we list explicitly just the first few coefficients in (\ref{N_F}-\ref{N_G})
of the most singular terms in $y\rightarrow0$ limit, which will be relevant
for quantum-correcting Maxwell's equation (\ref{maxeqn}),
\begin{eqnarray}
\lefteqn{ \frac{\partial}{\partial\beta}(q_0A_0) = \frac{(\beta\!-\!5)}{48} \qquad ,
\qquad \frac{\partial}{\partial\beta}(q_1A_1) = \frac{(\beta\!-\!5)}{96} \ , }
\\
\lefteqn{\frac{\partial}{\partial\beta}(q_0B_0) = - \frac{5}{4\beta^2}
	- \frac{(\beta\!-\!2)(9\beta\!-\!86)}{2304} }
\nonumber \\
& &	\hspace{2cm}+\frac{(\beta\!-\!5)}{48} 
	\Bigl[ \psi\Bigl( \frac{5}{2}\!+\!b_N \Bigr) 
		+ \psi\Bigl( \frac{5}{2}\!-\!b_N \Bigr)
		+ 2\gamma_E - \frac{5}{2} \, \Bigr]
\nonumber \\
& &	\hspace{2cm}+ \frac{(\beta\!-\!6)(\beta\!-\!4)}{96} \Bigl( \frac{-1}{2b_N} \Bigr)
	\Bigl[ \psi'\Bigl( \frac{5}{2}\!+\!b_N \Bigr)
		- \psi'\Bigl(\frac{5}{2}\!-\!b_N \Bigr) \Bigr] \, ,
\\
\lefteqn{ \frac{\partial}{\partial\beta}(q_1B_1) = 
	- \frac{43}{384} + \frac{29\beta}{640} - \frac{3\beta^2}{512} 
	+\frac{\beta^3}{3840} }
\nonumber \\
& & \hspace{2cm}+ \frac{(\beta\!-\!5)}{96} \Bigl[ \psi\Bigl( \frac{7}{2}\!+\!b_N \Bigr)
	+ \psi\Bigl( \frac{7}{2}\!-\!b_N \Bigr)
	+ 2\gamma_E - \frac{10}{3} \, \Bigr]
\nonumber \\
& & \hspace{2cm}+ \frac{(\beta\!-\!6)(\beta\!-\!4)}{192} \Bigl( \frac{-1}{2b_N} \Bigr)
	\Bigl[ \psi'\Bigl( \frac{7}{2}\!+\!b_N \Bigr)
		- \psi'\Bigl( \frac{7}{2}\!-\!b_N \Bigr) \Bigr] \, , \qquad
\end{eqnarray}
where $b_N=[25/4-\beta]^{1/2}$. Even though the rest of the coefficients can be calculated from
(\ref{q_n}-\ref{B_n}), and the remaining series in (\ref{N_F})
and (\ref{N_G}) summed into generalized hypergeometric functions,
these will give irrelevant contributions when used in (\ref{maxeqn}),
so we do not do it here.

The renormalized contribution to the structure functions is
\begin{eqnarray}
\lefteqn{F_{0d}^{\rm ren}(y) =
	\Bigl( \frac{2b\!-\!1}{b\!-\!2} \Bigr)^2 \Biggl\{
	 \frac{\kappa^2}{48\pi^2} 
		\frac{\ln(a)}{aa'} \partial^2 i\delta^4(x\!-\!x') 
	-  \Bigl( \frac{b\!-\!8}{b\!-\!2} \Bigr)
		\frac{\kappa^2H^2}{72\pi^2} i\delta^4(x\!-\!x') }
\nonumber \\
& &	\hspace{2cm} - \frac{\kappa^2H}{48\pi^2a}
		\partial_0 i\delta^4(x\!-\!x')
 +  \frac{\kappa^2}{384\pi^4}
		\frac{\partial^4}{aa'} 
		\Bigl[ \frac{\ln(\mu^2\Delta x^2)}{\Delta x^2} \Bigr]
\nonumber \\
& & \hspace{2cm} +  \Bigl( \frac{b\!-\!8}{b\!-\!2} \Bigr)
		\frac{\kappa^2H^2}{576\pi^4}
		\partial^2 \Bigl[ \frac{\ln(\mu^2\Delta x^2)}{\Delta x^2} \Bigr] 
	 - \frac{\kappa^2H^6}{6\pi^4}(aa')^2 \mathcal{N}_F(y) \Biggr\} \; ,
\label{F0ren}
\\
\lefteqn{G_{0d}^{\rm ren}(y) = 
	\Bigl( \frac{2b\!-\!1}{b\!-\!2} \Bigr)^2
		\Biggl\{  \frac{\kappa^2H^2}{24\pi^2}
		\Bigl[ 1-\ln(a) \Bigr] i\delta^4(x\!-\!x')  }
\nonumber \\
& &	\hspace{3.5cm}-\frac{\kappa^2H^2}{192\pi^4} \partial^2 
		\Bigl[ \frac{\ln(\mu^2\Delta x^2)}{\Delta x^2} \Bigr]
+ \frac{\kappa^2H^6}{12\pi^4} (aa')^2
	\mathcal{N}_G(y)  \Biggr\} . \qquad 
\label{G0ren}
\end{eqnarray}
Since the renormalized "0a" and "0b" contributions (\ref{F0aG0a}) 
and (\ref{F0bG0b})
vanish, the "0d" contribution above constitutes the full contribution to
the vacuum polarization structure functions from the spin 0 part.
Note that there also might appear finite contributions of the form
(\ref{DF}-\ref{DG}).

\section{Discussion}\label{discuss}

We have evaluated the one graviton loop contribution to the photon
vacuum polarization $i[\mbox{}^{\mu} \Pi^{\nu}](x;x')$ on de 
Sitter background in the 1-parameter family of exact, de Sitter 
invariant gauges (\ref{gauge}). The result is represented in 
terms of two structure functions $F(x;x')$ and $G(x;x')$, whose
relation to $i[\mbox{}^{\mu} \Pi^{\nu}](x;x')$ was defined in
equations (\ref{Piform}) and (\ref{ourrep}). Our graviton propagator 
has a gauge independent, de Sitter breaking spin two part and a 
gauge dependent but de Sitter invariant (for $b > 2$) spin zero 
part. Each part makes distinct contributions to the structure 
functions. The spin two structure functions are (\ref{F2ren}) 
and (\ref{G2ren}); the corresponding spin zero results are 
(\ref{F0ren}) and (\ref{G0ren}).

The point of this exercise was to check the conjecture 
\cite{Miao:2012xc} that the leading secular dependence of solutions 
to effective field equations such as (\ref{maxeqn}) might be gauge
independent. The full solutions certainly contain unphysical gauge
and field variable dependent information because the flat space
limit does \cite{Leonard:2012fs}. However, they also contain 
physical information because one can use them to construct the flat 
space S-matrix \cite{BjerrumBohr:2002sx}. What we need is a filter 
to distinguish physical effects from unphysical ones. The S-matrix 
provides this in flat space, but there is as yet no analog for 
cosmology. There is no question that such a filter exists because
astronomers are measuring {\it something}. Identifying what 
theoretical quantity represents these measurements is one of the
central problems of cosmological quantum field theory 
\cite{Woodard:2014jba}.

Note that the vacuum polarization will play an essential role {\it 
whatever} is the outcome. If the conjecture proves to be correct 
then one can extract physical information from the leading secular 
dependence of solutions to (\ref{maxeqn}). If the conjecture proves 
false then one must resort to some form of gauge invariant Green's 
function, which would inevitably consist of the expectation value 
of the field (which is what solving the effective field equations 
gives) plus some extra terms to filter out the gauge dependence 
\cite{Tsamis:1989yu}.

Unfortunately, we are not yet in a position to answer this fascinating
question. Had this computation produced the same kinds of terms, with 
different numerical coefficients, as occur with the noncovariant gauge 
\cite{Leonard:2013xsa}, it would not have been necessary to do much
work to solve equation (\ref{maxeqn}). In that case we would simply 
have read off the result for each term from the previous solutions 
\cite{Glavan:2013jca,Wang:2014tza}. Those analyses show that the
largest effects derive from the $\kappa^2 H^2 \ln(a) i\delta^4(x - x')$ 
part of $F(x;x)$. It is interesting to note that the coefficient of
this term is independent of the parameter $b$, although it does not 
agree with the noncovariant result. We find a coefficient of 
$+\frac{85}{72\pi^2}$ in equation (\ref{F2ren}) whereas the coefficient
was $+\frac1{8\pi^2}$ in equation (136) of the noncovariant gauge
analysis \cite{Leonard:2013xsa}. However, our computation also produced
some quantitatively different terms at the end of expressions 
(\ref{F2ren}), (\ref{G2ren}), (\ref{F0ren}) and (\ref{G0ren}). Each of 
those terms contains multiplicative factors of $(a a')^2$ which should
compensate for the measure of the $d^4x'$ integration. If these terms
acquire an extra logarithm they can contribute just as strongly as the 
local logarithms. So there seems no alternative to working out another 
set of complicated integrations for these new terms, multiplied by the 
classical field strengths for dynamical photons and for the response to 
charges and currents. Key questions are whether or not the leading 
secular effects depend on the gauge parameter $b$, and whether or not
they agree with the noncovariant gauge. It would also be interesting to 
work out the spin 0 structure functions for $b < 2$.

One novel feature of our computation is the need for a
noninvariant counterterm, despite our use of an invariant
regularization and a de Sitter invariant gauge. As was
explained in section 2.3, the problem arises from the time
ordering of the $h \partial A \partial A$ interaction and from 
the fact that the coincident graviton propagator contains a 
logarithmic divergence (the famous ``tail term'') which gives
rise to a factor of $1/(D-4)$ in dimensional regularization.
There is nothing we can do about the derivative interactions
of quantum gravity or the logarithmic divergence of the
coincident graviton propagator on de Sitter. We {\it might} 
impose a covariant ordering prescription on the interactions 
but this is problematic in view of the need to keep the effective 
field equation (\ref{maxeqn}) real and causal by using the 
Schwinger-Keldysh formalism \cite{Schwinger:1960qe,
Mahanthappa:1962ex,Bakshi:1962dv,Bakshi:1963bn,Keldysh:1964ud,
Chou:1984es,Jordan:1986ug,Calzetta:1986ey,Calzetta:1986cq}. 
\footnote{Note that our wish to use the in-in formalism is only
the {\it motivation} for employing time-ordering. Once that is
done, the ultraviolet divergences of the in-in formalism are the
same as those of the in-out formalism.}
The only alternative is to resort to the same noninvariant 
counterterm --- the term proportional to $\Delta C$ in 
expression (\ref{cterms}) --- that was needed with the
noncovariant gauge \cite{Leonard:2013xsa}.

One major spin-off from our work is the great simplification 
that was made in section (3.1) for representing the spin two
part of the graviton propagator. This should facilitate 
checking the gauge dependence of other one loop computations 
involving gravitons such as the fermion self-energy 
\cite{Miao:2005am,Miao:2006gj,Miao:2007az,Miao:2012bj} and 
the graviton self-energy \cite{Tsamis:1996qk,Mora:2013ypa}.

The photon propagator identities (\ref{photon1}-\ref{photon2}) 
and their graviton analogs (\ref{graviton1}) and (\ref{graviton2}) 
are strikingly similar. Equation (\ref{photon2}) expresses the
photon propagator as a longitudinal term plus the tensor 
$-\frac1{2 H^2} \partial_{\rho} \partial_{\sigma}' y(x;x')$ times 
the $B$-type propagator, whereas equation (\ref{graviton2}) gives 
the (de Sitter invariant part of the) graviton propagator as a 
collection of traces and gradients plus the tensor $\frac1{2 H^4}
\partial_{\mu} \partial_{(\rho}' y \times \partial_{\sigma)}' 
\partial_{\nu} y$ times the (invariant part of the) $A$-type 
propagator. So the tensor structure of the graviton propagator
is the square of the photon tensor structure, up to terms which
drop out when contracted into the appropriate polarizations. This
seems like the remarkable insight that on-shell gravitational 
scattering amplitudes are essentially the squares of gauge 
scattering amplitudes \cite{Sannan:1986tz,Kawai:1985xq,Bern:1998ug,
Bern:2005bb}. This has usually been thought to arise from 
simplifications that occur from going on shell but perhaps it is 
partly due to working in exact gauges.

\vskip 1cm

\centerline{\bf Acknowledgements}

We are grateful for conversation and correspondence on this subject
with S. Deser and K. E. Leonard. This work is part of the D-ITP
consortium, a program of the Netherlands Organization for Scientific
Research (NWO) that is funded by the Dutch Ministry of Education,
Culture and Science (OCW). It was also partially supported by Taiwan 
MOST grant 103-2112-M-006-001-MY3, the Focus Group on Gravitation 
of the Taiwan National Center for Theoretical Sciences, by NSF grant 
PHY-1205591, and by the Institute for Fundamental Theory at the 
University of Florida.

\subsection{Appendix A: Photon Propagator Functions}

The propagator for minimally coupled scalar with mass $M_S^2 = 
(D-2) H^2$ obeys equation (\ref{1stint}) and consists of a de Sitter
invariant function of $y(x;x')$ called $i\Delta_B(x;x) \equiv B(y)$.
It is closely related --- by equation (\ref{BCident}) --- to the
propagator for a minimally coupled scalar of $M_S^2 = 2(D-3) H^2$,
which is called $i\Delta_C(x;x') \equiv C(y)$. Their expansions are, 
\begin{eqnarray} 
\lefteqn{B(y)= \frac{H^{D-2}}{(4 \pi)^{\frac{D}2}} \Biggl\{ 
\frac{\Gamma(\frac{D}2)}{\frac{D}2 \!-\! 1} \Bigl(\frac{4}{y}
\Bigr)^{\frac{D}2 -1} } \nonumber \\
& & \hspace{2.5cm} - \sum_{n=0}^{\infty} \Biggl[ \frac{\Gamma(n 
\!+\! D \!-\! 2)}{\Gamma(n \!+\! \frac{D}2)} \Bigl(\frac{y}4 \Bigr)^n
- \frac{\Gamma(n \!+\! \frac{D}2)}{\Gamma(n \!+\! 2)} \Bigl(\frac{y}4 
\Bigr)^{n- \frac{D}2 +2} \Biggr]\!\Biggr\} . \qquad \label{B(y)} \\ 
\lefteqn{C(y)= \frac{H^{D-2}}{(4 \pi)^{\frac{D}2}} \Biggl\{ 
\frac{\Gamma(\frac{D}2)}{\frac{D}2 \!-\! 1} \Bigl(\frac{4}{y}
\Bigr)^{\frac{D}2 -1} } \nonumber \\
& & \hspace{-.5cm} \!+\! \sum_{n=0}^{\infty} \Biggl[ \frac{(n\!+\!1) 
\Gamma(n \!+\! D \!-\! 3)}{\Gamma(n \!+\! \frac{D}2)}
\Bigl(\frac{y}4 \Bigr)^n \!\!\!\!- \frac{(n \!-\! \frac{D}2 \!+\! 3) 
\Gamma(n \!+\! \frac{D}2 \!-\! 1)}{\Gamma(n \!+\! 2)} 
\Bigl(\frac{y}4 \Bigr)^{n - \frac{D}2 +2} \! \Biggr] \! \Biggr\} . 
\qquad \label{C(y)}
\end{eqnarray}
They obviously agree in $D=4$ spacetime dimensions. Our work requires
only the first two terms in the expansions of their derivatives,
\begin{eqnarray}
B'(y) & = & -\frac{H^{D-2} \Gamma(\frac{D}2)}{4 \pi^{\frac{D}2}} 
\Biggl\{\frac1{y^{\frac{D}2}} + \frac{(D\!-\!4)}{8 y^{\frac{D}2-1}}
+ O\Bigl( (D\!-\!4) y^0\Bigr)\Biggr\} \; , \qquad \label{Bprim} \\
C'(y) & = & -\frac{H^{D-2} \Gamma(\frac{D}2)}{4 \pi^{\frac{D}2}} 
\Biggl\{\frac1{y^{\frac{D}2}} + \frac{(D\!-\!6)(D\!-\!4)}{8 (D\!-\!2)
y^{\frac{D}2-1}} + O\Bigl( (D\!-\!4) y^0\Bigr)\Biggr\} \; . \qquad 
\label{Cprim}
\end{eqnarray}

\subsection{Appendix B: Spin Two Propagator Functions}

The propagator for a massless, minimally coupled scalar obeys
equation (\ref{DA}) and consists of a de Sitter invariant function 
of $y(x;x')$ plus a de Sitter breaking logarithm \cite{Onemli:2002hr},
\begin{equation}
i\Delta_A(x;x') = A(y) + k \ln(a a') \qquad , \qquad k \equiv
\frac{H^{D-2}}{(4 \pi)^{\frac{D}2}} \frac{\Gamma(D\!-\!1)}{
\Gamma(\frac{D}2)} \; .
\end{equation}
The expansion for $A(y)$ is, 
\begin{eqnarray}
\lefteqn{A(y) = \frac{H^{D-2}}{(4\pi)^{\frac{D}2}} \Biggl\{ 
\frac{\Gamma(\frac{D}2)}{\frac{D}2 \!-\! 1} \Bigl(\frac{4}{y}
\Bigr)^{\frac{D}2 -1} + \frac{\Gamma(\frac{D}2 \!+\!1)}{\frac{D}2 
\!-\! 2} \Bigl( \frac{4}{y}\Bigr)^{\frac{D}2-2} + A_1 } \nonumber \\
& & \hspace{1.1cm} + \sum_{n=1}^{\infty} \Biggl[
\frac{\Gamma(n \!+\! D \!-\! 1)}{n \, \Gamma(n \!+\! \frac{D}2)}
\Bigl(\frac{y}4 \Bigr)^n - \frac{\Gamma(n \!+\!  \frac{D}2 \!+\! 1)}{
(n \!-\! \frac{D}2 \!+\! 2) \Gamma(n \!+\! 2)} \Bigl(\frac{y}4
\Bigr)^{n - \frac{D}2 +2} \Biggr] \!\Biggr\} , \qquad \label{A(y)}
\end{eqnarray}
where the constant $A_1$ is,
\begin{equation}
A_1 = \frac{\Gamma(D \!-\!1)}{\Gamma(\frac{D}2)} \Biggl\{ -
\psi\Bigl(1 \!-\! \frac{D}2\Bigr) + \psi\Bigl( \frac{D \!-\! 1}2
\Bigr) + \psi(D \!-\! 1) + \psi(1)\Biggr\} .
\end{equation}

The full, $D$-dimensional series expansion for $i\Delta_{AAABB
}^{\rm inv}(x;x')$ has been derived \cite{Kahya:2011sy} but we here 
require only the de Sitter breaking part,
\begin{eqnarray}
\lefteqn{i\Delta_{AAABB}^{\rm brk} = \frac{k H^{-8}}{(D\!-\!2)^2} 
\Biggl\{\frac{\ln^3(4 a a')}{6 (D \!-\! 1)^2} + \Biggl[ \frac{\psi(
\frac{D-1}2)}{(D \!-\! 1)^2} -\frac{(D \!-\! \frac32) D}{(D \!-\! 2) 
(D \!-\! 1)^3} \Biggr] \ln^2(4 a a') } \nonumber \\
& & \hspace{-.1cm} + \Biggl[ \frac{\psi'(\frac{D-1}2) \!+\! 
2 \psi^2(\frac{D-1}2)}{(D \!-\! 1)^2} - \frac{2 (2D \!-\! 3) D 
\psi(\frac{D-1}2)}{(D \!-\! 2) (D\!-\!1)^3} + \frac3{(D \!-\! 2)^2} 
\Biggr] \ln(4 a a') \Biggr\} .\qquad \label{DAAABBbrk}
\end{eqnarray}

It is simple to act the scalar d'Alembertian on functions of
$\ln(a a')$,
\begin{equation}
\square f\Bigl( \ln(a a')\Bigr) = -H^2 \Biggl\{ (D\!-\!1) 
f'\Bigl( \ln(a a')\Bigr) + f''\Bigl( \ln(a a')\Bigr) \Biggr\} 
\; . \label{boxlog}
\end{equation}
Because $i\Delta_{AAABB}^{\rm inv} = i\Delta_{AAABB} - 
i\Delta_{AAABB}^{\rm brk}$ we can use relations 
(\ref{DAAABB}-\ref{DAABB}), (\ref{DAAABBbrk}) and (\ref{boxlog})
to conclude,
\begin{eqnarray}
\lefteqn{ \Bigl[\square \!-\! (D\!-\!2) H^2\Bigr] \square 
i\Delta^{\rm inv}_{AAABB}(x;x') = i\Delta_{AAB}(x;x') - 
\frac{k H^{-4}}{(D\!-\!2)(D\!-\!1)} } \nonumber \\
& & \hspace{.5cm} \times \Biggl\{\frac12 \ln^2(4 a a') 
 + \Bigl[ 2 \psi\Bigl(\frac{D\!-\!1}2\Bigr)
- \Bigl(\frac{D\!-\!1}{D\!-\!2}\Bigr) \Bigr] \ln(4 a a') +
{\rm constant}\Biggr\} \; , \qquad \label{AABinv}
\end{eqnarray}
where the constant is,
\begin{equation}
\psi'\Bigl(\frac{D\!-\!1}2\Bigr) \!+\! 2 \psi^2\Bigl(
\frac{D\!-\!1}2\Bigr) \!-\! \Bigl(\frac{D\!-\!1}{D\!-\!2}\Bigr) 
\psi\Bigl(\frac{D\!-\!1}2\Bigr) \!+\! \Bigl(\frac{D\!-\!1}{D\!-\!
2}\Bigr) \!+\! \frac{2D \!-\! 3}{(D\!-\!2)^2 (D\!-\!1)^2} \; .
\end{equation}
This completes the demonstration of equation (\ref{Jzero}).

Because our structure functions are at most quadratically divergent
we require only the leading two terms of $J(y)$ to be kept in $D$ 
dimensions. Rather than acting the derivatives of (\ref{AABinv})
on the complicated series expansion of $i\Delta_{AAABB}(x;x')$ 
that has been derived \cite{Kahya:2011sy} it is simplest to 
construct $J(y)$ by integrating the differential equation 
$i\Delta_{AAB}(x;x')$ obeys,
\begin{equation}
\square \, i\Delta_{AAB}(x;x') = i\Delta_{AB}(x;x') =
\frac{[i\Delta_{B}(x;x') \!-\! i\Delta_{A}(x;x')]}{(D\!-\!2)
H^2} \; . \label{AABeqn}
\end{equation}
From the expansions (\ref{B(y)}) and (\ref{A(y)}) one finds,
\begin{equation}
i\Delta_{AB}(x;x') = H^2 \Biggl\{ \frac{\alpha_1}{y^{\frac{D}2-2}}
+ \alpha_2 + \frac{\beta_1}{y^{\frac{D}2-3}} + \beta_2 y +
O\Bigl( (D\!-\!4) y^2\Bigr) \Biggr\} \; ,
\end{equation}
where the coefficients are,
\begin{eqnarray}
\alpha_1 & = & - \frac{H^{D-6}}{4 \pi^{\frac{D}2}} 
\frac{\Gamma(\frac{D}2)}{(D\!-\!2)(D\!-\!4)} \; , 
\label{alpha1} \\
\alpha_2 & = & \frac{H^{D-6}}{(4 \pi)^{\frac{D}2}} 
\frac{\Gamma(D\!-\!2)}{\Gamma(\frac{D}2)} \Biggl\{ 
\frac2{D\!-\!4} \nonumber \\
& & \hspace{3cm} + \psi\Bigl(3 \!-\! \frac{D}2\Bigr) \!-\!
\psi\Bigl(\frac{D\!-\!1}{2}\Bigr) \!-\! \psi(D\!-\!2) \!-\!
\psi(1)\Biggr\} , \qquad \label{alpha2} \\
\beta_1 & = & - \frac{H^{D-6}}{16 \pi^{\frac{D}2}} 
\frac{\Gamma(\frac{D}2 \!+\! 1)}{(D\!-\!2)(D\!-\!6)} \; , 
\label{beta1} \\
\beta_2 & = & - \frac{H^{D-6}}{(4 \pi)^{\frac{D}2}} 
\frac{\Gamma(D\!-\!2)}{2 \Gamma(\frac{D}2)} \; . 
\label{beta2}
\end{eqnarray}

Expression (\ref{boxlog}) defines how to integrate the de
Sitter breaking $\ln(a a')$ term on the right hand side of 
(\ref{AABeqn}). The de Sitter invariant analog is,
\begin{equation}
\square g(y) = H^2 \Bigl[ (4y\!-\!y^2) g''(y) + D(2\!-\!y)
g'(y)\Bigr] \; . \label{boxy}
\end{equation}
Equation (\ref{boxy}) suggests that we can solve $\square g(y) = 
h(y)$ as a double integral with respect to $y$,
\begin{equation}
\square g(y) = h(y) \qquad \Longrightarrow \qquad g(y) =
\frac1{H^2} \, I\Biggl[ \frac{I[(4y \!-\! y^2)^{\frac{D}2-1} 
h(y)]}{(4y \!-\! y^2)^{\frac{D}2}} \Biggr] \; . \label{formal}
\end{equation}
However, relation (\ref{formal}) is neither tractable in $D$
dimensions, nor even correct. The problem with tractability 
is obvious from the $D$-dependent powers of $(4y - y^2)$. The
problem with validity derives from the impossibility of avoiding
poles at either $y=0$ or $y=4$ that would introduce delta
functions into relation (\ref{boxy}) which are not present in
the desired source function $h(y)$. 

We solve the first problem by extracting the two leading powers 
of $y$ from the solution and then taking $D=4$ on the remainder,
\begin{eqnarray}
\lefteqn{ i\Delta_{AAB}(x;x') = -\frac{\alpha_1}{4 (D\!-\!6)} 
\frac1{y^{\frac{D}2-3}} + \frac{\alpha_2}{2 D} y -
\frac{[(D\!+\!4) \alpha_1 \!+\! 16 \beta_1]}{96 (D\!-\!8)}
\frac1{y^{\frac{D}2-4}} } \nonumber \\
& & \hspace{.2cm} + \frac{[\alpha_2 \!+\! 2 \beta_2]}{8
(D\!+\!2)} \, y^2  + \frac{k H^{-4}}{(D\!-\!1)(D\!-\!2)} \Bigl[
\frac12 \ln^2(a a') \!-\! \frac{\ln(a a')}{D\!-\!1}\Bigr] +
\Delta g(x;x') \; . \qquad \label{1stexp}
\end{eqnarray}
The remainder $\Delta g(x;x')$ obeys,
\begin{eqnarray}
\lefteqn{ \frac{\square}{H^2} \Delta g = 
\frac{(D\!+\!6) [ (D\!+\!4) \alpha_1 \!+\! 16 \beta_1]}{384
\, y^{\frac{D}2-4}} + \frac{(D\!+\!1) (\alpha_2 \!+\! 2 
\beta_2) y^2}{4 (D\!+\!2)} + O(D\!-\!4) , } \\
& & \hspace{4.3cm} \longrightarrow \frac{5 H^{-2}}{384 \pi^2} 
\Bigl[ \ln\Bigl( \frac{y}4\Bigr) \!-\! \frac{143}{60} \!+\! 2 
\ln(2) \!+\! 2 \gamma\Bigr] y^2 \; . \qquad \label{D4source}
\end{eqnarray}
We solve the second problem by first setting the lower limit 
to $y=4$ on the inner integral of (\ref{formal}), which means 
there are no poles at $y=4$. Then the poles at $y=0$ are 
cancelled by adding a constant times the $D=4$ limit of the $i
\Delta_A(x;x')$. Because expression (\ref{D4source}) consists
of terms proportional to $y^2$ and to $y^2 \ln(\frac{y}4)$ we
only need solutions for these two sources,
\begin{eqnarray}
\lefteqn{ \frac{\square}{H^2} g_1(x;x') = y^2 \qquad 
\Longrightarrow \qquad g_1(x;x') = -\frac35 y -\frac1{10} y^2
- \frac85 \ln(a a') \; , \label{g1} } \\
\lefteqn{ \frac{\square}{H^2} g_2(x;x') = y^2 \ln\Bigl(\frac{y}4
\Bigr) \; \Longrightarrow \; g_2(x;x') = -\frac85
\Bigl[ {\rm Li}_2\Bigl(\frac{y}4\Bigr) \!+\! \ln\Bigl(1 \!-\! 
\frac{y}4\Bigr) \ln\Bigl(\frac{y}4\Bigr)\Bigr] } \nonumber \\
& & \hspace{2cm} + \Biggl[ \frac{ \frac45 y}{4 \!-\!y} \!-\!
\frac35 y \!-\! \frac1{10} y^2\Bigr] \ln\Bigl( \frac{y}4\Bigr)
\!+\! \frac{67}{100} y \!+\! \frac{7}{100} y^2 \!+\! 
\frac{18}{25} \ln(a a') \; . \qquad \label{g2}
\end{eqnarray}
Here ${\rm Li}_{2}(z)$ is the dilogarithm function,
\begin{equation}
{\rm Li}_2(z) \equiv -\int_0^z \!\!\! dt \, 
\frac{\ln(1\!-\!t)}{t} \; . \label{dilog}
\end{equation}

The de Sitter breaking factors of $\ln(a a')$ in expressions
(\ref{1stexp}), (\ref{g1}) and (\ref{g2}) obviously do not 
belong in $J(y)$,
\begin{eqnarray}
\lefteqn{ J(y) = -\frac{\alpha_1}{4 (D\!-\!6)} 
\frac1{y^{\frac{D}2-3}} + \frac{\alpha_2}{2 D} y -
\frac{[(D\!+\!4) \alpha_1 \!+\! 16 \beta_1]}{96 (D\!-\!8)}
\frac1{y^{\frac{D}2-4}} + \frac{[\alpha_2 \!+\! 2 \beta_2]}{8
(D\!+\!2)} \, y^2 } \nonumber \\
& & \hspace{.2cm} + {\rm Constant} 
+ \frac{5 H^{-2}}{384 \pi^2} \Bigl[-\frac{143}{60} \!+\! 2
\ln(2) \!+\! 2 \gamma\Bigr] \!\times\! g_1(y) 
+ \frac{5 H^{-2}}{384 \pi^2} \!\times\! g_2(y)
\; . \qquad 
\label{Jexp}
\end{eqnarray}
Recall that the four $D$-dependent constants $\alpha_i$
and $\beta_i$ are given in expressions (\ref{alpha1}-\ref{beta2}).
The functions $g_1(y)$ and $g_2(y)$ are just the de Sitter 
invariant parts of (\ref{g1}) and (\ref{g2}),
\begin{eqnarray}
g_1(y) & \equiv & -\frac35 y - \frac1{10} y^2 \; , \qquad 
\label{g1(y)} \\
g_2(y) & \equiv & -\frac85 \Bigl[ {\rm Li}_2\Bigl(\frac{y}4\Bigr) 
\!+\! \ln\Bigl(1 \!-\! \frac{y}4\Bigr) \ln\Bigl(\frac{y}4\Bigr)\Bigr] 
\nonumber \\
& & \hspace{2.5cm} + \Biggl[ \frac{ \frac45 y}{4 \!-\!y} \!-\!
\frac35 y \!-\! \frac1{10} y^2\Bigr] \ln\Bigl( \frac{y}4\Bigr)
\!+\! \frac{67}{100} y \!+\! \frac{7}{100} y^2 \; . \label{g2(y)}
\qquad 
\end{eqnarray}

\subsection{Appendix C: Spin Zero Propagator Functions}

The spin-0 scalar structure function $i\Delta_{WNN}(x;x')$, 
introduced in (\ref{gravpropspin0}) is constructed out of two scalar propagators, 
$i \Delta_W(x;x')$ and $i\Delta_N(x;x')$ \cite{Kahya:2011sy}. They satisfy 
propagator equations for a massive minimally coupled scalar,
\begin{eqnarray}
\Bigl[ \square + DH^2 \Bigr] i\Delta_{W}(x;x') 
	&=& \frac{i\delta^D(x \!-\! x')}{\sqrt{-g(x)}} \; ,
\\
\Bigl[ \square - (\beta \!-\! D)H^2 \Bigr] i\Delta_{N}(x;x') 
	&=& \frac{i\delta^D(x \!-\! x')}{\sqrt{-g(x)}} \; .
\end{eqnarray}
The $W$-type scalar propagator has a tachyonic mass
$M_S^2=-DH^2$, and necessarily breaks de Sitter invariance.
We split it into a de Sitter invariant and a de Sitter breaking part,
\begin{equation}
i \Delta_W(x;x') = W(y) + \delta W(y,u,v) \; .
\label{Wsplit}
\end{equation}
The de Sitter invariant part is
\begin{eqnarray}
\lefteqn{W(y) = \frac{H^{D-2}}{(4\pi)^{\frac{D}2}} \Biggl\{ 
	\frac{\Gamma(\frac{D}2)}{\frac{D}2 \!-\! 1} 
		\Bigl(\frac{4}{y}\Bigr)^{\frac{D}2 -1} 
	+ \frac{\Gamma(\frac{D}2 \!+\!2)}
			{(\frac{D}2 \!-\! 2)(\frac{D}2 \!-\! 1)} 
		\Bigl( \frac{4}{y}\Bigr)^{\frac{D}{2}-2} }
\nonumber \\
& & \hspace{2.5cm}	
	+ \frac{\Gamma(\frac{D}{2} \!+\! 3)}
			{2(\frac{D}{2} \!-\! 3)(\frac{D}{2} \!-\! 2)} 
		\Bigl( \frac{4}{y}\Bigr)^{\frac{D}{2}-3}  
	+ W_1 + W_2 \Bigl( \frac{y \!-\! 2}{4} \Bigr)
\nonumber \\
& & \hspace{1.3cm} 
	+ \sum_{n=2}^{\infty} \Biggl[
	\frac{\Gamma(n \!+\! \frac{D}{2} \!+\! 2)(\frac{y}{4} )^{n-\frac{D}{2}+2}}
		{(n \!-\! \frac{D}{2} \!+\! 2)(n \!-\! \frac{D}{2} \!+\! 1)(n \!+\! 1)!}
 	- \frac{\Gamma(n \!+\! D)(\frac{y}{4} )^{n}}
		{n(n \!-\! 1) \Gamma(n \!+\! \frac{D}{2})}  \Biggr] \!\Biggr\} , \qquad 
\label{W(y)}
\end{eqnarray}
where
\begin{eqnarray}
W_1 &=& \frac{\Gamma(D \!+\! 1)}{\Gamma(\frac{D}{2} \!+\! 1)}
	\Biggl\{ \frac{D \!+\! 1}{2D} \Biggr\} ,
\\
W_2 &=& \frac{\Gamma(D \!+\! 1)}{\Gamma(\frac{D}{2} \!+\! 1)}
	\Biggl\{ \psi \Bigl(\!-\frac{D}{2}\Bigr) - \psi\Bigl(\frac{D \!+\! 1}{2}\Bigr)
	- \psi(D \!+\! 1) - \psi(1) \Biggr\} .
\end{eqnarray}
The de Sitter breaking part of the $W$-type propagator is
\begin{eqnarray}
 \delta W(y,u,v) = k \, \Biggl\{ (D \!-\! 1)^2 e^u
	+ \Bigl( \frac{D \!-\! 1}{2} \Bigr) (2 \!-\! y) u
	- 2\cosh(v) \Biggr\}  ,
\label{deltaW}
\\
 k = \frac{H^{D-2}}{(4\pi)^{\frac{D}{2}}}
	\frac{\Gamma(D\!-\!1)}{\Gamma(\frac{D}{2})} \; . \qquad
\end{eqnarray}
In general, the $N$-type propagator contains de Sitter breaking parts as well,
but for the choice of gauge in this work $b>2$ ($\beta>D$), it mass is
$M_S^2=(\beta\!-\!D)H^2>0$. Therefore, it is completely de Sitter invariant,
$i\Delta_N(x;x')=N(y)$,
\begin{eqnarray}
\lefteqn{N(y)= \frac{H^{D-2}}{(4 \pi)^{\frac{D}2}} \Biggl\{ 
	\frac{\Gamma(\frac{D}2)}{\frac{D}2 \!-\! 1} 
		\Bigl(\frac{4}{y} \Bigr)^{\frac{D}{2} -1} }
\nonumber \\
& &	\hspace{1cm} - \frac{\Gamma(\frac{D}{2}) \Gamma(1 \!-\! \frac{D}{2})}
		{\Gamma(\frac{1}{2} \!+\! b_N) \Gamma(\frac{1}{2} \!-\! b_N)} 
		\sum_{n=0}^{\infty} \Biggl[ 
	\frac{\Gamma(\frac{3}{2} \!+\! b_N \!+\! n) 
			\Gamma(\frac{3}{2} \!-\! b_N \!+\! n)}
		{\Gamma(3 \!-\! \frac{D}{2}+n) (n \!+\! 1)!} 
		\Bigl(\frac{y}4 \Bigr)^{n-\frac{D}{2}+2} 
\nonumber \\
& &	\hspace{4cm} -\frac{\Gamma(\frac{D\!-\!1}{2} \!+\! b_N \!+\! n) 
			\Gamma(\frac{D\!-\!1}{2} \!-\! b_N \!+\! n)}
		{\Gamma(\frac{D}{2}+n) n!} 
		\Bigl(\frac{y}4 \Bigr)^{n} \Biggr]\!\Biggr\} ,
\label{N(y)}
\end{eqnarray}
where
\begin{equation}
b_N = \sqrt{\frac{(D \!-\! 1)}{4} \Bigl[ (D \!-\! 1) + \frac{8}{2 \!-\! b} \Bigr]}
	= \sqrt{ \Bigl( \frac{D \!+\! 1}{2} \Bigr)^2 - \beta} \; .
\end{equation}
Note that even though the terms in the series in (\ref{N(y)})
cancel in $D=4$, the whole series is multiplied by a factor diverging
as $1/(D\!-\!4)$ giving a finite contribution in this limit.

The spin 0 scalar structure function in (\ref{gravpropspin0}) is solved for
by inverting (\ref{DWNN}-\ref{DW}), the solution of which is
\begin{equation}
i \Delta_{WNN}(x;x') 
	= \frac{i\Delta_{NN}(x;x')-i\Delta_{WN}(x;x')}{\beta H^2} \; ,
\label{WNNsolution}
\end{equation}
where
\begin{eqnarray}
i\Delta_{WN}(x;x') 
	&=& \frac{i\Delta_N(x;x') - i \Delta_W(x;x')}{\beta H^2} \; ,
\label{WNsolution}
\\
i \Delta_{NN}(x;x')
	&=& - \frac{1}{2b_N H^2} \frac{\partial}{\partial b_N} i\Delta_N(x;x')
	= \frac{1}{H^2} \frac{\partial}{\partial\beta} i\Delta_N(x;x') \; .
\label{NNsolution}
\end{eqnarray}
The functions in (\ref{WNNsolution}-\ref{NNsolution}) we can split into
de Sitter invariant and de Sitter breaking parts,
\begin{eqnarray}
i\Delta_{WNN}(x;x') &=& W\!N\!N(y) + \delta W\!N\!N(y,u,v) \; ,
\\
i \Delta_{WN}(x;x') &=& W\!N(y) + \delta W\!N(y,u,v) \; ,
\\
i \Delta_{NN}(x;x') &=& N\!N(y) + \delta N\!N(y,u,v) \; .
\end{eqnarray}
The de Sitter breaking parts receive contribution only from the
de Sitter breaking part of the $W$-type propagator (\ref{deltaW}),
since the $N$-type propagator is de Sitter invariant ($\delta N(y,u,v)=0$),
\begin{eqnarray}
\delta W\!N\!N(y,u,v) &=& \frac{\delta W(y,u,v)}{\beta^2 H^4} \; ,
\label{deltaWNN}
\\
\delta W\!N\!(y,u,v) &=& - \frac{\delta W(y,u,v)}{\beta H^2} \; ,
\label{deltaWN}
\\
\delta N\!N(y,u,v) &=& 0 \; .
\label{deltaNN}
\end{eqnarray}
The de Sitter invariant parts are
\begin{eqnarray}
W\!N\!N(y) &=& \frac{N\!N(y) - W\!N(y)}{\beta H^2} \; ,
\label{WNN}
\\
W\!N(y) &=& \frac{N(y)-W(y)}{\beta H^2} \; ,
\label{WN}
\\
N\!N(y) &=& \frac{\partial}{\partial\beta} \frac{N(y)}{H^2} \; .
\label{NN}
\end{eqnarray}

The de Sitter breaking part in (\ref{Wsplit}) of the W-type scalar
propagator is not its homogeneous part, but rather
it satisfies
\begin{equation}
\Bigl[ \frac{\square}{H^2} + D \Bigr] \delta W(y,u,v)
	= - \frac{k}{2}(D^2\!-\!1)(2\!-\!y) \equiv w(y) \; ,
\label{wDef}
\end{equation}
which can be calculated by acting with a d'Alembertian on 
(\ref{deltaW}). Furthermore,
\begin{equation}
\Bigl[ \frac{\square}{H^2} \!+\! D \Bigr] w(y) = 0 \; .
\label{wDef2}
\end{equation}
Using these relations (\ref{wDef}-\ref{wDef2}) together with 
(\ref{WNNsolution}-\ref{NNsolution})
we can derive useful identities for d'Alembertians acting on
de Sitter invariant functions (\ref{WNN}-\ref{NN}),
\begin{eqnarray}
\frac{\square}{H^2} W\!N\!N(y) &=& (\beta\!-\!D)W\!N\!N(y)
	+ \frac{W\!N(y)}{H^2} - \frac{w(y)}{\beta^2H^4} \; ,
\label{BoxWNN}
\\
\frac{\square}{H^2} W\!N(y) &=& (\beta\!-\!D)W\!N(y) 
	+ \frac{W(y)}{H^2} + \frac{w(y)}{\beta H^2} \; ,
\label{BoxWN}
\\
\frac{\square}{H^2}N\!N(y) &=& (\beta\!-\!D)N\!N(y)
	+ \frac{N(y)}{H^2} \; .
\label{BoxNN}
\end{eqnarray}

It is convenient to define a dimensionless function which is a part of the
infinite sum in the $N$-type propagator (\ref{N(y)}),
\begin{eqnarray}
\lefteqn{\overline{N}_i(y) = - \frac{\Gamma(\frac{D}{2}) \Gamma(1 \!-\! \frac{D}{2})}
		{\Gamma(\frac{1}{2} \!+\! b_N) \Gamma(\frac{1}{2} \!-\! b_N)} 
		\sum_{n=i}^{\infty} \Biggl[ 
	\frac{\Gamma(\frac{3}{2} \!+\! b_N \!+\! n) 
			\Gamma(\frac{3}{2} \!-\! b_N \!+\! n)}
		{\Gamma(3 \!-\! \frac{D}{2}+n) (n \!+\! 1)!} 
		\Bigl(\frac{y}4 \Bigr)^{n-\frac{D}{2}+2} }
\nonumber \\
& &	\hspace{4cm} -\frac{\Gamma(\frac{D\!-\!1}{2} \!+\! b_N \!+\! n) 
			\Gamma(\frac{D\!-\!1}{2} \!-\! b_N \!+\! n)}
		{\Gamma(\frac{D}{2}+n) n!} 
		\Bigl(\frac{y}4 \Bigr)^{n} \Biggr]\!\Biggr\} , \qquad
\label{Ni}
\end{eqnarray}
where the sum starts at $n=i$,
and similarly for the $N\!N$-type propagator,
\begin{equation}
\overline{N\!N}_i(y) = \frac{\partial}{\partial \beta} 
\frac{\overline{N}_i(y)}{H^2} \; . \label{NNi}
\end{equation}
The $D\rightarrow4$ limit of function (\ref{Ni}) is
\begin{eqnarray}
\lefteqn{\overline{N}_i(y) = \sum_{n=i}^{\infty}	
	\frac{\Gamma(\frac{3}{2}\!+b_N\!+\!n) \Gamma(\frac{3}{2}\!-\!b_N\!+\!n)}
		{\Gamma(\frac{1}{2}\!+\!b_N)\Gamma(\frac{1}{2}\!-\!b_N)(n\!+\!1)!n!}
	\Bigl( \frac{y}{4} \Bigr)^n }
\nonumber \\
& & \times \Biggl\{ \ln\Bigl( \frac{y}{4} \Bigr) + \frac{1}{n\!+\!1} -2\psi(n\!+\!2)
	+ \psi\Bigl( \frac{3}{2}\!+\!b_N\!+\!n \Bigr) 
	+ \psi\Bigl( \frac{3}{2}\!-\!b_N\!+\!n \Bigr) \Biggr\} , \qquad
\label{NiD=4}
\end{eqnarray}
where
\begin{equation}
\beta = \frac{4b\!-\!2}{b\!-\!2} \quad,  \qquad
	b_N = \sqrt{\frac{3(14\!-\!3b)}{4(2\!-\!b)}}
	= \sqrt{\frac{25}{4}-\beta} \; .
\end{equation}
No terms arise in this limit from expanding the $D$-dependence 
of $b_N$ or $\beta$
in (\ref{Ni}) (they all cancel). Therefore, the definition of $\overline{N\!N}_i$
(\ref{NNi}) is still valid where now (\ref{NiD=4}) is differentiated.

\end{document}